\documentclass{ws-ijmpa}

\usepackage[super]{cite}
\usepackage{xcolor}
\usepackage[verbose,hypertexnames=false]{hyperref}
\hypersetup{colorlinks=false,allbordercolors=blue,pdfborderstyle={/S/U/W 1}}

\usepackage{bm,color,xcolor,amsfonts,amsmath,amssymb,epsfig,esint,orcidlink}

\usepackage{mathtools,slashed,braket}

\newcommand{\vp}{\varphi}

\newcommand{\be}{\begin{equation}}
\newcommand{\ee}{\end{equation}}
\newcommand{\bea}{\begin{eqnarray}}
\newcommand{\eea}{\end{eqnarray}}
\newcommand{\beas}{\begin{eqnarray*}}
\newcommand{\eeas}{\end{eqnarray*}}
\newcommand{\ds}{\displaystyle}
\newcommand{\vep}{{\bm p}}
\newcommand{\vek}{{\bm k}}
\newcommand{\veq}{{\bm q}}

\newcommand{\vex}{{\bm x}}
\newcommand{\vey}{{\bm y}}
\newcommand{\vez}{{\bm z}}

\newcommand{\vegam}{{\bm \gamma}}

\newcommand{\Tr}{\textrm{Tr}}

\def\vec#1{\boldsymbol{#1}}

\newcommand{\gt}{\chi}

\newcommand{\q}{q}

\newcommand{\Tch}{T_{\rm ch}}

\newcommand{\power}{\alpha}
\newcommand{\muIR}{\mu_{\scalebox{0.6}{{\rm IR}}}}

\def\too{\mathop{\to}\limits_{\nu\to\infty}}

\graphicspath{{Figs.dir/}}

\begin{document}

\markboth{Alexey Nefediev}{Quark models: What can they teach us?}

\title{Quark models: What can they teach us?}

\author{Alexey Nefediev\orcidlink{0000-0002-9988-9430}}

\address{Helmholtz-Institut f\"ur Strahlen- und Kernphysik, Universit\"at Bonn, D-53115 Bonn, Germany\\
a.nefediev@uni-bonn.de}

\maketitle

\begin{abstract}
Quark models have a more than 60-year history and through this time they served as a powerful investigation and prediction tool in hadronic physics. In recent years, a lot of new experimental information has been arriving on hadrons that do not qualify as simple quark model states. Yet, quark models remain the cornerstone of the classification scheme for hadrons, provide valuable insights into various phenomena inherent in QCD, and facilitate gaining a clear and physically transparent picture of the underlying physics. In the spotlight of this review is a chiral quark model inspired by quantum field theory approach to confined quarks. The model is well suited for studies of spontaneous breaking of chiral symmetry in the vacuum of QCD as well as its implications in the spectrum of hadrons. It can also be employed to investigate chiral restoration at finite temperatures.
\end{abstract}


\ccode{PACS numbers: 12.38.Aw, 12.39.Ki, 11.30.Rd}

\tableofcontents

\section{Introduction}
\label{sec:int}

Quark models have a long history, with the first steps made already in the 50's of the XX century when the number of observed strongly interacting states (hadrons) became too large to regard them all as elementary particles. Then a natural expectation would be that only some found states are in fact elementary while all the others are composed of these elementary building blocks. One of the first promising attempts to proceed in this direction was undertaken by Sakata \cite{Sakata:1956hs} who suggested to regard the proton, neutron, and $\Lambda$-baryon as elementary particles. Successes of this model, which may look too na{\"i}ve in the contemporary paradigm, should not come as a big surprise. Indeed, on the one hand, the three particles selected as elementary are fermions, in line with the basic principles of the Standard Model (SM). On the other hand, these three baryons contain all three quark flavours ($u$, $d$, and $s$) known at that time. Just a few years later, in 1961, a classification scheme for hadrons based on $SU(3)$ flavour symmetry was proposed by Gell-Mann \cite{Gell-Mann:1961omu} and Ne'eman \cite{Neeman:1961jhl}. According to this scheme, all the hadrons known at that time could be classified according to the irreducible representations of this symmetry group. A clearcut outcome from this approach was the existence of a $\Omega^-$-baryon composed entirely of strange quarks. Its properties could be predicted in the framework of the suggested classification scheme, and indeed such a state was discovered experimentally in Brookhaven in 1964 \cite{Barnes:1964pd}.
The idea of elementary building blocks for hadrons was put forward in the same year by Gell-Mann \cite{Gell-Mann:1964ewy} and Zweig \cite{Zweig:1964ruk}, and the name ``quarks'' given to these hypothetical particles by Gell-Mann is widely used nowadays
contrary to the obsolete name ``aces'' proposed by Zweig. Since then and for almost 40 years, all strongly interacting states could be understood as either quark--antiquark mesons or 3-quark baryons, so the quark model was recognised as a relevant classification scheme for all hadrons. Remarkably, the existence of more complicated multiquark configurations than $\bar{q}q$ mesons and $qqq$ baryons was also anticipated by Gell-Mann and others, however, no clear experimental signature of such states had arrived by that time.

The existence of a fourth quark flavour, nowadays referred to as charm (or $c$ quark), was anticipated already in 1964 by Bjorken and Glashow \cite{Bjorken:1964gz} while the first theoretical conjecture based on its existence was put forward in 1970 \cite{Glashow:1970gm}, and the first hadron containing this new quark and known today under the name $J/\psi$ was observed in 1974 \cite{E598:1974sol,SLAC-SP-017:1974ind}. A next big step in completing the quark scheme was made by Kobayashi and Maskawa \cite{Kobayashi:1973fv} who theoretically predicted the existence of the fifth quark flavour, today known as either ``beauty'' or ``bottom'' (or simply $b$ quark), which was later confirmed experimentally by Lederman in 1977 \cite{E288:1977xhf}. The first hadron containing this new quark was $\Upsilon(1S)$ --- the lowest vector state in the spectrum of bottomonium, that is, a meson made of the bottom quark and its antiquark.
The last representative of the SM quark sector --- the ``top'' (or simply $t$)  quark --- was experimentally observed in 1994-1995 \cite{CDF:1995wbb,D0:1994wmk}. No doubt it was an essential landmark in the history of SM though by that time not many doubted of the existence of this fundamental brick of the matter.
To summarise, the quark sector of the SM comprises of six different types of quarks that are conventionally referred to as flavours: $u$ (``up''), $d$ (``down''), $s$ (``strange''), $c$ (``charm''), $b$ (``beauty''), and $t$ (``top''). The hierarchy of their masses is rather peculiar covering a broad range from just a few MeV for the lightest quarks to hundreds of GeV for the heaviest $t$ quark \cite{ParticleDataGroup:2024cfk},
\be
\begin{split}
&\mbox{light:}~~m_u\approx 2~\mbox{MeV},~~\quad m_d\approx 5~\mbox{MeV},~\quad m_s\approx 94~\mbox{MeV},\label{quarkmasses}\\
&\mbox{heavy:}~m_c\approx 1.3~\mbox{GeV},\quad m_b\approx 4.2~\mbox{GeV},\quad m_t\approx 160..170~\mbox{GeV},
\end{split}
\ee
where a given quark is conventionally regarded as light or heavy depending on whether its mass is less ($u$, $d$, $s$) or greater ($c$, $b$, $t$) than the intrinsic scale of QCD, $\Lambda_{\rm QCD}\simeq 300~\mbox{MeV}$.\footnote{A strict definition and precise determination of this scale is somewhat tricky --- see, for example, \cite{FlavourLatticeAveragingGroupFLAG:2024oxs} and the references therein.}

Strictly speaking, quantum field theory --- the cornerstone concept of the SM --- allows an uncontrolled light-quark pairs creation, so each hadron needs to be treated as a multicomponent object containing all possible combinations of quarks, antiquarks, and gluons with the given quantum numbers $J^{PC}$ that do not have an open colour. It seems to preclude using simple quantum mechanical approaches to hadrons and treating them as plain quark--antiquark or three-quark systems. In the meantime, even for hadrons composed of light quarks, higher Fock components
appear to be suppressed, so usually one can understand the properties of these hadrons restricting oneself to a single component with the least number of constituents only. Although no strict proof of this suppression exists, it is widely accepted that the quark model works surprisingly well even for light hadrons, though at a price of introducing an ill-defined concept of the quark constituent mass. As a very na{\"i}ve estimate for the constituent mass of the lightest $u$ and $d$ quarks one can simply share the mass of the nucleon around 1~GeV among the three quarks arriving at
\be
m_{u,d}^{\rm const}\simeq 300~\mbox{MeV}.
\label{mconst}
\ee
This value should be confronted with the tiny current quark masses in Eq.~\eqref{quarkmasses} to conclude on a non-trivial effect of ``dressing'' of the quarks with strong interactions that  takes place in QCD.
Meanwhile, if the given effect is taken as granted and the constituent quark mass is treated as somewhat adjustable parameter, then the resulting quark model appears quite successful in explaining and predicting the properties of hadrons. As an example, let us consider the simplest possible version of the quark model based on quantum mechanics. Then a quark--antiquark meson can be described by a centre-of-mass Schr{\"o}dinger equation,
\be
\left[2\sqrt{\vep^2+m^2}+V(r)\right]\psi(r)=M\psi(r),
\label{Salp}
\ee
where $m$ is the quark mass most naturally associated with the constituent one in Eq.~\eqref{mconst}, $\vep$ is the momentum of the quark in the centre-of-mass frame (therefore, $-\vep$ for the antiquark), and the potential $V(r)$ incorporates the spin-independent confining interaction and, in more advanced versions of the model, also spin-dependent terms. For the former, spin-independent contribution the Cornell potential,
\be
V_0(r)=\sigma r-\frac{\frac43\alpha_s}{r}+C_0,
\label{Cornell}
\ee
is said to provide a decent approximation, where $\sigma$ is known as the fundamental string tension, $\alpha_s$ is the strong coupling constant (its running with the scale is typically taken into account either explicitly or simply by considering several discrete flavour-dependent values of $\alpha_s$), $\frac43$ is the eigenvalue of the Casimir operator in the fundamental colour representation, and the constant $C_0$ providing an overall shift of the spectrum may be missing in some versions of the model. The form of the spin-dependent interaction can be deduced from the respective works of Eichten, Feynberg, \cite{Eichten:1980mw} and Gromes \cite{Gromes:1984ma}. It has to be noticed, however, that
the values of the quark masses employed in such calculations remain a stumbling block.
In particular, the above spin-dependent potentials are derived to the leading order ${\cal O}(1/m^2)$ implying that constituent quarks must be effectively heavy compared to the typical scale $\Lambda_{\rm QCD}$. This idea can be put on a somewhat more rigorous ground by using the einbein field formalism of \cite{Brink:1976uf}
and treating the values of the einbeins (inverse einbeins in the settings of the original work) as effective, level-dependent quark masses \cite{Simonov:1999qj}.
A comprehensive study of the spectrum of mesons in the models of the type \eqref{Salp} can be found in an early work \cite{Godfrey:1985xj} as well as in many later ones. Many sophisticated versions of the quark model were suggested an employed so far to a general success. As the final remark, let us mentioned that a lot of inspiration relevant for the discussion below can be gained from an early work \cite{Llewellyn-Smith:1969bcu}.

However, in 2003, after an almost 40-year history of success, the quark model was challenged by the discovery of the first unusual hadron in the spectrum of charmonium, $\chi_{c1}(3872)$ (aka $X(3872)$) \cite{Belle:2003nnu}. Although the established quantum numbers of this state $J^{PC}=1^{++}$ are compatible with its assignment as a generic $\bar{c}c$ charmonium, its observed properties are at odds with the latter. Since then, several dozen states in the hadronic spectrum of QCD with the nature and properties incompatible with a simple quark-antiquark or 3-quark assignment
have been discovered experimentally --- for a review see, for example, \cite{Brambilla:2019esw}. Such states are conventionally referred to as exotic ones as opposed to the ordinary mesons and baryons suggested by Gell-Mann and Zweig in 1964. It should be noted that the observation of such exotic states became possible after the high-energy experiments started their physical programme above the open-flavour threshold, that is, at the energies high enough to make possible light-quark pair creation (with each light quark confined then in a heavy--light meson), so that a hadron--hadron component of the wave function leaves its strong imprint in the properties of such exotic states. Thus treating exotic states as generic mesons or baryons within the quark model scheme meets severe difficulties with understanding their observed properties. Still the generic quark model approach remains the basis of the classification scheme for hadrons and provides important reference values for the properties of the hadrons with the given quantum numbers in case they had been pure quark-model states. Deviation of the measured properties of hadronic states from these reference values allows one to assess the exoticity of the object under study.

It has to be noted, in addition, that quark models based on quantum mechanics have to inevitably fail also for the lightest meson in the spectrum --- the pion. The problem can be immediately recognised by considering the mass difference between the pion and the $\rho$-meson. In a simple quantum mechanical quark model, this $^3S_1$-$^1S_0$ (in this simple estimate we disregard the $^3D_1$ component of the $\rho$-meson) splitting has to come entirely from the spin--spin interactions which are, however, not able to accommodate for the mass shift around 600~MeV for any phenomenologically adequate set of the parameters, so in such calculations the pion comes out to be unphysically heavy. More generally, the problem extends to the entire octet of the lightest pseudoscalar states including, in addition to the three pions, the four kaons and the $\eta$-meson. Indeed, these eight pseudoscalar mesons are at the same time the pseudo-Goldstone bosons of the spontaneous breaking of chiral symmetry (SBCS) in the vacuum of QCD and as such must be massless in the so-called strict $SU(3)$ chiral limit of the vanishing current quark masses, $m_u=m_d=m_s=0$. A natural question that arises then is whether or not it is possible to explore chiral physics in the framework of quarks models at all. The answer to this question is in positive if the employed quark model is based on quantum field theory and incorporates the quark spins as intrinsic degrees of freedom. Below we review in some detail one such models and discuss the progress achieved in its framework.

\section{Chiral symmetry and its realisation(s) in QCD}
\label{sec:chreal}

We start from a brief introduction to the physics of chiral symmetry, its spontaneous breaking in the vacuum of QCD, and the respective implications in the spectrum of hadrons.

A basic prerequisite for the classification scheme for hadrons suggested by Gell-Mann \cite{Gell-Mann:1961omu} and Ne'eman \cite{Neeman:1961jhl} in 1961 and implemented in the quark model by Gell-Mann \cite{Gell-Mann:1964ewy} and Zweig \cite{Zweig:1964ruk} in 1964 is the quark field,
\be
q^i\equiv
\begin{pmatrix}
q^1\\ q^2\\ q^3
\end{pmatrix}
=
\begin{pmatrix}
u\\ d\\ s
\end{pmatrix},
\label{qfield}
\ee
regarded as a vector transformed according to the fundamental representation (conventionally denoted as 3) of the flavour $SU(3)_f$ group. The vector composed of the fields of antiquarks,
\be
\bar{q}_i
\equiv
\begin{pmatrix}
\bar{q}_1\\ \bar{q}_2\\ \bar{q}_3
\end{pmatrix}
=
\begin{pmatrix}
\bar{u}\\ \bar{d}\\ \bar{s}
\end{pmatrix},
\label{qbarfield}
\ee
transforms according to the antifundamental representation $\bar{3}$. Then the flavour wave function of a hadron composed of several quark and antiquark fields transforms according to some reducible representation of the flavour group and as such can be decomposed into a direct sum of irreducible representations, for example,
\be
3\otimes\bar{3}=1\oplus 8,\quad 3\otimes 3\otimes 3=1\oplus 8\oplus 8\oplus 10,
\ee
for a generic meson and baryon, respectively.

If the difference between the masses of the $u$, $d$, and $s$ quarks can be disregarded (exact $SU(3)_f$ limit), then the Lagrangian of QCD is symmetric with respect to the infinitesimal transformation of the quark field \eqref{qfield},
\be
\delta\q^i=-i\gt^a\left(t^a\right)_j^i\q^j,\qquad i,j=1,2,3,\quad a=1..8,
\label{vec}
\ee
where $\chi$'s are 8 arbitrary constants and $t^a=\lambda^a/2$ (with $\lambda$'s for the 8 Gell-Mann flavour matrices) are $3\times 3$ traceless Hermitian matrices satisfying the $SU(3)$ algebra,
\be
[t^a,t^b]=if^{abc}t^c,
\ee
with $f^{abc}$ for the antisymmetric structure constants of the group. Then, according to the Noether theorem, a conserved current exists that can be identified with the flavoured vector current,
\be
j_\mu^a=\bar{\q}_i\gamma_\mu(t^a)^i_j\q^j,\qquad \partial^\mu j_\mu^a=0.
\label{jvec}
\ee
The corresponding charge,
\be
Q^a=\int d^3x j_0^a(x),
\label{Qa}
\ee
is a constant of motion since
\be
\frac{\partial Q^a}{\partial t}=\int d^3x \frac{\partial j_0^a}{\partial t}=
-\int d^3x\; \nabla {\bm j}^a=-\oint_{S_\infty} d{\bm S}\;{\bm j}^a=0,
\ee
where it was used that all fields vanish at the infinitely remote three-dimensional surface $S_\infty$. Therefore, the charge $Q^a$ commutes with the Hamiltonian of QCD, $[Q^a,H_{\rm QCD}]=0$.

If, in addition, the masses of the light quarks are negligible as compared to the other relevant mass scales in the problem, so it is legitimate to set $m_u=m_d=m_s=0$, then the theory possesses an additional global symmetry. Indeed, for a massless fermion, the notion of chirality (from the Greek word ``$\chi\epsilon\iota\rho$'' meaning an ``arm'') can be naturally introduced that is closely related to its helicity (the projection of the fermion spin onto its momentum): fermions with positive helicity can be regarded as right-handed while fermions with negative helicity are left-handed. Chirality can be conveniently defined through the auxiliary Dirac matrix $\gamma_5$,
\be
\gamma_5=\gamma^5=i\gamma^0\gamma^1\gamma^2\gamma^3,
\label{gamma5}
\ee
as
\be
\q=\q_R+\q_L,\qquad
\q_R=P_R\q,\qquad
\psi_L=P_L\psi,
\label{psiRL}
\ee
where the right- and left-hand projectors are
\be
P_R=\frac12(1+\gamma_5),
\qquad
P_L=\frac12(1-\gamma_5).
\label{PRL}
\ee
Since, in the limit of a massless fermion with $m=0$,
the matrix $\gamma_5$ commutes with the Dirac Hamiltonian
\be
{\cal H}_D={\bm\alpha}\cdot\vep+\beta m,\qquad {\bm\alpha}=\gamma_0{\bm\gamma},\qquad \beta=\gamma_0,
\label{DiracHam}
\ee
then so do the projectors in Eq.~\eqref{PRL}. Therefore, massless quarks preserve their chirality during the evolution in time and the theory acquires an additional symmetry with respect to the quark field transformation of the form
\be
\delta\q^i=-i\gt^{\prime a}\left(t^a\right)_j^i\gamma_5\q^j,
\ee
with 8 arbitrary constants $\chi^{\prime a}$.
Then an additional conserved current that arises according to the Noether theorem,
\be
j_{5\mu}^a=\bar{\q}_i\gamma_\mu\gamma_5(t^a)^i_j\q^j,
\qquad \partial^\mu j_{5\mu}^a=0,
\label{j5def}
\ee
is an axial vector and the corresponding axial charge,
\be
Q_5^a=\int d^3x j_{50}^a(x),
\label{Q5a}
\ee
becomes yet another constant of motion that commutes with the QCD Hamiltonian. It proves convenient to build the combinations
of the charges in Eqs.~\eqref{Qa} and \eqref{Q5a},
\be
Q_R^a=\frac12(Q^a+Q_5^a),\qquad Q_L^a=\frac12(Q^a-Q_5^a),
\label{QLR}
\ee
which both commute with the Hamiltonian of QCD and generate two independent $SU(3)$ algebras,
\be
[Q_R^a,Q_R^b]=if^{abc} Q_R^c,\qquad
[Q_L^a,Q_L^b]=if^{abc} Q_L^c,\qquad
[Q_R^a,Q_L^b]=0,
\label{algebra}
\ee
where the charges in Eq.~\eqref{QLR} were classified as right and left, respectively, according to the definition of the projectors in Eq.~\eqref{PRL}.
The symmetry of QCD $SU(3)_f\otimes SU(3)_f$ arising from the operator algebra in Eq.~\eqref{algebra} and known as chiral symmetry is approximate for massive (but sufficiently light) quarks and becomes exact in the limit of massless quarks. For this reason, the latter limit is referred to as the chiral limit of QCD.

The consideration above can be formally extended to an arbitrary number of quark flavours $N_f$. However, considering $N_f>3$ does not make any practical sense given that only 3 quark flavours can be regarded as light in comparison with the QCD scale $\Lambda_{\rm QCD}$ --- see the quark masses listed in Eq.~\eqref{quarkmasses}. On the other hand, if the splitting $m_s-m_q$, with $m_q=(m_u+m_d)/2$, can not be disregarded, then the number of light quark flavours is $N_f=2$. The corresponding approximate but quite accurate $SU(2)_f$ symmetry of QCD is known as isospin symmetry. It was suggested by Heisenberg in 1932 \cite{Heisenberg:1932dw} --- long before the invention of the $SU(3)_f$ flavour symmetry by Gell-Mann and Ne'eman in 1961.

Finally, it is instructive to consider the case of $N_f=1$ separately, which appears to be less trivial than one could na{\"i}vely expect. To approach this limit, let us first take an arbitrary number of quark flavours $N_f$ and notice that, in addition to the transformations in Eqs.~\eqref{Qa} and \eqref{Q5a}, also flavour-singlet ones,
\be
\delta\q^i=-i\gt\q^i,\qquad
\delta\q^i=-i\gt'\gamma_5\q^i,
\ee
with arbitrary constants $\chi$ and $\chi'$,
leave the Lagrangian of the theory intact. Consequently, the respective singlet vector and axial-vector currents are also conserved,
\be
\partial_\mu \Bigl[\bar{\psi}_i\gamma^\mu\psi^i\Bigr]=
\partial_\mu \Bigl[\bar{\psi}_i\gamma^\mu\gamma_5\psi^i\Bigr]=0,
\label{vecaxvrc}
\ee
which results in the $U(1)_V\otimes U(1)_A=U(1)_L\otimes U(1)_R$ symmetry of the theory. It has to be noticed, however, that both
conservation laws (Ward identities) in Eq.~\eqref{vecaxvrc} are classical, and only the first of them (vector) survives at the quantum level. The latter should not come as a surprise given that the conservation of the vector current is closely related to the charge conservation, which is a fundamental law of nature. In the meantime, the divergence of the axial-vector current acquires a non-vanishing contribution from quantum fluctuations known as the axial or ABJ (for the names of Adler, Bell, and Jackiw who observed it in 1969) anomaly \cite{Adler:1969gk,Bell:1969ts},
\be
\partial_\mu \Bigl[\bar{\psi}_i\gamma^\mu\gamma_5\psi^i\Bigr]=\frac{g^2N_f}{16\pi^2}\varepsilon^{\mu\nu\lambda\rho}\sum_{A=1}^{N_c^2-1}G_{\mu\nu}^AG_{\lambda\rho}^A,
\label{axveccurrconsNf}
\ee
with $g$ for the gauge coupling constant and $G_{\mu\nu}^A$ for the non-Abelian gauge field tensor (for a comprehensive review on anomalies in gauge theories see the book \cite{Bertlmann:1996xk}). Equation \eqref{axveccurrconsNf} implies that, at the quantum level, the singlet axial-vector current is never conserved even for massless fermions since the corresponding symmetry is explicitly broken by the mass-independent anomaly. Notice, however, that a typical scaling of the gauge coupling $g$ with the number of fundamental quark colours is $g\propto 1/\sqrt{N_c}$ \cite{tHooft:1973alw}, so the right-hand side in Eq.~\eqref{axveccurrconsNf} scales as ${\cal O}(1/N_c)$ and vanishes in the limit of an infinite number of colours, $N_c\to\infty$.
Therefore, to employ the simplest case of $N_f=1$ in the studies of chiral symmetry one needs to ensure that either gluonic fields are absent or the number of colours is infinite\footnote{It should be noted that chiral anomaly appears also in the presence of an Abelian field, with a famous example of its observable manifestation provided by the two-photon decay of the neutral pion. In what follows we assume that electromagnetic fields are absent.} --- both conditions accessible only in models. The limit $N_c\to\infty$ is also known to bring many additional simplifications since only certain subclasses of Feynman diagrams (the so-called planar diagrams) need to be summed \cite{tHooft:1973alw}.

Because of the phenomenon of confinement inherent in QCD that prevents coloured objects from existing in an open state, experimental investigations in QCD are limited to studies of colourless (``white'') hadronic states composed of quarks and gluons. Then various symmetries of the QCD Lagrangian, including chiral symmetry introduced above, should leave their imprints in the spectrum of hadrons. Indeed, for an arbitrary hadronic state $\ket{h}$ with the mass $M$ obtained as the eigenenergy of the QCD Hamiltonian for this state,
\be
H_{\rm QCD}\ket{h}=M\ket{h},
\ee
an opposite-parity hadron can be built with the help of a component of the axial charge operator (we suppress the flavour indices of the hadronic states for simplicity),
\be
\ket{h'}=Q_5^a\ket{h}.
\label{hQh}
\ee
Then, using that, in the strict chiral limit of QCD, $Q_5^a$ is a constant of motion and as such commutes with $H_{\rm QCD}$, one readily finds that
\be
H_{\rm QCD}\ket{h'}=H_{\rm QCD}Q_5^a\ket{h}=
Q_5^aH_{\rm QCD}\ket{h}=MQ_5^a\ket{h}=M\ket{h'},
\label{mpm}
\ee
that is, chiral symmetry entails that $\ket{h}$ and its opposite-parity counterpart $\ket{h'}$ must be degenerate in mass. Since the consideration above applies to any hadronic state in the spectrum of QCD then (na{\"i}vely) chiral symmetry should imply that the spectrum of hadrons consists of degenerate pairs of opposite-parity states. Such type of chiral symmetry realisation in the theory is known as the Wigner--Weyl mode. It is easy to verify, however, that the experimentally observed spectrum of hadrons does not demonstrate this feature as a general pattern \cite{ParticleDataGroup:2024cfk}.
An alternative scenario for chiral symmetry realisation in QCD is referred to as the Nambu--Goldstone mode. In this mode, relation \eqref{hQh} does not hold since, unlike in the Wigner--Weyl mode, the axial charge operator does not annihilate the QCD vacuum but creates a pseudoscalar state degenerate in mass with the vacuum, that is, a massless pseudoscalar particle. This particle is known as the Goldstone boson and the phenomenon of SBCS takes place in the vacuum \cite{Goldstone:1961eq}. In this scenario, the vacuum is not chirally symmetric
and the hadronic states built on top of it lack this symmetry, too. Then the updated symmetry scheme for QCD looks like
\be
U(N_f)_L\otimes U(N_f)_R
\mathop{\longrightarrow}^{\rm Anomaly} U(1)_V\otimes SU(N_f)_L\otimes SU(N_f)_R\mathop{\longrightarrow}^{\mbox{\tiny SBCS}} U(1)_V\otimes SU(N_f)_V,
\label{symschNf2}
\ee
and a smoking gun for this scenario is the existence of massless (very light beyond the exact chiral limit) pseudoscalar particles, one per each generator of the broken symmetry group --- see, for example, the proof in \cite{Bernstein:1974rd}. Hereinafter we refer to them all as pions ($\pi$'s) and require that
\be
\braket{0|j_{5\mu}^a(x=0)|\pi^b(p)}=if_\pi p_\mu\delta^{ab},\quad a,b=1..N_f^2-1,
\label{sbcs}
\ee
with a non-vanishing constant $f_\pi$.
The latter relations entails that
\be
\braket{0|\partial^\mu j_{5\mu}^a|\pi^b}=f_\pi m_\pi^2\delta^{ab},
\label{sbcs2}
\ee
which is conjectured to be a consequence of a more general operator relation,
\be
\partial ^\mu j_{5\mu}^a=f_\pi m_\pi^2\pi^a,
\label{PCAC}
\ee
often referred to as a partial conservation of the axial-vector current (PCAC) \cite{Adler:1964um,Weinberg:1966kf}. Among many consequences, condition \eqref{sbcs} entails the so-called Goldberger--Treiman relations \cite{Goldberger:1958vp} between the masses of the hadrons $h$ and $h'$ exchanging a pion and the pion coupling to these hadrons $g_{hh'\pi}$. Let us employ general symmetry-related arguments to derive this relation for two heavy--light mesons consisting of a heavy (static) antiquark and a light quark (see, for example, \cite{Nefediev:2006bm}). Their masses can be presented in the form
\be
M=m_{\bar{Q}}+E+\ldots,\quad M'=m_{\bar{Q}}+E'+\ldots,
\ee
where in both cases the ellipsis denotes contributions from the heavy quark kinetic energy and spin-dependent interactions suppressed as ${\cal O}(1/m_{\bar{Q}})$ \cite{Eichten:1980mw,Gromes:1984ma} as well as higher terms in the $1/m_{\bar{Q}}$ expansion. We now evaluate the matrix element of the axial-vector current between the chosen hadronic states,
\be
\langle h'|j^a_{5\mu}|h\rangle=\langle h'|j^a_{5\mu}|h\rangle_{\not{\pi}}-
\frac{4Mq_\mu f_\pi g_{hh'\pi}}{q^2-m_\pi^2+i\epsilon} \left(h^{'\dag}\frac{\tau^a}{2}h\right),
\label{A}
\ee
where $q_\mu$ is the pion momentum, $\tau^a$ stand for the isospin Pauli matrices and, to the leading order in the heavy mass, we set $M_{h'}=M_h=M$. The subscript ``$\not{\pi}$'' denotes the contribution free of the pion pole, and
$h'$, $h$ are isospin vectors representing the flavours of the corresponding mesons.
On the other hand, the matrix element in Eq.~(\ref{A}) can be parameterised in the most general form as
\be
\langle h'|j^a_{5\mu}|h\rangle = \left[(P'_\mu+P_\mu)G_A(q^2)
-(P'_\mu-P_\mu)G_S(q^2)\right]\left(h^{'\dag}\frac{\tau^a}{2}h\right),
\label{A2}
\ee
with $P'_\mu$ and $P_\mu$ for the 4-momenta of the hadrons $h'$ and $h$, respectively. Conservation of the axial-vector current in the strict chiral limit implies that the left-hand side of Eq.~\eqref{A2} vanishes after contraction with the pion momentum $q^\mu=P^{\prime\mu}-P^\mu$. Then, to the leading order in the heavy mass, one has
\be
2M(E'-E)G_A- q^2G_S=0.
\label{GT}
\ee
Therefore, from Eq.~(\ref{A}) one can conclude that, for $m_\pi=0$ and in the limit $q^2\to 0$, it is
possible to have a non-vanishing axial charge, $G_A(0)\neq 0$, if $G_S$ is identified with the pion pole contribution,
\be
\lim_{q^2\to 0}G_S(q^2)\to\frac{4Mf_\pi g_{hh'\pi}}{q^2}.
\ee
Then Eq.~(\ref{GT}) turns to the sought Goldberger--Treiman relation,
\be
\frac12(E'-E)G_A=f_\pi g_{hh'\pi}.
\label{GTh}
\ee
In what follows the latter relation will be revisited in the framework of a microscopic approach to hadrons $h$ and $h'$ provided by a chiral quark model.

Finally, it is commonly believed that the regimes of chiral symmetry realisation in QCD change if the temperature of the system increases, so chiral symmetry gets restored in the vacuum above some finite critical temperature $\Tch$ \cite{Weinberg:1974hy,Linde:1975sw,Gross:1980br,Pisarski:1983ms}. Then, while chiral symmetry is realised according to the Nambu--Goldstone scenario below $\Tch$, the theory is in the Wigner--Weyl mode at $T>\Tch$. Since a real experiment in heavy ion collisions provides only indirect evidences
for the regime change, numerical lattice calculations have to be regarded as a source of ``experimental'' information on hot QCD. In particular, it was concluded that a smooth analytic crossover takes place at the chiral restoration temperature \cite{Aoki:2006we}.

\section{Bare and dressed particles}
\label{sec:BV}

Before we proceed to a detailed investigation
of chiral symmetry, its spontaneous breaking and restoration in a suitable chiral quark model, let us discuss the effect of the interaction on the properties of fermions and bosons in general terms and introduce a convenient investigation tool known as the Bogoliubov-Valatin transformation invented about 70 years ago in relation with the theory of superconductivity \cite{Bogolyubov:1947zz,Bogolyubov:1958km,Valatin:1958ja}.

\subsection{A toy model for fermions}
\label{sec:Bogferm}

Consider first a toy model for fermions with the Hamiltonian
\be
H=\epsilon (b_0^\dagger b_0-d_0 d_0^\dagger)+\Delta(b_0^\dagger d_0^\dagger+d_0b_0),
\label{Htoy}
\ee
where $b_0$ and $d_0$ are the annihilation operators for ``bare'' (that is, defined disregarding the interaction) particles and $\epsilon$ and $\Delta$ are the two constants parameterising the diagonal (proportional to $\epsilon$) and anomalous (proportional to $\Delta$) parts of the Hamiltonian, respectively. The energy of the trivial (empty) vacuum state $\ket{0}_0$ annihilated by the operators $b_0$ and $d_0$ can be evaluated as
\be
E_{\rm vac}^{(0)}={}_0\braket{0|H|0}_0=- {}_0\braket{0|\epsilon\, d_0d_0^\dagger|0}_0=-\epsilon.
\label{vac0}
\ee
The Hamiltonian in Eq.~\eqref{Htoy} is diagonal in the limit of $\Delta=0$. However, for $\Delta\neq 0$, it has a non-vanishing matrix element between the vacuum $\ket{0}_0$ and a fermion--antifermion state $b_0^\dagger d_0^\dagger\ket{0}_0$, which implies that the effective degrees of freedom in the system are not the bare fermions but some quasiparticles (``dressed'' fermions) incorporating the effect of the fermions self-interaction due to the $\Delta$-term in the Hamiltonian~\eqref{Htoy}. Then a convenient way to proceed is the Bogoliubov-Valatin transformation that consists in reformulating the theory in terms of the effective fermions annihilated by the ``dressed'' operators,
\be
b=ub_0+vd_0^\dagger,\qquad d=ud_0-vb_0^\dagger,
\label{bandddef}
\ee
with some amplitudes $u$ and $v$. Then, to ensure that the new fermionic operators possess the standard anticommutation relations,
\be
\{bb^\dagger\}=\{dd^\dagger\}=1,
\ee
the amplitudes $u$ and $v$ should obey a constraint,
\be
u^2+v^2=1,
\ee
that can be trivially satisfied by choosing $u=\cos\theta$ and $v=\sin\theta$ with some (yet unspecified) angle $\theta$. To keep a closer contact with the chiral physics discussed in the next sections, we refer to this parameter as the ``chiral angle''.

It is straightforward then to arrange a normal ordering of the Hamiltonian in Eq.~\eqref{Htoy} in terms of the operators $b$ and $d$ and their Hermitian conjugates,
\be
H=H_0+:H_2:,
\ee
where
\be
H_0=-\epsilon\cos2\theta-\Delta\sin2\theta,
\ee
\be
:H_2:=(\epsilon\cos2\theta+\Delta\sin2\theta)(b^\dagger b+d^\dagger d)+
(\Delta\cos2\theta-\epsilon\sin2\theta)(b^\dagger d^\dagger+db).
\ee

Now there exist two equivalent ways to proceed. On the one hand, one can require that the Hamiltonian $:H_2:$ should be diagonal, that is, the coefficient at the anomalous Bogoliubov term $b^\dagger d^\dagger+db$ should vanish. The constraint that emerges is a non-linear equation for the chiral angle $\theta$,
\be
\Delta\cos2\theta-\epsilon\sin2\theta=0,
\label{mgtoy}
\ee
with the solution
\be
\theta=\frac12\arctan\frac{\Delta}{\epsilon}.
\ee

Alternatively, one can require that the energy of the new vacuum annihilated by the operators in Eq.~\eqref{bandddef},
\be
E_{\rm vac}={}\braket{0|H|0}=H_0,
\label{vac00}
\ee
should be a minimum, that is,
\be
\frac{\partial H_0}{\partial\theta}=0.
\ee
It is easy to see that this way one readily re-derives Eq.~\eqref{mgtoy}.

The new vacuum state normalised to unity and annihilated by the dressed fermion and antifermion operators $b=Ub_0U^\dagger$ and $d=Ud_0U^\dagger$ is
\be
\ket{0}=U\ket{0}_0=\left(\cos\frac{\theta}{2}+b_0^\dagger d_0^\dagger \sin\frac{\theta}{2}\right)\ket{0}_0,
\label{newvactoy}
\ee
where
\be
U=e^{Q^\dagger-Q},\qquad Q^\dagger=\frac12\theta\; b_0^\dagger d_0^\dagger.
\ee
The energy of the new vacuum is
\be
E_{\rm vac}=-\sqrt{\epsilon^2+\Delta^2}<E_{\rm vac}^{(0)},
\label{vac1}
\ee
so it is energetically favourable for the system to proceed from the trivial vacuum with the energy in Eq.~\eqref{vac0} to the vacuum in Eq.~\eqref{newvactoy} possessing a lower vacuum energy \eqref{vac1}.

\subsection{A toy model for bosons}

The situation for bosons is ideologically similar to that for fermions though its treatment somewhat differs in details. Consider a toy Hamiltonian,
\be
H=h_1a_0^\dagger a_0+\frac12h_2(a_0^\dagger a_0^\dagger+a_0a_0),\quad a_0\ket{0}_0=0,
\label{Haa}
\ee
with $|h_2|<|h_1|$, that admits a bosonic Bogoliubov-Valatin transformation from bare to dressed bosons annihilated by the operator $a$,
\be
a=ua_0-va_0^\dagger,
\qquad a\ket{0}=0,
\label{BVa}
\ee
with a convenient parameterisation,
\be
u=\cosh\theta,\qquad v=\sinh\theta,
\ee
to ensure the standard bosonic commutation relation for the operators $a$ and $a^\dagger$,
\be
[aa^\dagger]=u^2-v^2=1.
\label{uvnorm}
\ee
The Hamiltonian in Eq.~\eqref{Haa} normally ordered in terms of the dressed operators $a$ and $a^\dagger$ reads
\be
H=H_0+:H_2:,
\ee
with
\be
H_0=-\frac12h_1+\frac12\left(h_1\cosh2\theta+h_2\sinh2\theta\right),
\ee
\be
:H_2:=(h_1\cosh2\theta+h_2\sinh2\theta)a^\dagger a+
\frac12(h_1\sinh2\theta+h_2\cosh2\theta)(a^\dagger a^\dagger+aa).
\ee
Then, in a full analogy to the case of fermions, one can either minimise the vacuum energy $H_0$ with respect to the angle $\theta$ or require that the coefficient at the anomalous term $a^\dagger a^\dagger+aa$ in $:H_2:$ should vanish. In both cases one arrives at the same equation for the angle $\theta$,
\be
h_1\sinh2\theta+h_2\cosh2\theta=0,
\ee
with the solution
\be
\theta=-\frac12\operatorname{artanh}\frac{h_2}{h_1}.
\ee

Then the Hamiltonian \eqref{Haa} takes a diagonal form in terms of the dressed operators,
\be
H=\frac12\left(\sqrt{h_1^2-h_2^2}-h_1\right)+\sqrt{h_1^2-h_2^2}\;a^\dagger a,
\ee
and the new vacuum has the energy,
\be
E_{\rm vac}=\braket{0|H|0}=\frac12\left(\sqrt{h_1^2-h_2^2}-h_1\right)<0,
\ee
that is lower than the energy of the trivial vacuum, $E_{\rm vac}^{(0)}={}_0\langle 0|H|0\rangle_0=0$.

Thus, from the two toy models studied above, one is led to conclude that (i) the empty vacuum may be unstable and it is energetically favourable for the system to proceed to a nontrivial vacuum that possesses a lower energy, (ii) the new vacuum is not empty but contains ``condensed'' pairs of bare particles and antiparticles, and (iii) the Bogoliubov-Valatin transformation provides a convenient framework for studies of such systems.

\section{Spontaneous breaking and restoration of chiral symmetry in the vacuum}

\subsection{Chiral quark model}
\label{sec:model}

We start from a brief summary of Sec.~\ref{sec:chreal}. In the exact chiral limit of QCD for $N_f$ massless quark flavours, the Lagrangian is symmetric with respect to the transformations from the chiral group,
\be
SU(N_f)_L\otimes SU(N_f)_R,
\label{ch}
\ee
that can be realised in the spectrum of hadrons either in the Nambu--Goldstone or Wigner--Weyl way. In the former case, the symmetry in Eq.~\eqref{ch} is spontaneously broken by the vacuum down to $SU(N_f)_V$ and the spectrum of hadronic states built on top of this broken vacuum lacks implications of chiral symmetry.
As an inevitable consequence of this scenario, $N_f^2-1$ massless pseudoscalar particles --- Goldstone bosons --- must appear. Since quarks in QCD are confined and only their bound states can be observed experimentally, these Goldstone bosons are to be identified as the lowest pseudoscalar mesons in the spectrum of the theory. If quarks are not strictly massless but appear to be light compared with the typical scale of strong interactions $\Lambda_{\rm QCD}$, chiral symmetry in Eq.~\eqref{ch} survives as an approximate but rather accurate symmetry of the theory. The pseudoscalar bosons related to its spontaneous breaking (on top of the slight explicit breaking by the quark masses) are not strictly massless in this case either, though their masses are ``unnaturally'' small compared to the other, ``ordinary'', hadrons.
For this reason, they are often referred to as pseudo-Goldstone bosons. For $N_f=3$ they can be identified with the three pions ($\pi^0$, $\pi^\pm$), four kaons ($K^0$, $\bar{K}^0$, $K^\pm$), and one $\eta$-meson. $SU(3)_f$ symmetry is stronger explicitly broken by the strange quark mass, $m_s\gg (m_u+m_d)/2$, than the isospin symmetry $SU(2)_f$ is broken by the non-vanishing mass difference $m_d-m_u$, so it should not come as a surprise that the physical kaons and $\eta$ are heavier than the pions. Nevertheless, all eight pseudoscalar mesons above would become massless in the strict $SU(3)_f$ chiral limit of $m_u=m_d=m_s=0$. Therefore,
these pseudo-Goldstone modes play a distinguished role in strong interactions, especially at low energies where they represent the most relevant degrees of freedom in the effective low-energy theory for QCD known as the Chiral Perturbation Theory --- for further details see, for example, the book \cite{Meissner:2022cbi} and the references therein. On the contrary,
in the Wigner--Weyl scenario, chiral symmetry is manifest in the vacuum and, therefore, in the spectrum of hadrons, too. As a consequence, opposite-parity hadronic states come degenerate in the mass. A change in the regime of chiral symmetry realisation in QCD takes place if the system is heated up to some critical temperature $\Tch$ or the baryon density is high enough (see, for example, the review \cite{Hatsuda:1994pi}).

Unfortunately, direct experimental studies of the phenomena related to chiral symmetry breaking and restoration are obviously impeded while a ``numerical'' experiment on the lattice is susceptible to technical complications that result in various truncations. Under such circumstances, a natural question to be asked is whether or not an insight into these properties of QCD can be gained in the quark model framework. As should be clear from the consideration above, the notion of spin is crucially important for addressing chiral physics, so quark models based on quantum mechanics and briefly mentioned in the Introduction have to inevitably fail. Therefore, a suitable quark model should be based on quantum field theory and incorporate spinning quarks. On the other hand, it may have gluons effectively integrated out (if the effects related to deconfinement are not addressed) but, importantly, must contain confinement as an intrinsic feature. To elaborate more on this important issue consider a physical system consisting of a quark $q$ of the mass $m$ interacting with a static antiquark source $\bar{Q}$ with $M_{\bar{Q}}\to\infty$. Following \cite{Simonov:1997et,Brambilla:1997kz}, let us study the Green function of this system in Euclidean space,
\be
S_{q\bar Q}(x,y)=\frac{1}{N_c}\int D{\q}D{\q^\dagger}DA_{\mu}
\Bigl[\q^\dagger(x) S_{\bar Q} (x,y|A)\q(y)\Bigr]e^{-S[q,q^\dagger,A]},
\label{SqQ}
\ee
with $S_{\bar Q} (x,y|A)$ for the propagator of the static antiquark (for convenience placed at the origin) and $S[q,q^\dagger,A]$ in the exponent for the action of the theory. The presence in the system of a static antiquark with a straight-line propagation trajectory along the 4th (temporal) direction in the Euclidean space allows one to considerably simplify the approach to the system by fixing a particular gauge for the gluonic field. Indeed, by choosing a modified Fock--Schwinger gauge \cite{Balitsky:1985iw},
\be
\vec{x}\vec{A}(x_4,\vec{x})=0,\quad A_4(x_4,\vec{0})=0,
\label{FSg}
\ee
one can ensure that the gluonic field always vanishes at the straight-line trajectory of the static antiquark, so its Green function takes a simple $A$-independent form,
\be
S_{\bar Q}(x,y|A)= i\frac{1-\gamma_4}{2} \theta (x_4-y_4) e^{-M_{\bar{Q}}(x_4-y_4)}+i\frac{1+\gamma_4}{2}\theta(y_4-x_4)e^{-M_{\bar{Q}}(y_4-x_4)}.
\label{SQ}
\ee
Then the integration over the gluons in Eq.~\eqref{SqQ} can be easily performed yielding
\be
S_{q\bar Q}(x,y)=\frac{1}{N_c}\int D{\q}D{\q^\dagger}e^{-S_{\rm eff}[\q,\q^\dagger]}\q^\dagger(x) S_{\bar Q} (x,y)\q(y),
\ee
with $S_{\rm eff}[\q,\q^\dagger]$ for an effective actions for the light quark moving in the field of the static antiquark source,
\be
\begin{split}
&S_{\rm eff}[\q,\q^\dagger]=\int d^4x\q^\dagger_{\alpha}(x)(-i\hat\partial -im)\q^{\alpha}(x)\\
&+\sum_{n=1}^\infty\frac{1}{n!}
\int dx_1\ldots dx_n
J_{\alpha_1;\mu_1}^{\beta_1}(x_1)\ldots J_{\alpha_n;\mu_n}^{\beta_n}(x_n)
\braket{
A_{\beta_1;\mu_1}^{\alpha_1}(x_1)\ldots A_{\beta_n;\mu_n}^{\alpha_n}(x_n)
},
\label{Seff}
\end{split}
\ee
where $J_{\alpha;\mu}^\beta(x)=\bar{q}_\alpha(x)\gamma_\mu q^\beta(x)$,
all $\alpha$'s and $\beta$'s are fundamental colour indices, and
$\braket{A_{\beta_1;\mu_1}^{\alpha_1}(x_1)\ldots A_{\beta_n;\mu_n}^{\alpha_n}(x_n)}$ is an irreducible correlator of the gluonic fields of the order $n$. Notice that the lowest correlator vanishes due to the Lorentz and gauge invariance of the vacuum, $\langle {A_{\mu}}^{\alpha}_{\beta}\rangle=0$.
Then a Gaussian dominance of the vacuum is assumed, which implies that only the second-order correlator $\braket{AA}$ is retained while all higher terms, starting from $\braket{AAA}$, are neglected \cite{Dosch:1987sk,Nachtmann:1991ua,Simonov:1999gk,DiGiacomo:2000irz}. In this approximation,
the Casimir scaling becomes exact implying that the ratio of two potentials between static colour sources in different representations of the $SU(N_c)$ group is given by the ratio of the corresponding
Casimir operators \cite{Ambjorn:1984dp,Shevchenko:2000du}. A high accuracy of this approximation was verified by lattice calculations \cite{Campbell:1985kp,Bali:2000un}.
As a result, the only term retained in the sum in Eq.~\eqref{Seff} contains the bilocal (Gaussian) correlator of the form
\be
\braket{A_{\beta;\mu}^\alpha(x)A_{\delta;\nu}^\gamma(y)}=\delta_{\delta}^{\alpha}\delta_{\beta}^{\gamma}K_{\mu\nu}(x,y),
\ee
where the Fierz formula $(t_c^A)_\beta^\alpha (t_c^A)_\delta^\gamma=\frac12
\delta_\delta^\alpha\delta_\beta^\gamma-\frac{1}{2N_c}\delta_\beta^\alpha\delta_\delta^\gamma$
for the generators of the fundamental representation of the colour group $SU(N_c)$ was employed (we use subscript $c$ here to avoid confusion with the flavour $SU(N_f)$ group) and only the first term, not suppressed in the limit $N_c\to\infty$, was retained.\footnote{Thus, hereinafter a large-$N_c$ limit is implied.\label{Nc1}}
Then the effective action in Eq.~\eqref{Seff} can be written in the form
\be
\begin{split}
S_{\rm eff}[\q,\q^\dagger]&=\int d^4x\;\q^\dagger_{\alpha}(x)(-i\hat\partial -im)\q^{\alpha}(x)\\
&+\frac{1}{2}\int d^4x d^4y\; \q^\dagger_{\alpha}(x)\gamma^\mu\q^{\beta}(x)K_{\mu\nu}(x,y)\q^\dagger_{\beta}(y)\gamma^\nu\q^{\alpha}(y),
\label{Seff2}
\end{split}
\ee
which leads to a nonlinear Schwinger--Dyson-type equation \cite{Simonov:1997et,Brambilla:1997kz},
\be
(-i\hat{\partial}_x-im)S(x,y)+\int d^4zK_{\mu\nu}(x,z)\gamma_{\mu}S(x,z)\gamma_{\nu}S(z,y)=\delta^{(4)}(x-y),
\label{DS4}
\ee
for the colour trace of the light-quark Green's function $S(x,y)=\frac{1}{N_c}\langle\q^{\beta}(x)\q^\dagger_{\beta}(y)\rangle$.

A central object of Eq.~\eqref{DS4} is the quark interaction kernel $K_{\mu\nu}(x,y)$. We notice that a specific property of the entire family of the Fock--Schwinger-like (or so-called radial) gauges is the possibility to express the gluonic field in terms of the gluonic field tensor and, as a result, express correlators of the gluonic fields through gauge invariant correlators of the field tensors \cite{Leupold:1996hx}. In the gauge \eqref{FSg}, such a relation reads
\begin{eqnarray}
\ds A^A_4(x_4,\vec x)&=&\int_0^1 d\alpha x_i F^A_{i4}(x_4,\alpha\vec{x}),\nonumber\\[-3mm]
\label{AF}\\[-3mm]
\ds A^A_i(x_4,\vec x)&=&\int_0^1\alpha x_k F^A_{ki}(x_4,\alpha\vec{x}) d\alpha,\quad i=1,2,3,\nonumber
\end{eqnarray}
so that the kernel $K_{\mu\nu}$ can be finally expressed in terms of the Gaussian field strength
correlator $\langle F^A_{\mu\nu}(x)F^B_{\lambda\rho}(y)\rangle$ that admits a convenient parameterisation \cite{DiGiacomo:2000irz}\footnote{For recent developments in the Field Correlators Method see \cite{Lukashov:2025pvd}.},
\be
\langle F^A_{\mu\nu}(x)F^B_{\lambda\rho}(y)\rangle=\frac{\delta^{AB}}{N_c^2-1}D(x-y)
(\delta_{\mu\lambda}\delta_{\nu\rho}-\delta_{\mu\rho}\delta_{\nu\lambda}),
\label{15}
\ee
where only the term responsible for confinement is retained on the right-hand side. The decrease of the profile function $D(z_0,|\vez|)$ in all directions of the Euclidean space is governed by the gluonic correlation length $\lambda$ that is quite small. Indeed, lattice calculations give the value of $\lambda\approx 0.2\div 0.3$ fm \cite{Campostrini:1984xr,Campostrini:1986hy,Bali:1997aj,Koma:2006fw} while more recent theoretical estimates provide an even smaller value around or less than 0.1~fm \cite{Badalian:2008kc,Simonov:2018cbk}, so the so-called string limit of $\lambda\to 0$ can be employed. In this limit, the profile function takes a simple form
$D(\tau,r)=2\sigma\delta(\tau)\delta(r)$ trivially satisfying the normalisation condition for the Wilson loop area law with the string tension \cite{DiGiacomo:2000irz}
\be
\sigma=2\int_0^\infty d\tau\int_0^\infty dr D(\tau,r).
\label{sigma}
\ee

Then, by virtue of Eqs.~\eqref{AF} and \eqref{15}, the non-vanishing components of the quark kernel $K_{\mu\nu}(x,y)=K_{\mu\nu}(\tau,\vex,\vey)$, with $\tau=x_4-y_4$, are found to be
\be
\begin{split}
&K_{44}(\tau,\vex,\vey)=(\vex\vey)\int_0^1d\alpha\int_0^1 d\beta D(\tau,|\alpha\vex-\beta\vey|),\\
&K_{ik}(\tau,\vex,\vey)=((\vex\vey)\delta_{ik}-y_ix_k)\int_0^1\alpha d\alpha\int_0^1 \beta d\beta
D(\tau,|\alpha\vex-\beta\vey|).
\end{split}
\label{kern1}
\ee

In what follows, the spatial part of the quark kernel will be disregarded for simplicity. Then the only remaining (temporal) component reads
\be
K(\tau,\vex,\vey)\equiv K_{44}(\tau,\vex,\vey)=2\sigma\delta(\tau)(\vex\vey)\int_0^1d\alpha\int_0^1 d\beta\;\delta(|\alpha\vex-\beta\vey|).
\label{kern2}
\ee
It is easy to see that the kernel $K(\tau,\vex,\vey)$ does not vanish only for collinear vectors $\vex$ and $\vey$ and, in this case \cite{Nefediev:2007pc},
\be
K(\tau,\vex,\vey)=\delta(\tau)\sigma (|\vex|+|\vey|-|\vex-\vey|)\equiv\delta(\tau)K(\vex,\vey).
\label{K4v4}
\ee

The kernel in Eq.~\eqref{K4v4} possesses several remarkable properties. On the one hand, it is instantaneous (proportional to a $\delta$-function in the relative time). On the other hand, it vanishes for non-collinear vectors $\vex$ and $\vey$. Both above features are quite natural in the string limit of $\lambda\to 0$ when the interaction between quarks is provided by an infinitely thin string-like object formed by nonperturbative gluons. To proceed to a model suitable for theoretical investigation we relax the requirement of collinearity. In addition, the structure of the kernel in Eq.~\eqref{K4v4} admits a straightforward generalisation from linear confinement $\sigma r$ to an arbitrary confining potential $V(r)$. Then the quark interaction kernel ultimately reads\footnote{In \cite{Bicudo:1998bz}, the coordinate dependence of the interaction kernel in Eq.~\eqref{K4v4} was approximated by $(\vex\vey)$ that corresponds to a particular assumption for the profile function $D$ in Eq.~\eqref{kern1}. Then the decomposition $(\vex\vey)=\frac12\left(\vex^2+\vey^2-(\vex-\vey)^2\right)$ in the spirit of Eq.~\eqref{KV} can be applied to arrive at a harmonic oscillator confining potential $V(r)\propto r^2$.}
\be
K(\vex,\vey)=V(|\vex|)+V(|\vey|)-V(|\vex-\vey|).
\label{KV}
\ee

Notice that the actual form of the kernel in Eq.~\eqref{KV} is specific for the gauge choice in Eq.~\eqref{FSg}. Indeed, in this gauge, the static antiquark decouples from the system, so the resulting Eq.~\eqref{DS4} takes the form of a single-particle equation for the light quark. However, it is not the case since the latter equation describes the evolution of the light quark interacting with the static antiquark source placed at the origin. Since strong interactions are blind to the quark flavour, the kernel \eqref{KV} simultaneously incorporates both dressing of the light quark with nonperturbative gluons (encoded in the translationally invariant term $V(|\vex-\vey|)$) and the interaction of the dressed light quark with the static antiquark (as given by the translationally non-invariant terms $V(|\vex|)$ and $V(|\vey|)$). For further details of disentangling the above two contributions see \cite{Nefediev:2007pc,Kalashnikova:2017ssy}. A similar study of a heavy--light quark--antiquark system in the two-dimensional 't~Hooft model for QCD \cite{tHooft:1974pnl} can be found in \cite{Kalashnikova:1997tb,Kalashnikova:2001df}.

Now, when the interquark interaction kernel is identified, we introduce a model in Minkowski space often referred to in the literature as the Generalised Nambu--Jona-Lasinio (GNJL) model,\footnote{Hereinafter the fundamental colour indices for the quark fields are suppressed to simply notations. They will be restored in some formulae below.\label{foot:Nc}}
\be
\begin{split}
H_{\rm GNJL}=
\int d^3x\;\q^\dagger(\vex,t)&\left(-i\vec{\alpha}\cdot
{\bm\nabla}+\beta m\right)\q(\vex,t)\\
&+\frac{1}{2} \int d^3x\; d^3y\;\rho^A(\vex)K_{AB}^{(0)}(|\vex-\vey|)\rho^B(\vey),
\label{GNJL}
\end{split}
\ee
where for simplicity we consider only one (light) quark flavour (see also the discussion in the end of this subsection) and include the interaction of two quark colour charge densities,
$\rho^A=\q^\dag\frac{\lambda_c^A}{2}\q$ and $\rho^B=\q^\dag\frac{\lambda_c^B}{2}\q$, taken at the spatial points $\vex$ and $\vey$, respectively, and interacting via an instantaneous confining kernel,
\be
K_{AB}^{(0)}(|\vex-\vey|)=\delta_{AB}V_0(|\vex-\vey|),\quad A,B=1..N_c^2-1.
\label{Kab}
\ee
Then the confining potential in the fundamental representation for the quarks introduced in Eq.~\eqref{KV} reads
\be
V(r)=C_FV_0(r),
\label{Vlin}
\ee
with $C_F=(N_c^2-1)/(2N_c)$ for the eigenvalue of the fundamental Casimir operator.
It should be noted that the model in Eq.~\eqref{GNJL} can also be related to QCD in the Coulomb gauge \cite{Christ:1980ku,Szczepaniak:2001rg,Feuchter:2004mk,Reinhardt:2017pyr,Nguyen:2024ikq}, if only the Coulomb part of the interaction,
\be
H_C=\frac{g^2}2\int d^3\;x d^3 y\;J^{-1}\rho^A(\vex)F_{AB}(\vex,\vey) J\rho^B(\vey),
\label{coul}
\ee
is retained, the Faddeev-Popov determinant $J$ is set to unity, and the colour charge density
$\rho^A(\vex)$ is assumed to be saturated by the quarks only, without gluons. Then it is sufficient to parameterise the ``Coulombic'' interaction kernel $F_{AB}(\vex,\vey)$ in a suitable form --- for example, through a confining potential in Eq.~\eqref{Kab}.

The GNJL model in Eq.~\eqref{GNJL} does not provide a systematic approach to QCD but proves useful in studies of the phenomenon of SBCS. The model has a long and rich history --- see, for example, \cite{Amer:1983qa,LeYaouanc:1983it,LeYaouanc:1983huv,LeYaouanc:1984ntu,Adler:1984ri,Kocic:1985uq,Bicudo:1989sh,Bicudo:1989si,Bicudo:2002eu,Llanes-Estrada:1999nat,Bicudo:2003cy,Nefediev:2004by,Alkofer:2005ug,Wagenbrunn:2007ie,Bicudo:2008kc} and references therein. Importantly, an explicit form of the quark interaction potential does not change the qualitative argument, so it can either be kept in the most general form as $V_0(r)$ or a suitable particular
shape can be assumed to facilitate numerical calculations. It is only crucial that the potential is confining and the kernel in Eq.~\eqref{Kab} introduces a physical scale with the dimension of mass. These properties are evident for the most popular choice for $V_0(r)$ as a power-like function of the distance between the quarks,
\be
V_0(r)=K_0^{\power+1}r^\power,
\label{Vconf}
\ee
with $K_0$ for the aforementioned physical scale with the dimension of mass. The powers range $-1<\power<3$ was studied in the literature, with $\power\to-1$ approaching the colour Coulomb potential (for similar studies of the non-confining Coulomb potential with $
\alpha=-1$ see \cite{Bicudo:2008kc}) and $\power\geqslant 3$ (more generally, $\power>d$, with $d$ for the number of spatial dimensions of the space-time) resulting in intractable infrared singularities. A logarithmic potential arises in the limit of $\power\to 0$ if an appropriate modification of the potential is made,
\be
V_0(r)\to V^{\rm mod}_0(r)=\lim_{\power\to 0}\frac{K_0}{\power}\Bigl[(K_0r)^\power-1\Bigr]=K_0\log(K_0r).
\ee
For $\alpha=2$, the Fourier transform of the potential to the momentum space is given by the Laplas operator applied to a 3D $\delta$-function. As a result, all integral equations that will be derived and discussed below turn to second-order differential equations that are easier tractable technically. For this reasons, the harmonic oscillator potential with $\alpha=2$ is a popular choice in the literature. However, the most phenomenologically adequate choice of the confining interaction is provided by the linear function with $\power=1$. In this case, the GNJL model shares many common features with the 't~Hooft model for QCD in two dimensions \cite{tHooft:1974pnl,Bars:1977ud,Li:1986gf,Kalashnikova:2001df,Glozman:2012ev}.\label{tHooft}

Notice that GNJL is intrinsically a large-$N_c$ model (in particular, see footnotes \ref{Nc1} and \ref{Nc2}) although in practical calculations one typically sets $N_c=3$. Then it is assumed that the potential in Eq.~\eqref{Vlin} reaches a finite limit as $N_c\to\infty$. For instance, for the potential in Eq.~\eqref{Vconf} it implies that $K_0$ scales as $N_c^{1/(\power+1)}$ for large $N_c$.

It is instructive to notice that the action in Eq.~\eqref{Seff2} resembles in spirit an early quantum-field-theory-inspired quark model suggested by Nambu and Jona-Lasinio (NJL model) in 1961 \cite{Nambu:1961tp,Nambu:1961fr}. There are, however, several essential differences. On the one hand, the quark interaction kernel in the NJL model is provided by a $\delta$-function with a dimensionless coefficient in front of it. For this reason, the model lacks confinement and all dimensional quantities in it acquire their dimensions from the regulator that needs to be introduced to the model to tame its large-momentum behaviour. On the contrary, in the effective action in Eq.~\eqref{Seff2} and Hamiltonian \eqref{GNJL}, the interaction kernel is explicitly confining and provides a mass scale (through the string tension parameter $\sigma$ or, for the potential of a generic power-like form, through the parameter $K_0$). On the other hand, the Lorentz nature of the 4-fermion interaction in the NJL model is given by a particular, tuned to be consistent with chiral symmetry of the full Lagrangian, mixture of the scalar and pseudoscalar terms with the structure in terms of the Dirac matrices being $\hat{1}\otimes\hat{1}$ and $\gamma_5\otimes\gamma_5$, respectively. Notice, however, that the interquark interaction in QCD is mediated by vector gluons, so the Lorentz nature of the interaction in Eq.~\eqref{Seff2}, given by $\gamma^\mu\otimes\gamma^\nu$, has a closer contact to QCD. Nevertheless, the NJL model appeared quite successful in explaining the effect of SBCS in the vacuum of QCD and related phenomena, so it was a very popular subject in the literature for many years ---
for a comprehensive review and various developments of the model see, for example, \cite{Bernard:1987sg,Vogl:1991qt,Klevansky:1992qe,Hatsuda:1994pi,Buballa:2003qv,Volkov:2005kw}.

We stress that the consideration in this section can not be regarded as a rigorous derivation or proof of validity of the GNJL quark model in Eq.~\eqref{GNJL}. Instead, the goal of the discussion above was to establish a connection between the models of this kind and full QCD and give the reader an idea of approximations and simplifications that need to be done to arrive at this relatively simple and tractable but still realistic model. Thus, following the famous claim by George Box that ``all models are wrong, but some are useful,'' below we shall argue that this model is indeed useful for understanding the phenomenon of SBCS in the QCD vacuum and providing its ``microscopic'' picture. In particular, it allows one to address the dual nature of the chiral pion and the way it appears in the model as a Goldstone boson of SBCS and the lightest quark--antiquark state in the spectrum of hadrons at the same time. Meanwhile, one final remark concerning this model is in order here.
It should be clear from the discussion above that the model lacks explicit gluonic degrees of freedom. This shortcoming of the model gives two advantages in the studies addressed in this review. On the one hand, the effects of chiral symmetry breaking and, specifically,
the chiral restoration transition at finite temperatures can be studied without connection with the deconfinement phenomenon that is also believed to take place in full QCD at some finite temperature $T_d$. Also, once the employed quark model
does not include coupling of quarks to gluonic and electromagnetic fields, the chiral anomaly does not show up, so it is sufficient to set $N_f=1$ and study the chiral group $U(1)_L\otimes U(1)_R$ and its spontaneous breaking down to $U(1)_V$, with a generalisation to the case of $N_f\geqslant 2$ being a straightforward task.

\subsection{Gap in the spectrum and chirally broken vacuum}
\label{sec:BCS}

Now, when the stage is set up, all necessary aspects of the technique are discussed, and relevant disclaimers are made, we can turn to investigations of chiral symmetry implications in QCD employing the chiral quark model in Eq.~\eqref{GNJL}. We shall mainly follow the lines of \cite{Bicudo:1989sh,Bicudo:1989si,Nefediev:2004by}. The central idea of the entire approach is an expectation that the self-interaction of quarks described by the confining kernel in the Hamiltonian \eqref{GNJL} will lead to an instability of the trivial (empty) vacuum, so it will be energetically favourable for the system to proceed to an alternative vacuum state that lacks chiral symmetry. Indeed, such instability for a generic confining potential was demonstrated in the pioneering works by the theoretical group in Orsay \cite{Amer:1983qa,LeYaouanc:1983huv} and later confirmed by many other authors. Below we discuss the aforementioned effect in some detail.

In the absence of interaction, $K_{AB}^{(0)}(|\vex-\vey|)=0$, the Hamiltonian in Eq.~\eqref{GNJL} describes ``free'' quarks obeying a free Dirac equation, with its solution presented in a standard form,
\be
q^{(0)}(\vex,t)=\sum_{s=\pm1/2}\int\frac{d^3p}{(2\pi)
^3}e^{i\vep\vex}\Bigl[e^{-iE_p^{(0)}t}b_s^{(0)}(\vep)u_s^{(0)}(\vep)
+e^{iE_p^{(0)}t}d_s^{(0)\dagger}(-\vep)v_s^{(0)}(-\vep)\Bigr].
\label{psi0}
\ee
Here $b^{(0)}$ and $d^{(0)}$ are the quark and antiquark operators annihilating the trivial vacuum $\ket{0}_0$, $E_p^{(0)}=\sqrt{\vep^2+m^2}$ is the free dispersion law, and the free quark and antiquark bispinor amplitudes read
\be
\begin{split}
u_s^{(0)}(\vep)&=\frac{1}{\sqrt{2}}\left[\sqrt{1+\sin\vp_p^{(0)}}+
\sqrt{1-\sin\vp_p^{(0)}}\;(\vec{\alpha}\hat{\vep})\right]u_s(0),\\
v_{-s}^{(0)}(-\vep)&=\frac{1}{\sqrt{2}}\left[\sqrt{1+\sin\vp_p^{(0)}}-
\sqrt{1-\sin\vp_p^{(0)}}\;(\vec{\alpha}\hat{\vep})\right]v_{-s}(0),
\label{uandv0}
\end{split}
\ee
where the rest-frame bispinors are defined as
\be
u_s(0)=
\begin{pmatrix}
w_s \\ 0
\end{pmatrix}
,\qquad
v_{-s}(0)=-i\gamma^2 u_s^*(0)=
\begin{pmatrix}
0\\
i\sigma_2 w_s^*
\end{pmatrix},
\label{spinors}
\end{equation}
with $\gamma^2(\sigma_2)$ for the second Dirac(Pauli) matrix, $s=\pm 1/2$ for the quark helicity, and $(w_s)_{s'}=\delta_{ss'}$ for the rest-frame two-component spinor. For future convenience, we introduced the angle $\vp_p^{(0)}$ (in what follows referred to as the chiral angle) such that
\be
\sin\vp_p^{(0)}=\frac{m}{E_p^{(0)}},\quad \cos\vp_p^{(0)}=\frac{p}{E_p^{(0)}}.
\label{vpfree}
\ee
In particular, in the chiral limit of $m=0$, one has $\vp_p^{(0)}\equiv 0$.
Notice that the factor $1/\sqrt{2E_p^{(0)}}$, that often appears explicitly in the spectral decomposition in Eq.~\eqref{psi},
is included here in the definition of the bispinors in Eq.~\eqref{uandv0}, so the latter are normalised as
\be
\bar{u}_s^{(0)}(\vep)u_s^{(0)}(\vep)=-\bar{v}_s^{(0)}(-\vep)v_s^{(0)}(-\vep)=\frac{m}{E_p^{(0)}}.
\ee

Then, once the quarks start to interact through the confining kernel $K_{AB}^{(0)}(|\vex-\vey|)=0$, the effect of their ``dressing'' can be conveniently described in terms of the chiral angle $\vp_p$ that now deviates from its free form in Eq.~\eqref{vpfree}. Therefore,
\be
q(\vex,t)=\sum_{s=\pm1/2}\int\frac{d^3p}{(2\pi)
^3}e^{i\vep\vex}\Bigl[e^{-iE_pt}b_s(\vep)u_s(\vep)\\
+e^{iE_pt}d_s^\dagger(-\vep)v_s(-\vep)\Bigr],
\label{psi}
\ee
with
\be
\begin{split}
u_s(\vep)&=\frac{1}{\sqrt{2}}\left[\sqrt{1+\sin\vp_p}+
\sqrt{1-\sin\vp_p}\;(\vec{\alpha}\hat{\vep})\right]u_s(0),\\
v_{-s}(-\vep)&=\frac{1}{\sqrt{2}}\left[\sqrt{1+\sin\vp_p}-
\sqrt{1-\sin\vp_p}\;(\vec{\alpha}\hat{\vep})\right]v_{-s}(0).
\label{uandv}
\end{split}
\ee
The dressed quark dispersion law $E_p$ in Eq.~\eqref{psi} also departs from the free formula $\sqrt{\vep^2+m^2}$.
By convention, the chiral angle is defined to vary in the range
$-\frac{\pi}{2}<\vp_p\leqslant \frac{\pi}{2}$ with the boundary
conditions $\vp(0)=\frac{\pi}{2}$ and $\vp_p(p\to\infty)\to 0$. It is easy to verify that the free angle in Eq.~\eqref{vpfree} satisfies these constraints. The actual shape of the chiral angle $\vp_p$ should follow from some equation that guarantees that the dressed quarks are energetically favourable for the system. Then the standard approach is reminiscent of the one explained in detail in Sec.~\ref{sec:BV} that can be naturally discussed in terms of the Bogoliubov-Valatin transformation for the quarks. Thus, we notice that the Hamiltonian in Eq.~\eqref{GNJL} can be normally ordered in terms of the dressed quark/antiquark creation and annihilation operators $b$ and $d$ and then splits into three parts,
\be
H_{\rm GNJL}=H_0\;+:H_2:+:H_4:,
\label{H024}
\ee
where the subscript 0, 2, or 4 indicates the power of the quark operators contained in the respective term. In agreement with natural expectations, the vacuum energy $H_0$ is calculated as a sum over the individual contributions from the zero-point oscillations and reads
\be
H_0=\sum_{\rm colour}\sum_{\rm spin}\sum_p\left(-\frac12 E_p\right)=-N_cV \int\frac{d^3p}{(2\pi)^3}E_p,
\label{massgap}
\ee
where, for a time being, we explicitly restored the 3D volume of space $V$. For stability of the vacuum its energy should be a minimum, which is guaranteed by the condition
\be
\frac{\delta H_0[\vp_p]}{\delta\vp_p}=0.
\ee
After a straightforward calculation one can derive the corresponding equation for the chiral angle, known as the gap or mass-gap equation, in the form
\be
A_p\cos\vp_p-B_p\sin\vp_p=0,
\label{mge}
\ee
where the auxiliary functions read\footnote{Note that here $V(\vep)$ is the confining ``gluon propagator''
that differs from the Fourier transform of the potential $V(r)$ by an overall sign. This sign difference can be traced down to the opposite signs of the generators in the fundamental and antifundamental colour representations and is clearly seen in the explicit form of the kernel in Eq.~\eqref{KV} that provides different signs for the selfinteraction of the light quark and its interaction with the static antiquark. See also Subsec.~\ref{sec:IR}.\label{gluonprop}}
\be
\begin{split}
A_p&=m+\frac12\int\frac{d^3k}{(2\pi)^3}V(\vep-\vek)\sin\vp_k,
\label{AB}\\
B_p&=p+\frac12\int \frac{d^3k}{(2\pi)^3}\;
(\hat{\vep}\hat{\vek})V(\vep-\vek)\cos\vp_k.
\end{split}
\ee
In terms of the the same functions the dressed quark dispersion law reads
\be
E_p=A_p\sin\vp_p+B_p\cos\vp_p.
\label{Ep}
\ee
It is easy to verify that, in the no-interaction limit, Eqs.~\eqref{mge} and \eqref{Ep} provide the free solution for the chiral angle in Eq.~\eqref{vpfree} and the free dispersion law $E_p^{(0)}=\sqrt{\vep^2+m^2}$, respectively.

As explained in Sec.~\ref{sec:BV}, an alternative way to arrive at the mass-gap equation \eqref{mge} would be to require that the part of the Hamiltonian $:H_2:$ in Eq.~\eqref{H024} should be diagonal in terms of the quark creation and annihilation operators. Then it is easy to see that $:H_4:$ scales as $1/\sqrt{N_c}$ and, therefore, it is suppressed in the large-$N_c$ limit. This observation is the final step in diagonalisation of the model in the single-quark sector,
\be
H_{\rm GNJL}=E_{\rm vac}+\sum_{\alpha=1}^{N_c}\sum_{s=\pm 1/2}\int
\frac{d^3 p}{(2\pi)^3} E_p[b^\dagger_{\alpha s}({\bm p}) b^\alpha_s({\bm p})+d^\dagger_{\alpha s}(-{\bm p}) d^\alpha_s(-{\bm p})]+\ldots,
\label{H2diag}
\ee
where the ellipsis stands for the $N_c$-suppressed terms and the summation over the quark colours is made explicit for completeness.

The physical fermionic vacuum $\ket{0}$ that minimises the vacuum energy is (see also the discussion of the simple fermionic model in Sec.~\ref{sec:Bogferm})
\be
|0\rangle=U\ket{0}_0=e^{Q^\dagger-Q}|0\rangle_0,
\ee
where
\be
Q^\dagger=\frac12\sum_{p}\vp_p C_p^\dagger,\quad
C_p^\dagger=\sum_{\alpha=1}^{N_c}\sum_{s,s'=\pm1/2}b^{(0)\dagger}_{\alpha s}(\vep)\Bigl[({\bm \sigma}\hat{{\bm p}})i\sigma_2\Bigr]_{ss'}d^{(0)\dagger}_{\alpha s'}(\vep),
\label{S00}
\ee
with $\sigma$'s for the Pauli matrices. The operator
$C_p^\dag$ creates a quark-antiquark pair with the relative momentum $2\vep$ and the quantum numbers of the vacuum $0^{++}$.
Then the dressed quark and antiquark operators take the form
\be
b=Ub^{(0)}U^\dagger,\quad d=Ud^{(0)}U^\dagger,
\label{bandddressed}
\ee
so that
\be
b\ket{0}=Ub^{(0)}U^\dagger U\ket{0}_0=Ub^{(0)}\ket{0}_0=0
\ee
and, similarly, $d\ket{0}=0$.
Expanding the exponent in the operator $U$ and taking into account the Pauli exclusion principle for the quarks and antiquarks, one straightforwardly arrives at a simple representation for the physical vacuum,
\be
|0\rangle=\mathop{\prod}\limits_{p}\left[\sqrt{w_{0p}}+
\frac{1}{\sqrt{2}}\sqrt{w_{1p}}C^\dagger_p
+\frac12\sqrt{w_{2p}}C^{\dagger 2}_p\right]|0\rangle_0,
\label{nv}
\ee
where the coefficients,
\be
w_{0p}=\cos^4\frac{\vp_p}{2},\quad
w_{1p}=2\sin^2\frac{\vp_p}{2}\cos^2\frac{\vp_p}{2},\quad
w_{2p}=\sin^4\frac{\vp_p}{2},
\ee
obey the condition $w_{0p}+w_{1p}+w_{2p}=1$ (providing that the physical vacuum is normalised to unity, $\braket{0|0}=1$) and admit an interpretation as the probabilities of having no, one, and two $q\bar{q}$ pairs with the given relative momentum. The chiral angle plays a role of the radial wave function of such pairs. The vacuum $\ket{0}$ is orthogonal to the trivial one,
\be
\braket{0|0}_0=\exp\left[\sum_p \log\left(\cos^2\frac{\vp_p}{2}\right)\right]
=\exp\left[V\int\frac{d^3p}{(2\pi)^3}\ln\left(\cos^2\frac{\vp_p}{2}\right)\right]
\mathop{\longrightarrow}\limits_{V\to \infty}0.
\ee

The picture of the quark--antiquark pairs condensed in the physical vacuum resembles the Bardeen, Cooper, and Schrieffer (BCS) theory of superconductivity. For this reason, the physical vacuum $\ket{0}$ is often tagged as the BCS vacuum and the diagonalisation of the Hamiltonian in Eq.~\eqref{GNJL} in the quark sector is known as the BCS level of the model. It is also often referred to in the literature as a random phase approximation (RPA) --- see \cite{Bohm:1951zz,Pines:1952zz} for the original papers on this method applied to collective phenomena in the electron gas. The vacuum expectation value of a chirally non-symmetric operator $\bar{q}q$ takes a non-vanishing value in the BCS vacuum,
\be
\braket{\bar{q}q}=-\frac{N_c}{\pi^2}\int_0^{\infty}dp\;p^2\sin\vp_p\neq 0,
\label{chircond}
\ee
thus signalling that chiral symmetry is spontaneously broken if $\vp_p\neq 0$.\footnote{Here we assume the chiral limit of $m=0$ to avoid a divergent contribution from $\vp_p^{(0)}$ in Eq.~\eqref{vpfree}.} To ensure it is indeed the case we build the axial charge operator defined in Eq.~\eqref{Q5a} and, for $N_f=1$, express it in terms of the dressed quark operators as
\be
\begin{split}
Q_5=\sum_{\alpha=1}^{N_c}
\sum_{s,s'=\pm1/2}
\int\frac{d^3p}{(2\pi)^3}
\Bigl[&\cos\vp_p({\bm\sigma} \hat{\vep})_{ss'}\left(b_{\alpha s}^\dagger(\vep) b_{s'}^\alpha(\vep)
+d_{\alpha s}^\dagger(-\vep) d_{s'}^\alpha(-\vep)\right)\\
+&\sin\vp_p(i\sigma_2)_{ss'}\left(b_{\alpha s}^\dagger(\vep) d^{\dagger}_{\alpha s'}(-\vep)+d_s^\alpha(-\vep) b_{s'}^\alpha(\vep)\right)\Bigr].
\label{ChiralChar}
\end{split}
\ee
It is easy to see that, because of the second,
anomalous, term on the right-hand side and as required by the necessary and sufficient condition of SBCS in Eq.~\eqref{sbcs}, the axial charge operator in Eq.~\eqref{ChiralChar} has a non-vanishing matrix element between the vacuum and a pseudoscalar state with the spin wave function $i\sigma_2$ and the radial wave function (up to normalisation) $\sin\vp_p$. This state should be identified with the Goldstone boson of SBCS --- its detailed discussion will be provided in Sec.~\ref{sec:pion} below.

\subsection{Mass-gap equation and its solutions}

For the power-like confining potentials in Eq.~\eqref{Vconf}, the mass-gap equation \eqref{mge} takes the form \cite{Bicudo:2003cy}
\be
\begin{split}
&p^2(p\sin\vp_p-m\cos\vp_p)\\
&=K_0^{\alpha+1}\Gamma(\alpha+1)\sin\frac{\pi\alpha}{2}
\int_{-\infty}^{\infty}
\frac{dk}{2\pi}\left\{\frac{pk\sin[\vp_k-\vp_p]}{|p-k|^{\alpha+1}}+
\frac{\cos\vp_k\sin\vp_p}{(\alpha-1)|p-k|^{\alpha-1}}\right\},
\end{split}
\label{mg2}
\ee
where $\Gamma(\alpha+1)$ is the Euler Gamma function and the absolute value of the
momentum $p$ is
formally extended to $p<0$ as
$\cos\vp_{-p}=-\cos\vp_p$ and $\sin\vp_{-p}=\sin\vp_p$. The latter extension rule comes out automatically if one parameterises the chiral angle through an even function of the momentum $M_p$ as
\be
\sin\vp_p=\frac{M_p}{\omega_p},\quad\cos\vp_p=\frac{p}{\omega_p},
\label{Mpdef}
\ee
where
\be
\omega_p=\sqrt{p^2+M_p^2}.
\label{omegadef}
\ee
Then it is easy to arrive at a simple formula
\be
M_p=p\tan\vp_p
\label{Mpdef2}
\ee
that will play an essential role for a finite-temperature extension of the model.
The formulae \eqref{Mpdef}-\eqref{Mpdef2} are no more than a natural generalisation of the relations in Eq.~\eqref{vpfree} from the free to dressed quarks, so $M_p$ can be interpreted as an effective mass of the quark that incorporates the effects from both the current quark mass and SBCS.

As was explained above, the mass-gap equation \eqref{mg2} takes a particularly simple form for the harmonic oscillator confining potential that corresponds to $\alpha=2$,
\be
p^2(p\sin\vp_p-m\cos\vp_p)=\frac12K_0^3\left[p^2\vp''_p+2p\vp_p'+\sin2\vp_p\right].
\label{diffmge}
\ee
This second-order differential equation was numerically studied by many authors in the past
\cite{Amer:1983qa,LeYaouanc:1983it,LeYaouanc:1983huv,LeYaouanc:1984ntu,Bicudo:1989sh,Bicudo:1989si,Bicudo:1989sj,Bicudo:1993yh,Bicudo:1998mc}. In particular, it was noticed in
\cite{LeYaouanc:1983it,LeYaouanc:1983huv} that, in addition to the nodeless ground-state solution,
Eq.~\eqref{diffmge} possesses multiple excited solutions with a different number of nodes and different slopes of the chiral angle at the origin. The existence of such excited solutions was later confirmed for the mass-gap equation \eqref{mg2} for any power $\alpha$ \cite{Bicudo:2003cy}, for a phenomenology-driven potential in Eq.~\eqref{Cornell} \cite{Bicudo:2002eu}, in Coulomb-gauge QCD \cite{Garnacho-Velasco:2024wdb}
truncated and parameterised using the approach in \cite{Llanes-Estrada:2004edu}.

It is instructive to study the asymptotics of the chiral angle at large and small momenta.
In particular, the low-momentum behaviour of the chiral angle,
\be
\vp_p(p\to 0)=\frac{\pi}{2}-\frac{p}{M_p(0)}+\ldots,
\label{slor}
\ee
can be straightforwardly obtained expanding the relation in Eq.~\eqref{Mpdef2} in the limit of $p\to 0$ and bearing in mind that $\vp_p(0)=\pi/2$.
For the power-like confining potential in Eq.~\eqref{Vconf}, $M_p^{(n)}(0)$ for the $n$-th excited solution of the mass-gap equation \eqref{mg2} can be found employing a quasiclassical approach \cite{Bicudo:2003cy},
\be
M_p^{(n)}(0)=M_p(0)e^{-C_\alpha \pi n},\quad M_p(0)\equiv M_p^{(0)}(0),\quad
C_\alpha=\left[\frac{\sqrt{\pi}\Gamma\left(\frac{4-\alpha}{2}\right)}{2^\alpha\Gamma
\left(\frac{1+\alpha}{2}\right)}\right]^{1/\alpha},
\label{Calpha}
\ee
so it decreases with the excitation number $n$ and the effect of SBCS predictably fades out in the excited vacua.
On the other hand, in the large-momentum limit, the chiral angle behaves as \cite{Bicudo:2003cy}
\be
{\vp_p}_{|{m=0}}\mathop{\approx}\limits_{p\to\infty}-
\frac{\pi}{N_c}\Gamma(\alpha+2)K_0^{\alpha+1}\sin\left(\frac{\pi\alpha}{2}\right)
\frac{\langle\bar{q}q\rangle}{p^{\alpha+4}},
\label{asym}
\ee
where the chiral condensate is calculated as given in Eq.~\eqref{chircond}.

The excited solutions for the chiral angle were regarded as metastable \cite{Bicudo:2019ryc} replicas of the BCS vacuum and some attempts to prescribe them a physical interpretation were made in the literature \cite{Nefediev:2002nw,Antonov:2010qt,Antonov:2010ap,Ribeiro:2011zza}.
Meanwhile, in what follows, only the nodeless ground-state solution with $n=0$ and the corresponding chirally broken BCS vacuum will be discussed. To simplify notations, the corresponding superscript $^{(0)}$ will be omitted everywhere.

A final comment on the nature of the chiral angle is in order here. Since SBCS is an intrinsically quantum effect (see, for example, discussions in \cite{Glozman:2004gk}), then $M_p$ in Eq.~\eqref{Mpdef2} has to vanish (we assume here that the current quark mass $m=0$) in the formal classical limit of $\hbar\to 0$. Therefore, the chiral angle collapses to zero in this limit, in agreement with its natural interpretation as the wave function of the quark--antiquark pairs condensed in the chirally broken vacuum --- see Eq.~\eqref{S00} above as well as further details and a discussion of different regimes for the chiral angle contained in \cite{Glozman:2005tq}.

\subsection{Infrared-divergent and infrared-finite quantities}
\label{sec:IR}

The consideration in the previous section may leave a false impression that the model in Eq.~\eqref{GNJL} supports dressed quarks that can be observed in a free state. However, it is not the case, and the role of the confining interaction term $:H_4:$ in the decomposition in Eq.~\eqref{H024} will be highlighted and discussed in detail in Sec.~\ref{sec:BS} below. Meanwhile, one can observe that dressed quarks are still confined already at the BCS level.
To this end, let us for concreteness stick to the linear confining potential,
\be
V(r)=\sigma r,
\label{Vlin2}
\ee
with $\sigma=C_FK_0^2$ for the fundamental string tension in the notations of Eqs.~\eqref{Vlin} and \eqref{Vconf} above.
The Fourier transform of the potential \eqref{Vlin2} is singular at small momenta,\footnote{We use tilde for the Fourier transform of the potential in Eq.~\eqref{Vlin2} that differs by the overall sign from the confining ``gluon propagator'' $V(\vep)$ introduced in Eq.~\eqref{AB} above --- see also footnote \ref{gluonprop}.}
\be
\tilde{V}(p)=-\frac{8\pi\sigma}{p^4},
\ee
which leads to infrared divergences in the integrals in Eq.~\eqref{AB} (for a detailed treatment of this problem and related useful numerical recipes see \cite{Stadler:2024dph}). To cure this problem, let us introduce an infrared regulator $\muIR$, for example, following the prescription in \cite{Alkofer:2005ug,Wagenbrunn:2007ie,Bicudo:2010qp},
\be
\tilde{V}_{\rm reg}(p)=-\frac{8\pi\sigma}{(p^2+\muIR^2)^2}.
\label{FV}
\ee
Then, passing over back to the coordinate space, we find
\be
V_{\rm reg}(r)=\int \frac{d^3 p}{(2\pi)^3} \tilde{V}_{\rm reg}(p) e^{i\vec p \vec r}=-\frac{\sigma}{\muIR}e^{-\muIR r}\mathop{=}_{\muIR\to 0}-\frac{\sigma}{\muIR}+\sigma r+\ldots,
\label{div}
\ee
where the ellipsis stands for the terms suppressed in the limit $\muIR\to 0$. Therefore, the regularised potential in Eq.~\eqref{div} inherits the coordinate dependence from the original potential in Eq.~\eqref{Vlin2} but also acquires a constant contribution that diverges in the infrared limit. This, however, should not cause any trouble given that the interquark potential is not observable. Furthermore, other non-observable quantities also acquire similar singular contributions that cancel against each other in observables \cite{Glozman:2008fk}. For example, after the above regularisation with $\muIR$ \cite{Wagenbrunn:2007ie}, one finds
\begin{align}
A_p=\frac{\sigma}{2\muIR}\sin\varphi_p+A_p^{\rm fin},\nonumber\\[-3mm]
\label{ABir}\\[-3mm]
B_p=\frac{\sigma}{2\muIR}\cos\varphi_p+B_p^{\rm fin},\nonumber
\end{align}
with $A_p^{\rm fin}$ and $B_p^{\rm fin}$ for the infrared-finite contributions. It is easy to verify then that the single-quark dispersion law in Eq.~\eqref{Ep} takes the form
\be
E_p=\frac{\sigma}{2\muIR}+\Bigl(A_p^{\rm fin}\sin\vp_p+B_p^{\rm fin}\cos\vp_p\Bigr)=\frac{\sigma}{2\muIR}+E_p^{\rm fin}
\label{Epir}
\ee
and, therefore, also contains a singular
term in the limit $\muIR\to 0$.

\begin{figure}[t!]
\centering
\includegraphics[width=0.99\textwidth]{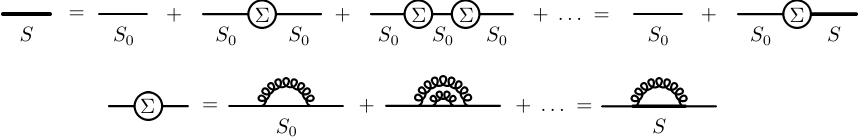}
\caption{The Dyson series in Eq.~\eqref{Ds} for the dressed quark propagator $S$ in Eq.~\eqref{Feynman}, with the bare propagator $S_0$ in Eq.~\eqref{S0T0} and the quark mass operator $\Sigma$ in Eq.~\eqref{Sigma0} taken in the rainbow approximation. The thin and thick solid lines are for the bare and dressed quark propagator, respectively, and the curly line is for the confining ``gluon propagator'' --- see footnote \ref{gluonprop}. Adapted from \cite{Glozman:2024dzz}.}
\label{fig:dyson}
\end{figure}

\begin{figure}[t]
\centering
\includegraphics[width=0.48\textwidth]{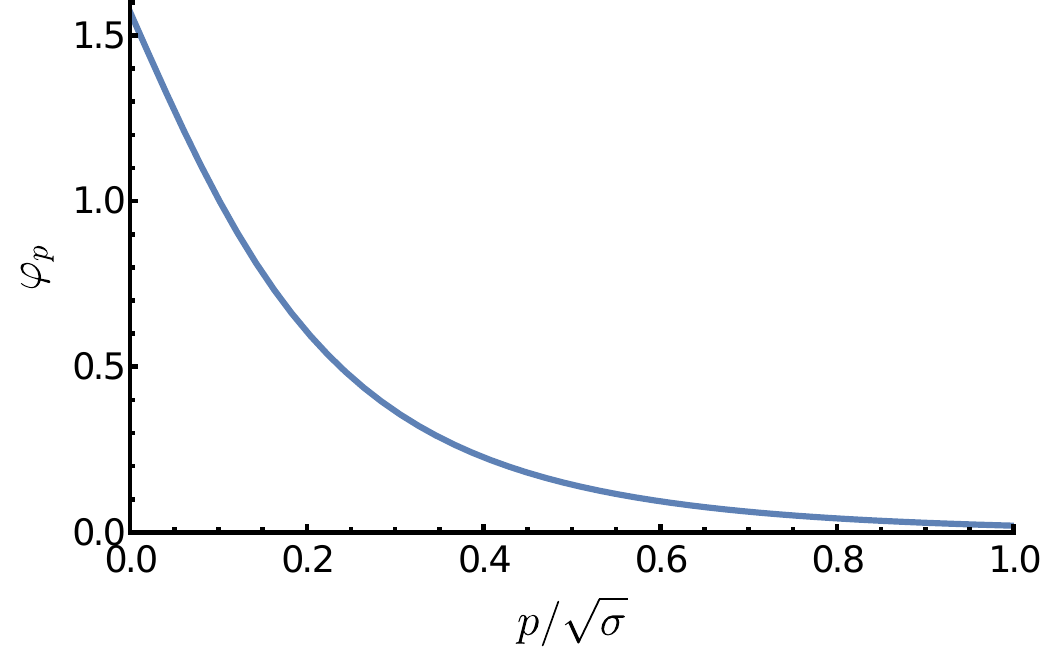}
\includegraphics[width=0.5\textwidth]{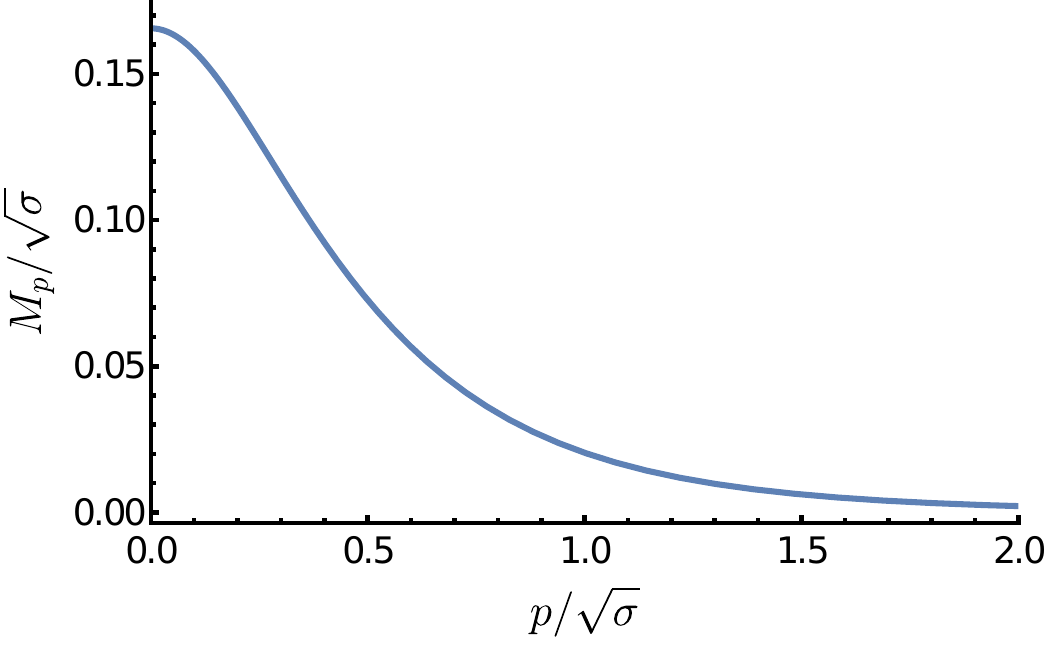}
\caption{The profile of the chiral angle --- solution to the mass-gap equation in Eq.~\eqref{mge} (left plot) and the effective quark mass $M_p$ (right plot) for the linear confinement in Eq.~\eqref{Vlin2}.}
\label{fig:vp}
\end{figure}

The Green function of the dressed quark,
\be
S(x-y)=\braket{0|Tq(x)\bar{q}(y)|0},
\label{Sxy}
\ee
can be derived in the usual way employing the spectral decomposition in Eq.~\eqref{psi} and the Fermi algebra of the quark operators, so the Fourier transform of $S(x-y)$ reads
\be
S(p_0,\vep)=\frac{\Lambda_+(\vep)\gamma_0}{p_0-E_p+i\epsilon}+
\frac{{\Lambda_-}(\vep)\gamma_0}{p_0+E_p-i\epsilon},
\label{Feynman}
\ee
where the dressed positive- and negative-frequency projectors are defined as
\be
\Lambda_+(\vep)=\sum_s u_s(\vep)\otimes u^\dagger_s(\vep),\quad
\Lambda_-(\vep)=\sum_s v_{-s}(-\vep)\otimes v^\dagger_{-s}(-\vep)
\label{Lambdasdef}
\ee
and, after a simple algebra, they can be found in the form
\be
\Lambda_\pm(\vep)=T_pP_\pm T_p^\dagger=\frac12[1\pm\gamma_0\sin\vp_p\pm({\bm\alpha}\hat{\vep })\cos\vp_p],
\label{Lpm}
\ee
with $P_\pm=\frac12(1\pm\gamma_0)$ for the bare projectors, $T_p$ for the Foldy operator,
\be
T_p=\exp\left[-\frac12({\bm\gamma\hat{\vep}})\left(\vp_p-\frac{\pi}2\right)\right],
\label{Tpdef}
\ee
and $\hat{\vep}$ for the unit vector in the direction of the momentum $\vep$. Notice that alternatively the inverse dressed propagator,
\be
S^{-1}(p_0,\vep)=S_0^{-1}(p_0,\vep)-\Sigma(\vep),
\label{Sm1}
\ee
with the free propagator,
\be
S_0^{-1}(p_0,\vep)=\gamma_0 p_0-{\bm\gamma}\vep-m,
\label{S0T0}
\ee
can be found as solution of the equation derived through the summation of the Dyson series,
\be
S=S_0+S_0\Sigma S_0+S_0\Sigma S_0\Sigma S_0+\ldots=S_0+S_0\Sigma S,
\label{Ds}
\ee
where the mass operator,
\be
i\Sigma(\vep)=\int\frac{d^4k}{(2\pi)^4}V(\vep -\vek)\gamma_0 S(k_0,{\vek})\gamma_0,
\label{Sigma0}
\ee
is calculated in the ``rainbow'' approximation for the interaction --- see Fig.~\ref{fig:dyson}.\footnote{In this approximation, only planar diagrams with non-crossing ``gluon'' lines are summed. It is justified in the large-$N_c$ limit where non-planar diagrams are suppressed by inverse powers of $N_c$.\label{Nc2}} In this approach, the mass-gap equation \eqref{mge} and the expression for the dressed quark dispersion law in Eq.~\eqref{Ep} appear from a selfconsistency condition for Eqs.~\eqref{Ds} and \eqref{Sigma0} above. Whichever way is chosen to derive the Green function in Eq.~\eqref{Feynman}, it takes the same form and possesses the quark and antiquark poles at $\pm E_p$. However, since the quantity $E_p$ in Eq.~\eqref{Epir} is infrared-divergent, the isolated quark Green function vanishes, which can be interpreted as the fact that, in agreement with natural expectations, single quarks are removed from the spectrum of the model. Meanwhile, it is easy to see with the help of Eq.~\eqref{ABir} that the infrared divergence goes away from the mass-gap equation \eqref{mge}, so the chiral angle is an infrared-finite quantity and, therefore, so are the effective quark energy $\omega_p$ and the effective mass $M_p$ defined in Eqs.~\eqref{omegadef} and \eqref{Mpdef2}, respectively. In Fig.~\ref{fig:vp} we visualise the profile of the solution to the mass-gap equation \eqref{mge} with the linear potential in Eq.~\eqref{Vlin2} and the corresponding profile of the effective quark mass $M_p$.

\begin{figure}[t]
\centering
\includegraphics[width=0.7\textwidth]{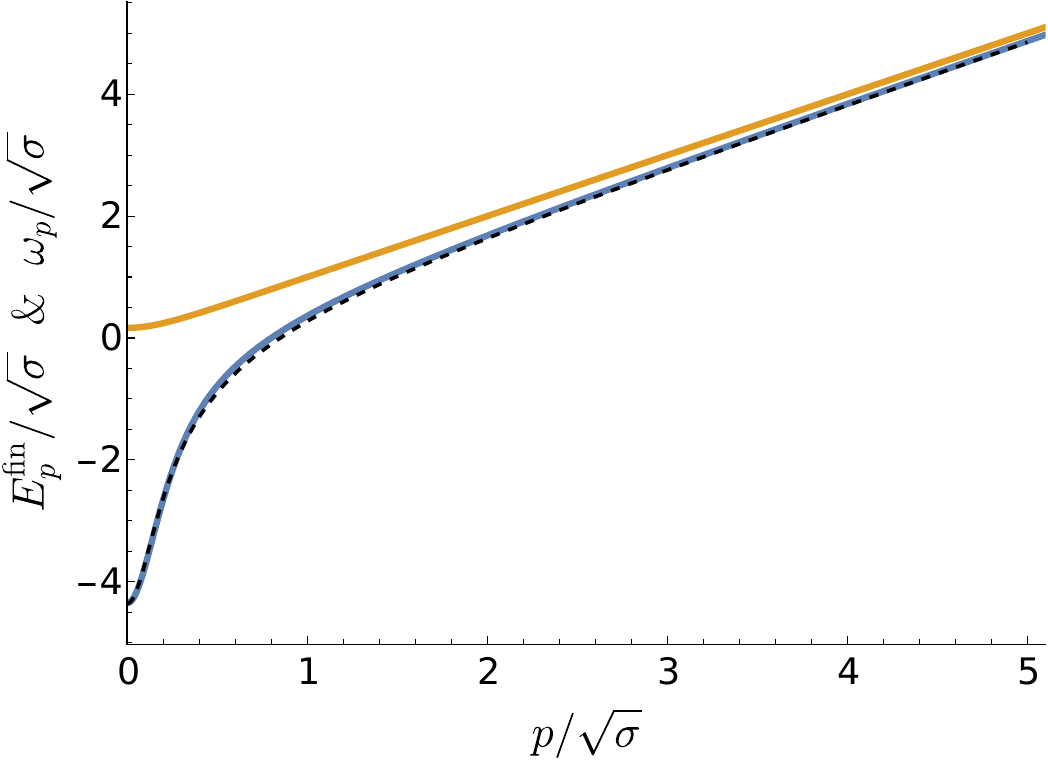}
\caption{The behaviour of the minimally subtracted dressed quark dispersion law $E_p^{\rm fin}$ defined in Eq.~\eqref{Epir} (the blue solid line) and the effective quark energy $\omega_p$ in Eq.~\eqref{omegadef} (the yellow solid line) for the linear confinement in Eq.~\eqref{Vlin2}. The black dashed line shows the approximation $(E_p^{\rm fin})_{\rm app}$ in Eq.~\eqref{Epfit}.}
\label{fig:Eomega}
\end{figure}

It is easy to ensure that the infrared-divergent contributions exactly cancel against each other in the sum,
\be
2E_p+V_{\rm reg}(r)=2E_p^{\rm fin}+V(r),
\ee
where Eqs.~\eqref{Vlin2}, \eqref{div}, and \eqref{Epir} were used. Then one would be tempted to na{\"i}vely
conclude that a simple quark model in Eq.~\eqref{Salp} should be relevant to address SBCS and related issues. However, it is not the case and, to see it explicitly, let us study the shape of the dressed quark dispersion law in Eq.~\eqref{Epir}. First of all, it should be noted that a subtraction of the infrared-divergent contribution is generally a scheme-dependent procedure, which makes $E_p^{\rm fin}$ \emph{per se} an ill-defined object. Thus, for a time being and for the presentation purposes only, let us stick to a particular form of $E_p^{\rm fin}$ that comes as a result of the regularisation procedure in Eq.~\eqref{FV} and the minimal subtraction when only the divergent piece $\sigma/(2\muIR)$ in Eq.~\eqref{Epir} is removed.
The behaviour of such $E_p^{\rm fin}$ as function of momentum is shown with the blue solid curve in Fig.~\ref{fig:Eomega}. For comparison, the effective quark energy $\omega_p$ is also shown in Fig.~\ref{fig:Eomega} (the yellow solid curve), and a crucial difference between the two objects can be observed: while $\omega_p$ predictably stays positive for all momenta, the dispersion law $E_p^{\rm fin}$ becomes negative at small $p$'s --- the property that was noticed and discussed in some detail already in an early work on the two-dimensional 't~Hooft model in the axial gauge \cite{Bars:1977ud}. The shape of $E_p^{\rm fin}$ in Fig.~\ref{fig:Eomega} can be well approximated by a simple formula (see the black dashed curve),
\be
(E_p^{\rm fin})_{\rm app}=\omega_p-0.75\frac{\sigma}{\omega_p},
\label{Epfit}
\ee
motivated by the investigation in \cite{Bicudo:2002eu}, and the fitted numerical coefficient in Eq.~\eqref{Epfit} appears to be fairly close to the approximate value $8/(3\pi)\approx 0.85$ following for $\alpha=1$ from the general expression
for the power-like confining potential \eqref{Vconf} derived employing a quasiclassical approach in \cite{Bicudo:2002eu}.

The small-momentum behaviour of the dispersion relation in Eq.~\eqref{Epfit} entails several important consequences. First of all, a negative contribution from the second term on the right-hand side in Eq.~\eqref{Epfit} is crucial to compensate for the positive contributions from the confining potential and the quark kinetic energies. Then a Salpeter-like equation,\footnote{Hereinafter, only infrared regularised quantities are considered, so the corresponding tag ``reg'' will be omitted for simplicity.}\label{negativeEp}
\be
\Bigl[2E_p+V(r)\Bigr]\psi=M\psi,
\label{Salp2}
\ee
can indeed have a nontrivial solution with a vanishing (or very small beyond the chiral limit) mass $M$ that corresponds to the Goldstone boson of SBCS. In particular, it will be explicitly demonstrated below that the bound state equation for the lightest pseudoscalar quark--antiquark meson takes the form of Eq.~\eqref{Salp2} with $M=0$ and (up to normalisation) $\psi=\sin\vp_p$ --- see also Eq.~\eqref{ChiralChar} and the discussion below it. Furthermore, the resulting bound state equation for the pion appears to be identical to the mass-gap equation \eqref{mge}, so one can naturally identify the quark--antiquark pion as the Goldstone boson of SBCS. Notice that no similar solution is possible for Eq.~\eqref{Salp} since the operator on its left-hand side is positively defined.

It should not be a surprise that the negative contribution to $E_p$ comes from small momenta where the chiral angle takes its largest values (see Fig.~\ref{fig:vp}) and, therefore, the mixing between the bare quarks and antiquarks as provided by the Bogoliubov transformation in Eq.~\eqref{bandddressed} reaches it maximum. The present discussion of the dressed quark dispersion law is also crucial for the extension of the GNJL model to finite temperatures. Indeed, since neither $E_p$ nor its finite part $E_p^{\rm fin}$ are well defined objects, they are not suited for this purpose. On the contrary, the effective quark energy $\omega_p$ is well defined and has no problem with changing its sign at different momenta. We get back to this discussion in Sec.~\ref{sec:finT} below.

To conclude this section, we notice that many alternative infrared regularisation schemes could be employed to tame the large-$r$ behaviour of the confining interquark interaction. For example, in \cite{Bicudo:2002eu}, the regularised potential in coordinate space was chosen in a form free from infrared singular contributions (\emph{cf} Eq.~\eqref{div}),
\be
V_{\rm reg}(r)=V(r)e^{-\muIR r}.
\label{divalt}
\ee
For the linearly rising potential in Eq.~\eqref{Vlin2}, it corresponds to
\be
\tilde{V}_{\rm reg}(p)=-\frac{8\pi\sigma}{(p^2+\muIR^2)^2}\left(1-\frac{4\muIR^2}{p^2+\muIR^2}\right).
\label{FValt}
\ee
Employing this regularisation scheme or, alternatively, the principal value prescription --- a popular choice in the studies of the two-dimensional model for QCD in the Coulomb gauge (see, for example, \cite{Bars:1977ud,Kalashnikova:1997tb,Kalashnikova:2001df}) --- one formally avoids explicitly infrared-divergent contributions. Obviously, it does not solve the problem of an ill-defined quark dispersion law but just sweeps it under the rug.
On the contrary, the regularisation schemes like the one in Eq.~\eqref{FV} \cite{Alkofer:2005ug,Wagenbrunn:2007ie,Bicudo:2010qp} may be useful to reveal explicit infrared divergences and thus allow one to unambiguously distinguish between physical infrared-finite and not observable infrared-infinite quantities. Meanwhile, we stress that observable quantities (for example, the spectrum of the quark--antiquark mesons studied below) can not depend on the chosen regularisation scheme and must approach the same finite limit for $\muIR\to 0$ irrespective of the particular way the regulator $\muIR$ was introduced.

\subsection{Chiral symmetry restoration at finite temperatures}
\label{sec:finT}

The picture of SBCS in the vacuum drawn above raises a natural question on the behaviour of the model at finite temperatures. A standard way of introducing the temperature to the theory is to proceed to the Euclidean space-time and perform compactification of the temporal axis with the period $\beta=1/T$, imposing anti-periodic boundary conditions on the quark field. Then the modified Green function in Eq.~\eqref{Sxy} reads
\be
S(\tau-\tau',\vex-\vey)=-\mbox{Tr}\braket{0|\rho\; Tq(\tau,\vex)\bar{q}(\tau',\vey)|0},\quad \rho=\frac{e^{-\beta (H-\mu N)}}{\mbox{Tr}e^{-\beta (H-\mu N)}},
\ee
where $\mu$ stands for a chemical potential and $N$ is the operator counting the number of fermions. All quantities depending on the Euclidean time $\tau$ can then be presented as infinite sums over the Matsubara frequencies\cite{Matsubara:1955ws}
$\omega_n=(n+\frac12)\frac{\pi}{\beta}$ with $n=0,1,\ldots$ and, given an instantaneous nature of the interaction, these sums can be evaluated explicitly. For the GNJL model, the outlined programme was performed in \cite{Kocic:1985uq} and a modified mass-gap equation in the form of Eq.~\eqref{mge} was derived with
the substitution
\begin{align}
A_p\to\tilde{A}_p&=m+\frac12\int\frac{d^3k}{(2\pi)^3}(1-n_k-\bar{n}_k)V(\vep-\vek)\sin\vp_k,\nonumber\\[-2mm]
\label{tildeAB}\\[-2mm]
B_p\to \tilde{B}_p&=p+\frac12\int \frac{d^3k}{(2\pi)^3}(1-n_k-\bar{n}_k)
(\hat{\vep}\hat{\vek})V(\vep-\vek)\cos\vp_k,\nonumber
\end{align}
where $n_p$ and $\bar{n}_p$ are the temperature-dependent Fermi--Dirac functions,
\be
n_p=\left(1 + e^{(\varepsilon_p-\mu)/T}\right)^{-1},
\quad
\bar{n}_p=\left(1 + e^{(\varepsilon_p+\mu)/T}\right)^{-1}.
\label{nngen}
\ee
In the last equation, the quantity $\varepsilon_p$ is the energy of the dressed quark or antiquark, and in \cite{Kocic:1985uq} it was identified with the dressed quark dispersion law $E_p$ in Eq.~\eqref{Ep}. However, although such identification indeed follows straightforwardly from the derivation in \cite{Kocic:1985uq}, it meets obvious difficulties. Indeed, as was discussed in Sec.~\ref{sec:IR} the single-quark energy $E_p$ is infrared divergent, so taking a formal infrared limit of $\muIR$ one arrives at $n_p=\bar{n}_p=0$ and this way kills the temperature dependence entirely. This undesired feature of the GNJL model should probably be not surprising given its obvious similarity to the 't~Hooft model for QCD in two-dimensions. There exists a vast literature on the temperature behaviour of the 't~Hooft model with the general conclusion that no chiral phase transition is possible in it even in the limit $T\to\infty$ --- for a pedagogical and comprehensive review see \cite{Schon:2000qy} and references therein. Notice also that the identification $\varepsilon_p=E_p$ meets obvious difficulties in view of the small-momentum behaviour of $E_p$ discussed in detail above and demonstrated in Fig.~\ref{fig:Eomega}. Indeed, even the regularised and finite in the infrared limit quantity $E_p^{\rm fin}$ necessarily becomes negative at $p\to 0$ (the feature of $E_p$ often referred to in the literature as a tachyonic behaviour) thus driving the Fermi--Dirac functions in Eq.~\eqref{nngen} ill-defined. On the other hand, interpreting the sign of the energy $E_p$ in the Dirac spirit (that is, assuming that the positive and negative energies correspond to fermions and antifermions, respectively) and bearing in mind that dressed quarks are effective degrees of freedom that are certain mixtures of the bare quarks and antiquarks (as encoded in the Bogoliubov transformation in Eq.~\eqref{bandddressed}), one is led to conclude that heating the theory and reformulating it in terms of the effective degrees of freedom may not be trivially interchangeable procedures. It is instructive to notice that truncated QCD in the Coulomb gauge was studied at finite temperatures in \cite{Quandt:2018bbu} employing compactification of one spatial (rather than temporal) direction in space. This way the authors avoided complications related to the proper definition of the energy in the Fermi--Dirac thermal distributions. As a result, they observed a chiral restoration at some finite temperature $\Tch$. Notice that a similar in spirit approach to the two-dimensional `t~Hooft model
\cite{Lenz:1991sa,Thies:1992qx} also led to a chiral phase transition at some finite critical length thus demonstrating an obvious tension with the aforementioned conclusion of no chiral restoration in the model at any temperature $T$. Indeed, since Lorentz invariance of the `t~Hooft model is manifest, there should be no difference between conclusions arrived at through the compactification of the temporal or spatial axis. The puzzle is resolved in
\cite{Schon:2000vd} where it is demonstrated that, if the model is put into a finite-size spatial box, in addition to the gluons that provide a linearly rising potential between quarks, zero mode gluons need to be taken into account, which changes the behaviour of the model under compression and restores the no-go result obtained at finite temperatures through the compactification of the temporal dimension. This unfortunate conclusion directly projected onto the GNJL model implies that its straightforward extension to finite temperatures may not meet the goal, that is, it does not allow one to study QCD at finite $T$'s and, in particular, the properties of hadrons below and above the chiral restoration transition. The lesson from the 't~Hooft model would be that gluons may need to be treated more rigorously when proceeding from full QCD in the Coulomb gauge to a simplified model of confined quarks. It still remains to be seen how this programme can be realised in practice while below we employ a more pragmatic approach and extend the GNJL model to finite temperatures in such a way that the resulting system demonstrates features commonly believed to be inherent in QCD such as chiral restoration at some finite temperature $\Tch$.
To this end, we employ a real-time formalism and evaluate the dressed quark propagator in Eq.~\eqref{Sxy}, taking into account non-vanishing thermal averages of the creation and annihilation operators,
\be
\braket{b_{\alpha s}^\dagger(\vep)b_{s'}^\beta(\vep)}=n_p\delta_\alpha^\beta\delta_{ss'},\quad
\braket{d_{\alpha s}^\dagger(\vep)d_{s'}^\beta(\vep)}=\bar{n}_p\delta_\alpha^\beta\delta_{ss'},
\ee
with $n_p$ and $\bar{n}_p$ introduced in Eq.~\eqref{nngen}. The quark energy $\varepsilon_p$ in them will be defined later. The result reads \cite{Glozman:2024xll,Glozman:2024dzz}
\begin{align}
S(p_0,\vep;T)&=S(p_0,\vep;T=0)
+2\pi i\Bigl[n_p\Lambda_+(\vep)\delta(p_0-E_p)
-\bar{n}_p\Lambda_-(\vep)\delta(p_0+E_p)\Bigr]\gamma_0\nonumber\\
&=\mbox{v.p.}\left(
\frac{\Lambda_+(\vep)\gamma_0}{p_0-E_p}+
\frac{{\Lambda_-}(\vep)\gamma_0}{p_0+E_p}
\right)\label{ST}\\
&-i\pi\Bigl[(1-2n_p)\Lambda_+(\vep)\delta(p_0-E_p)
-(1-2\bar{n}_p)\Lambda_-(\vep)\delta(p_0+E_p)\Bigr]\gamma_0,\nonumber
\end{align}
where v.p. stands for the principal value prescription, the explicit form of the zero-temperature dressed fermionic Green function $S(p_0,\vep;T=0)\equiv S(p_0,\vep)$ in Eq.~(\ref{Feynman}) was used, and the projectors $\Lambda_\pm(\vep)$ were previously introduced in Eq.~\eqref{Lambdasdef} and evaluated in Eq.~(\ref{Lpm}). It follows from the definition in Eq.~\eqref{nngen} that both $n_p$ and $\bar{n}_p$ tend to the same limit 1/2 as the temperature increases, so the temperature ``eats up'' the imaginary part of the Green function in Eq.~\eqref{ST}.

To proceed, we require that the Green function in Eq.~\eqref{ST} reproduces a well known result,\cite{Asakawa:1989bq}
\be
S_0(p_0,\vep;T)=(\slashed{p}-m+i\epsilon)^{-1}
+2\pi i(\slashed{p}+m)\delta(p^2-m^2)\Bigl[\theta(p_0)n_p^{(0)}+\theta(-p_0)\bar{n}^{(0)}_p\Bigr],
\label{S0T}
\ee
with
\be
n_p^{(0)}=\left(1 + e^{(\sqrt{p^2 + m^2}-\mu)/T}\right)^{-1},\quad
\bar{n}_p^{(0)}=\left(1 + e^{(\sqrt{p^2 + m^2}+\mu)/T}\right)^{-1},
\ee
in the no-interaction limit and provides a well-defined and physically meaningful extension for all zero-temperature expressions derived above. Then it is easy to notice that the above constraints are readily met if one sets $\varepsilon_p=\omega_p$, with the effective quark energy $\omega_p$ defined in Eq.~\eqref{omegadef} and depicted in Fig.~\ref{fig:Eomega}. This way we eventually arrive at the expressions similar to those in Eq.~\eqref{tildeAB} but with
\be
n_p=\left(1 + e^{(\sqrt{p^2 + M_p^2}-\mu)/T}\right)^{-1},\quad
\bar{n}_p=\left(1 + e^{(\sqrt{p^2 + M_p^2}+\mu)/T}\right)^{-1}.
\label{nnnew}
\ee
The suggested scheme is ideologically close to the one employed in \cite{Quandt:2018bbu} and can be regarded as a kind of variational approach to the model at finite temperatures.

\begin{figure}[t!]
\centering
\includegraphics[width=0.49\textwidth]{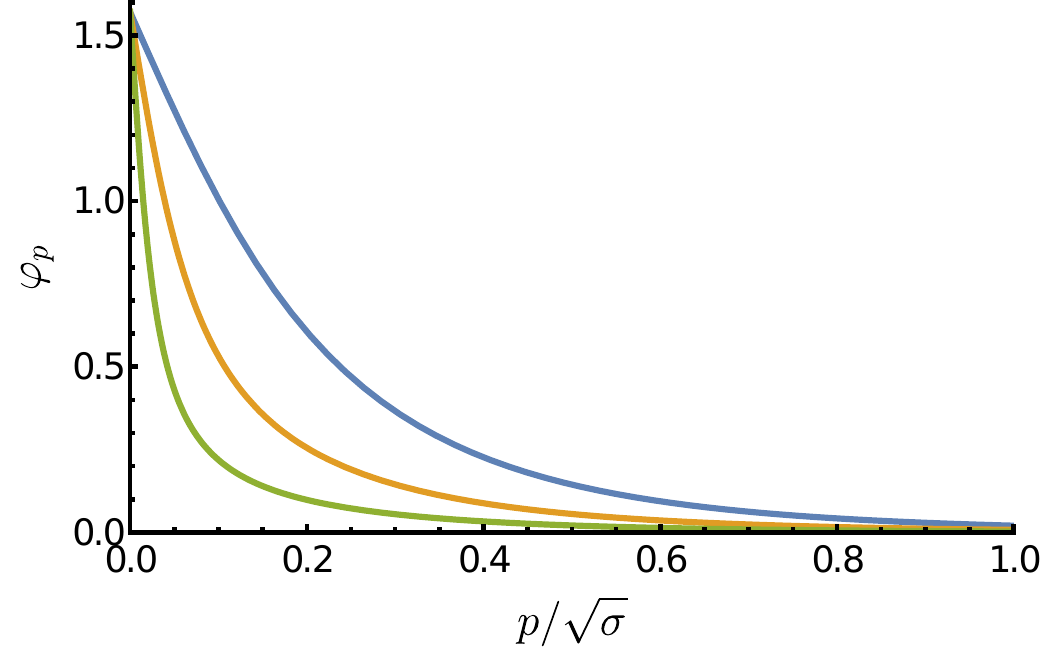}
\includegraphics[width=0.49\textwidth]{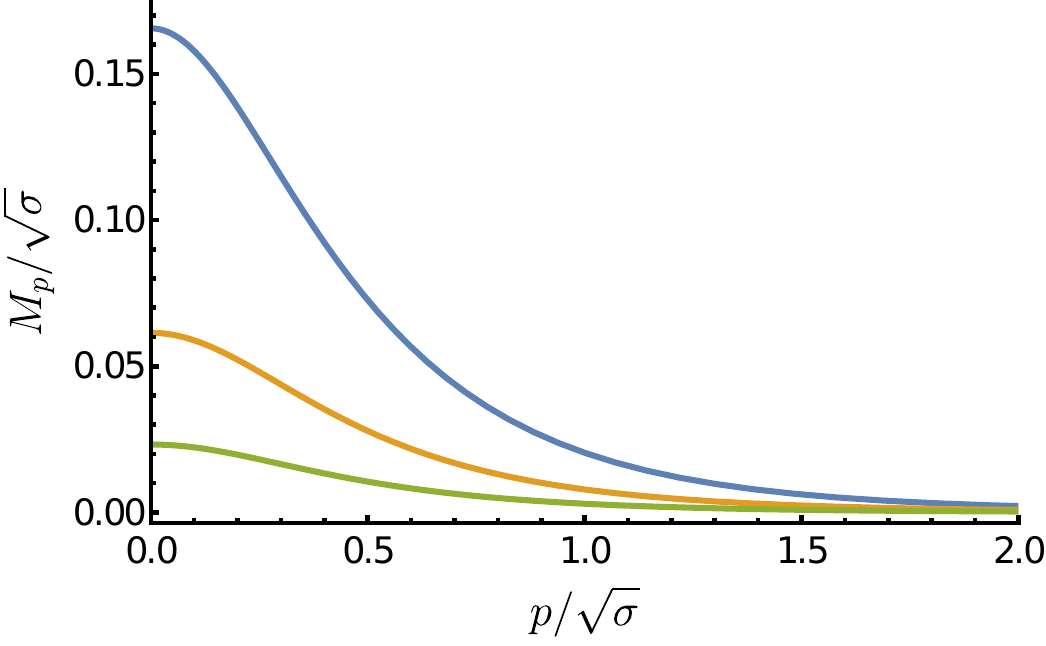}\\
\includegraphics[width=0.49\textwidth]{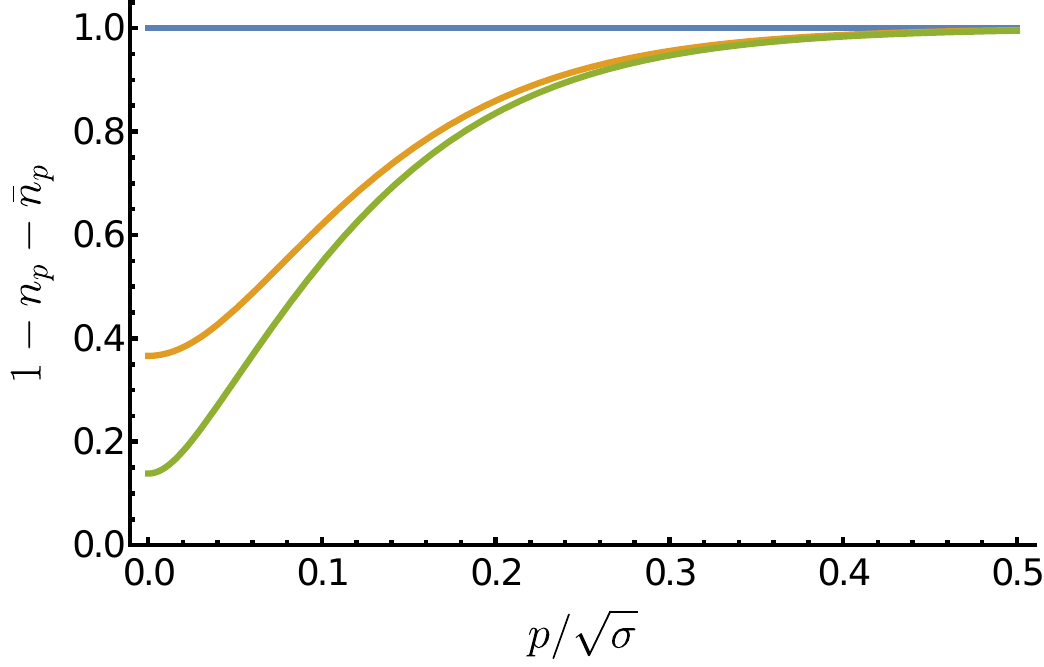}
\includegraphics[width=0.45\textwidth]{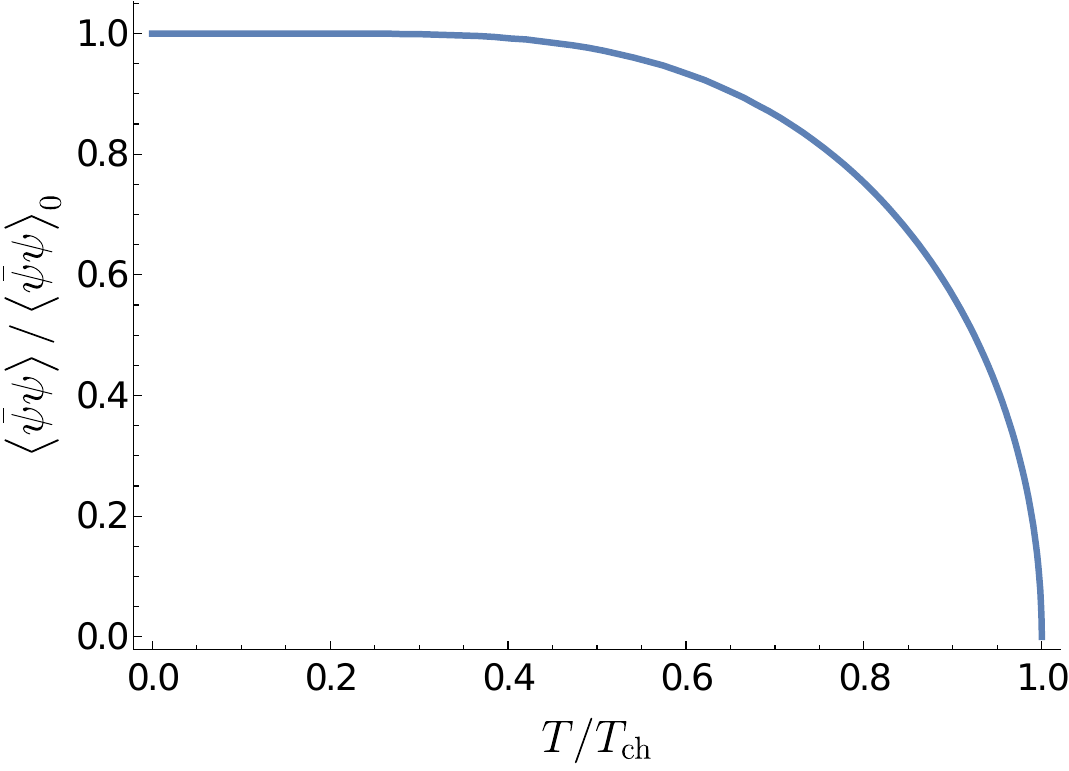}
\caption{The temperature dependence of (i) the chiral angle (upper left plot), (ii) the effective quark mass (upper right plot), and (iii) the damping factor $1-n_p-\bar{n}_p$ in Eq.~\eqref{tildeAB} (lower left plot) for a vanishing chemical potential ($\mu=0$) and $T=0$ (the blue curve), $T=0.95\Tch$ (the yellow curve), and
$T=0.99\Tch$ (the green curve). The lower right plot: the temperature dependence of the chiral condensate. The value of $\Tch$ is quoted in Table~\ref{tab:res}. The confining potential is chosen in the linear form in Eq.~\eqref{Vlin2} and all dimensional quantities are provided in the appropriate units of the string tension $\sigma$. The current quark mass $m$ is set to zero. Adapted from \cite{Glozman:2024xll}.}
\label{fig:rest}
\end{figure}

It should be noted that, by virtue of easily verifiable relations,
\be
n_p\Lambda_+(\vep)-\bar{n}_p\Lambda_-(\vep)=
\frac12(n_p+\bar{n}_p)[\Lambda_+(\vep)-\Lambda_-(\vep)]+\frac12(n_p-\bar{n}_p)[\Lambda_+(\vep)+\Lambda_-(\vep)]
\label{rels}
\ee
and
\be
\Lambda_+(\vep)+\Lambda_-(\vep)=1,\quad
\Lambda_+(\vep)-\Lambda_-(\vep)=\gamma_0\sin\vp_p+({\bm\alpha}\hat{\vep})\cos\vp_p,
\ee
the mass operator in Eq.~\eqref{Sigma0} takes the form,
\be
\Sigma(\vep ;T)=[\tilde{A}_p-m]+(\boldsymbol {\gamma}\hat{\vep })[\tilde{B}_p-p]+\gamma_0\tilde{C}_p,
\label{SigmaT}
\ee
and thus acquires an additional term on the right-hand side
proportional to the Dirac matrix $\gamma_0$, with the coefficient
\be
\tilde{C}_p=-\frac12\int\frac{d^3k}{(2\pi)^3}(n_k-\bar{n}_k)V(\vep-\vek).
\label{tildeC}
\ee
However, this extra term appears only for the chemical potential $\mu\neq 0$ when $n_p\neq \bar{n}_p$ \cite{Kocic:1985uq} while in what follows the case of a finite temperature but vanishing chemical potential will be studied, so we have $\tilde{C}_p=0$. Meanwhile, both $n_p$ and $\bar{n}_p$ will be kept separately in all other equations for generality.
Thus, an infrared-finite thermal mass-gap equation takes the form
\be
\tilde{A}_p\cos\vp_p-\tilde{B}_p\sin\vp_p=0,
\label{mgeT}
\ee
with the modified functions $\tilde{A}_p$ and $\tilde{B}_p$ introduced in Eq.~\eqref{tildeAB}. Then the effect of the temperature on the solution of the mass-gap equation \eqref{mgeT} can be easily anticipated from the behaviour of the factor $1-n_p-\bar{n}_p$ that appears under the momentum integral in Eq.~\eqref{tildeAB} and accompanies the confining potential. The temperature dependence of the Fermi--Dirac thermal functions in Eq.~\eqref{nnnew} suggests that the interaction between the quark and antiaquark is damped
with the temperature rise. Most importantly, the damping should be the strongest for small interquark momenta, that is, in the region crucial for the effect of SBCS --- see Fig.~\ref{fig:vp}. In other words, thermal quark--antiquark pairs produced by the temperature occupy the states with the lowest energies. As a result, these low-lying states become unavailable for the Cooper-like pairs that, therefore, fail to condense in the vacuum. This way the temperature suppresses SBCS via the Pauli blocking effect. In Fig.~\ref{fig:rest}, we show the temperature dependences of the chiral angle, effective quark mass, damping factor, and chiral condensate that were obtained numerically for the linear confining potential in \cite{Glozman:2024xll,Glozman:2024dzz}. All plotted quantities indeed demonstrate the pattern just outlined and, in particular, the chiral condensate vanishes at some finite temperature $\Tch$ thus hinting that the vacuum becomes chirally symmetric at $T>\Tch$. It has to be noted that the chiral condensate is no more than the lowest-order vacuum expectation value (VEV) that breaks chiral symmetry, so a vanishing chiral condensate does not necessarily imply chiral restoration since chiral symmetry may still be broken by higher-order VEV's. Thus, it is more crucial for chiral restoration that, at $T>\Tch$, the thermal mass-gap equation \eqref{mgeT} possesses only a trivial solution with $\vp_p\equiv 0$, so the second term on the right-hand side of Eq.~\eqref{ChiralChar} vanishes identically. Consequently, the criterion for SBCS in Eq.~\eqref{sbcs} fails and chiral restoration in the vacuum (and, therefore, in the spectrum built on top of it) takes place.

\begin{table}[t!]
\tbl{The results of the numerical calculations in the GNJL model with linear confinement \cite{Glozman:2024xll,Glozman:2024dzz}. All values in the units of $\sqrt{\sigma}$ are predicted by the model; the values in MeV are obtained by fixing the string tension $\sigma$ to provide a phenomenological value of the chiral condensate quoted in the third row. \label{tab:res}}
{\begin{tabular}{@{}cccc@{}} \toprule
 & $|\braket{\bar{q}q}_0|^{1/3}$ & $M_p(0)$ & $T_c$\\
 \colrule
In units of $\sqrt{\sigma}$ & 0.231 & 0.166 & 0.084 \\
\colrule
In MeV & 250 (fixed) & 180 & 90\\
\botrule
\end{tabular}}
\end{table}

The results of numerical calculations obtained in \cite{Glozman:2024xll,Glozman:2024dzz} for linear confinement in Eq.~\eqref{Vlin2}  are collected in Table~\ref{tab:res}. In the chiral limit of massless quarks, all dimensional quantities in the model are proportional to the string tension $\sigma$ to the appropriate power as the only dimensional parameter of the model. Then parameter-free predictions of the model come as relations between physical quantities after the exclusion of the parameter $\sigma$. In particular, the model predicts \cite{Glozman:2024xll}
\be
\Tch\approx 0.36\,|\braket{\bar{q}q}_0|^{1/3}.
\label{Tch}
\ee
If the chiral condensate on the right-hand side of the latter relation is fixed to a phenomenologically adequate value $(-250~\mbox{MeV})^3$, then the chiral restoration temperature is calculated to be around 90 MeV, as quoted in Table~\ref{tab:res}. It appears in a qualitative agreement with (i) $\Tch \simeq 132^{+3}_{-6}$~MeV extracted on the lattice for QCD with massless $u$ and $d$ quarks and a physical strange quark mass \cite{HotQCD:2019xnw} and (ii) the chiral restoration temperature obtained in \cite{Quandt:2018bbu} through the compactification of one spatial direction in Euclidean space.
Furthermore, the effective quark mass  predicted by the model to be
\be
M_p(0)\approx 0.72\, |\braket{\bar{q}q}_0|^{1/3},
\ee
takes a numerical value around 180~MeV (see Table~\ref{tab:res}) that is fairly close to the phenomenological value of the constituent quark mass in Eq.~\eqref{mconst}. This way the employed chiral quark model provides a natural explanation for the constituent quark mass, that appears due to the effect of SBCS, and provides a decent numerical estimate for it.

The behaviour of the chiral condensate on the temperature deserves a dedicated comment. It was found in \cite{Glozman:2024dzz} that, in approaching the critical temperature $\Tch$ from below, the chiral condensate can be approximately described by a simple function,
\be
\braket{\bar{q}q}_{T\to\Tch}/\braket{\bar{q}q}_0
=2.39(1-T/\Tch)^{0.54},
\ee
where the numerical coefficients were found from a fit to the numerical solution. Therefore, the obtained numerical dependence is consistent with the behaviour
\be
\braket{\bar{q}q}_{T\to\Tch}/\braket{\bar{q}q}_0\propto \sqrt{1-T/\Tch}
\label{chT}
\ee
that is natural for the phase transition order parameter and hints toward a distinct chiral phase transition (rather than a smooth crossover) as expected in the large-$N_c$ limit \cite{Cohen:2023hbq}.
On the other hand, the curve for the temperature dependence of the condensate in Fig.~\ref{fig:vp} looks inconsistent with the well-known expression obtained employing the chiral perturbation theory at small temperatures, $T\ll f_\pi$, \cite{Gasser:1987ah}
\be
\braket{\bar{q}q}_{T\to\Tch}/\braket{\bar{q}q}_0=1-T^2/(8f_{\pi}^2)+\ldots,
\label{chc}
\ee
where the ellipsis stands for higher terms in the expansion parameter $T/f_\pi$ (see, for example, \cite{Hofmann:2020ism}). This failure of the GNJL model should not come as a surprise given that, as will be shown below, the pion decay constant $f_\pi$ scales as $\sqrt{N_c}$ --- see Eq.~\eqref{fpidef}. Then, since all $N_c$-suppressed contributions were disregarded, the dependence in Eq.~\eqref{chc} lies beyond the precision of the BCS approximation for GNJL, so a very slow temperature dependence of the chiral condensate at small $T$'s demonstrated by the corresponding plot in Fig.~\ref{fig:vp} comes as a natural consequence of the approximations made.

As the final remark that concludes this section devoted to the GNJL model studied at the BCS level, we note that the numerical value of the parameter $\sqrt{\sigma}\simeq 1$~GeV fixed from the chiral condensate and employed in the estimates of dimensional quantities above, approximately two times exceeds the values usually assumed in quark model calculations with linear confinement. It should not come a surprise either given that, in realistic versions of quark models, a Cornell potential in Eq.~\eqref{Cornell} should be employed rather than linear confinement alone.
However, inclusion of the Coulomb potential and a constant term, that turns to a $\delta$-function in momentum space, require a suitable smearing and, therefore, has to be treated with care when solving the mass-gap equation. In the meantime, such more sophisticated potential can be arranged in a self-consistent way, with all its parameters and predictions lying within the domain of phenomenologically adequate values --- see, for example. \cite{Bicudo:2002eu} Thus, our aiming at qualitative understanding of SBCS and phenomena related to it justifies our sticking to a simplified version of the interquark potential or treating it in most general terms without specifying its particular form.

\section{Quark--antiquark bound states and their properties}
\label{sec:BS}

\subsection{General consideration and posed questions}
\label{sec:genquest}

\begin{figure}[t!]
\centering
\includegraphics[width=0.7\textwidth]{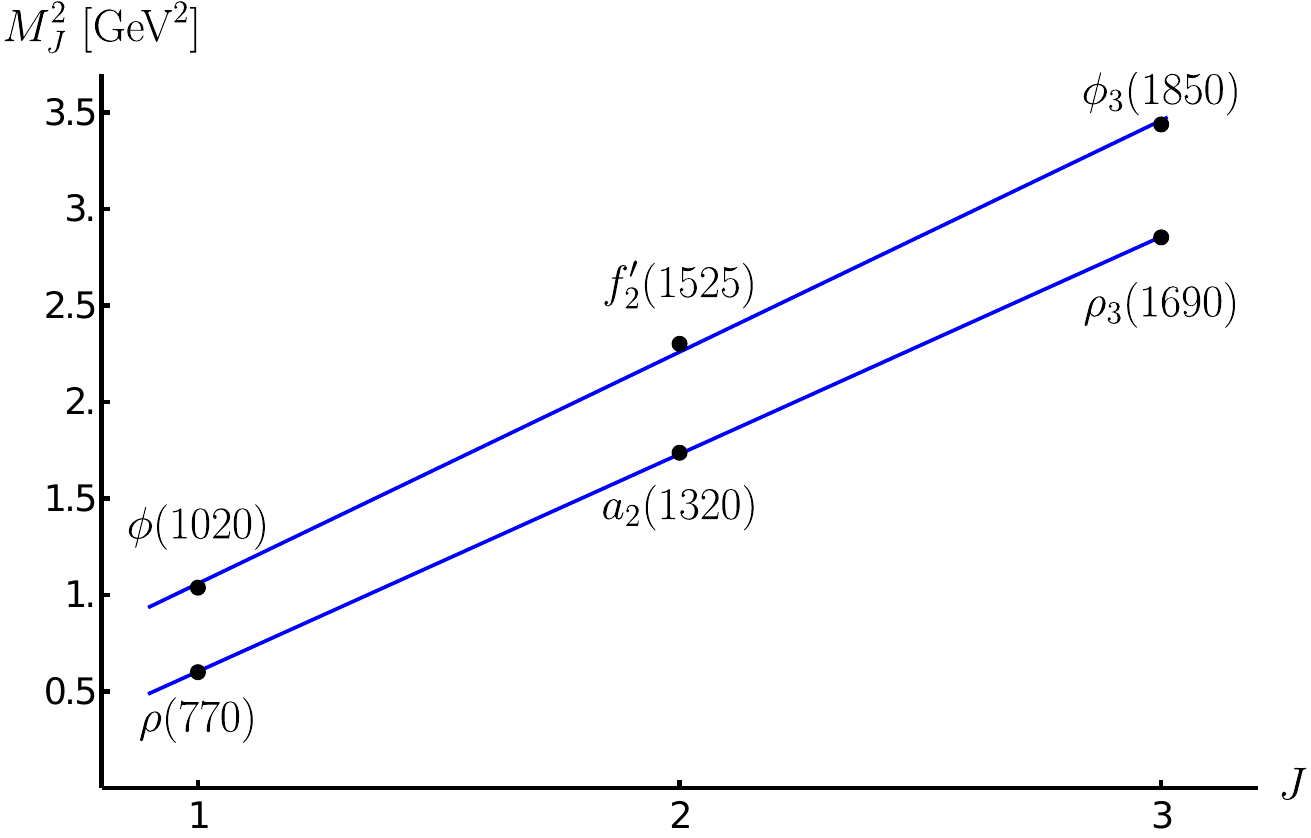}
\caption{The Regge trajectories for the experimentally measured masses (black dots) of several low-lying $\bar{q}q$ (with $q$ for the $u$ or $d$ quark) mesons with isospin $I=1$ and isoscalar $\bar{s}s$ mesons \cite{ParticleDataGroup:2024cfk}. Linear fits are shown as blue lines.}
\label{fig:regge}
\end{figure}

QCD, as the theory of strong interactions, should explain the spectrum and properties of hadrons as observable states composed of quarks, antiquarks, and gluons. Then the main purpose of quark models, as approximations for QCD, is to provide a simplified picture of the phenomena related to hadrons suitable for analytical and simple (as compared to the first principle calculations on the lattice) numerical studies. In particular, a straightforward prediction
of the constituent quark model in Eq.~\eqref{Salp} is the form of the dependence $J(M_J)$ with $J$ for the total spin of the meson and $M_J$ for its mass. This dependence is known as a Regge trajectory \cite{Regge:1959mz,Collins:1971ff} and its linear behaviour,
\be
J=\alpha' M_J^2+\alpha_0,
\label{Reggetrajectory}
\ee
is widely discussed in the literature
\cite{Chew:1962mpd,Kobzarev:1992wt,Olsson:1996jg} --- see also Fig.~\ref{fig:regge} for an example. The dependence in ~Eq.~\eqref{Reggetrajectory}
can be readily derived theoretically
for the linear confinement in Eq.~\eqref{Vlin2}, as the most phenomenologically justified choice,
and the slope $\alpha'$ can be related to the string tension parameter $\sigma$.\footnote{For a generic power-like confining potential in Eq.~\eqref{Vconf}, the power of $M_J$ in Eq.~\eqref{Reggetrajectory} is $1+1/\alpha$.}
However, two comments are in order here. On the one hand, non-vanishing masses of the quarks are expected to slightly spoil the linearity on the trajectories for small $J$'s. On the other hand, the theoretically predicted slope of the trajectories somewhat changes if a simple confining potential is superseded by a more sophisticated construction that effectively takes into account the degrees of freedom inherited from gluons. In the latter case, the interaction becomes momentum-dependent and the corresponding object that emerges is often referred to as the flux tube \cite{Isgur:1984bm} or QCD string \cite{Dubin:1994vn,Dubin:1995vw,Allen:2000sd}. Also, the form of the Salpeter equation \eqref{Salp} raises (but, as will be argued below, can not properly answer) a long-standing question on the Lorentz nature of confinement. Formulated in simple terms it implies that, if the potential is added to the quark mass it is regarded as scalar confinement while the potential added to the energy or momentum of the quark implies vector confinement. Na{\"i}vely, this definition entails that the confining potential in Eq.~\eqref{Salp} has to be regarded as a time-like vector interaction that is vulnerable to the famous Klein paradox \cite{Klein:1929zz}. This paradox consists in the observation that, under certain circumstances (including a pure time-like vector interaction growing with the distance), the flux of the reflected wave in relativistic scattering may exceed the flux of the incident wave --- for a discussion of this paradox in the Dirac equation see \cite{Nikishov:1970br}. Then one is forced to conclude that the Salpeter equation \eqref{Salp} is ill-starred and as such can not be used in phenomenological studies of hadrons. This conclusion is, however, misleading since as a matter of fact the form of Eq.~\eqref{Salp} does not imply a time-like rising potential between quarks in the corresponding Dirac-like equation --- see, for example, a detailed discussion in \cite{Nefediev:2005dw}. We are revisiting this issue below and demonstrate that, contrary to na{\"i}ve expectations, Eq.~\eqref{Salp} arises due to SBCS and corresponds to an effective scalar confining interaction.

As was already argued above, a simple quantum mechanical model in Eq.~\eqref{Salp} and its possible extensions that do not include quark spins as intrinsic degrees of freedom inevitably miss features tightly linked to chiral symmetry and its spontaneous breaking in the vacuum of QCD. Then a more sophisticated chiral quark model (GNJL being a natural choice) is to be invoked to address the following questions:
\begin{itemize}
\item How does the chiral pion appear as the lightest quark--antiquark meson and the Goldstone boson of SBCS at the same time?
\item What properties of the spectrum of hadrons in QCD are related to SBCS and what implications of this phenomenon can (or can not) be seen in this spectrum?
\item What emergent symmetries are inherent in the spectrum of hadrons when chiral symmetry is restored in the vacuum at $T>\Tch$?
\end{itemize}

Employing the experience gained so far, one can sketch possible answers to the questions just raised. It was demonstrated above that the mass-gap equation \eqref{mge} can be written in the form of a Salpeter-like equation \eqref{Salp2} for a boson with the radial wave function in momentum space (up to the overall normalisation and corrections in the small quark mass beyond the strict chiral limit) given by $\sin\vp_p$. A natural identification for this boson would be the lightest quark--antiquark meson with the quantum numbers $0^{-+}$, that is, the pion. Below we shall demonstrate that this identification is correct. On the other hand, SBCS in the vacuum entails that opposite-parity hadronic states do not form degenerate doublets. It is easy to see that this feature is inherent in the quark model based on Eq.~\eqref{Salp}. Indeed, the spatial parity of a quark--antiquark state is calculated as $P=(-1)^{L+1}$, with $L$ for the angular momentum in the quark--antiquark system. Then the opposite-parity states
need to have $L$'s different by a unity. Since spins of the quarks are not intrinsic degrees of freedom of the model, then $\Delta L=1$ implies $\Delta J=1$, with $J$ for the total spin of the meson. In the meantime, the Regge behaviour of the spectrum in Eq.~\eqref{Reggetrajectory} predicts that a unit shift in the total spin $J$ implies a constant (independent on $J$) shift in $\Delta M_J^2$. Then a clear prediction of this simple quark model is a pair of Regge trajectories for opposite-parity states that are parallel to each other for all values of the spin $J$. This result can be scrutinised employing a chiral quark model \eqref{GNJL} to observe an obvious tension between the predictions of the two models. Indeed, as argued on general grounds in \cite{Glozman:2004gk}, SBCS is a purely quantum effect\footnote{Notice that, in the GNJL model, the interaction terms on the right-hand side of both expressions for $A_p$ and $B_p$ in Eq.~\eqref{AB} are proportional to the Planck constant $\hbar$ and vanish in the formal classical limit of $\hbar\to 0$ \cite{Glozman:2005tq}.} that should, therefore, have no implications in the quasiclassical part of the hadronic spectrum with $J\gg 1$. Then hadrons are expected to approximately fill the multiplets inherent in the Weyl--Wigner realisation of chiral symmetry in nature and, in particular, opposite-parity states with the same radial excitation number should become gradually degenerate in mass with the growth of $J$. Moreover, the rate of this effective chiral restoration with $J$ should be fast enough, so that the two sibling Regge trajectories for $M_J^2$ merge in the regime with $J\gg 1$\footnote{Note that the Regge behaviour in Eq.~\eqref{Reggetrajectory} implies that $\Delta M_J\propto 1/\sqrt{J}\to 0$ for $J\gg 1$, which has nothing to do with the effective chiral symmetry restoration in the spectrum.}. Below we shall demonstrate that it is indeed the case for heavy--light mesons and provide a microscopic picture of the phenomenon \cite{Kalashnikova:2005tr}.

Finally, in the previous section, the GNJL model was demonstrated to describe a chiral restoration at some finite temperature $\Tch$ predicted selfconsistently within the model itself. Importantly, the driving force for the chiral restoration is not deconfinement but rather the effect of Pauli blocking that results in populating
the low-lying levels necessary for SBCS with thermal quarks and antiquarks thus leaving less room to the condensation of the BCS-like $q\bar{q}$ pairs in the vacuum. Importantly, in the given model, quarks remain confined at all temperatures including $T>\Tch$ when chiral symmetry is restored in the vacuum. Then relevant degrees of freedom of the model in this regime are chirally symmetric but confined hadrons \cite{Glozman:2022lda,Lowdon:2022xcl,Bala:2023iqu,Cohen:2024ffx}.

\subsection{Bound state equation for quark--antiquark mesons}
\label{sec:BSqq}

\begin{figure}[t!]
\centering
\includegraphics[width=0.7\textwidth]{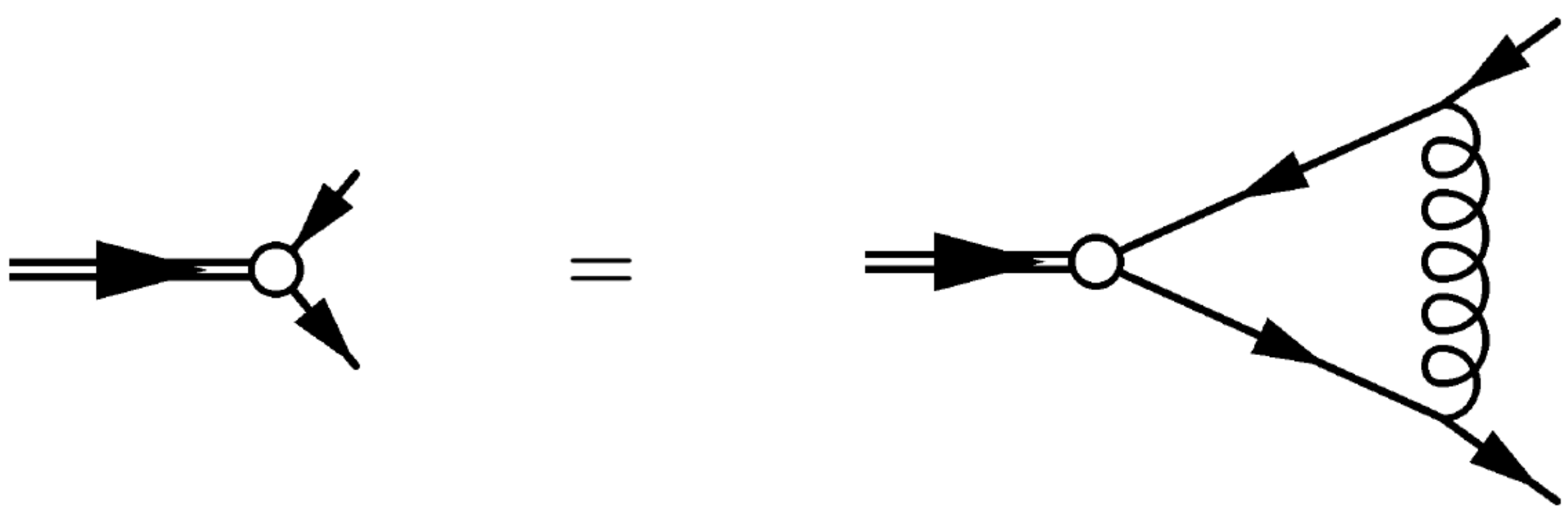}
\caption{The Bethe--Salpeter equation \eqref{BSeq} derived in the ladder approximation for the interquark interaction. The single, double, and curly line correspond to the quark (antiquark), meson, and the confining interaction, respectively. Adapted from \cite{Glozman:2024dzz}.}
\label{fig:bseq}
\end{figure}

In this section, we proceed to a derivation of the bound state equation for quark--antiquark mesons.
A generic quark--antiquark meson in its rest frame can be described by a matrix amplitude $\chi(\vep;M)$, with $\vep$ for the momentum of the quark ($-\vep$ for the antiquark) and $M$ for the mass of the meson, that obeys a Bethe--Salpeter equation schematically depicted in Fig.~\ref{fig:bseq}. Notice that, as before, only planar diagrams are retained resulting in the ladder approximation for the meson--quark--antiquark vertex, which is justified in the employed large-$N_c$ limit. Then the equation that emerges for the matrix amplitude $\chi(\vep;M)$ reads
\be
\chi({\vec p};M)=-i\int\frac{d^4q}{(2\pi)^4}V(\vep-\veq)\gamma_0 S(q_0+M/2,\veq)
\chi(\veq;M)S(q_0-M/2,\veq)\gamma_0.
\label{BSeq}
\ee
This bound state equation valid at $T=0$ was intensively studied in the literature --- see, for example, \cite{LeYaouanc:1984ntu,Bicudo:1989si,Nefediev:2004by,Wagenbrunn:2007ie} as well as many others. To generalise it to finite temperatures we employ the experience gained in Sec.~\ref{sec:finT} and substitute $S(p_0,\vep)$ by $S(p_0,\vep;T)$ from Eq.~\eqref{ST} that can be presented as a sum,
\be
S(p_0,\vep;T)=S(p_0,\vep)+\Delta S(p_0,\vep;T).
\label{ST2}
\ee
The further derivation follows the standard lines well documented in the literature, so we refrain from a detailed discussion here and only briefly mention its most essential steps.

First of all, we notice that the interaction $V(\vep-\veq)$ is instantaneous, so the mesonic Bethe--Salpeter amplitude $\chi(\vep;M)$ in Eq.~\eqref{BSeq} does not depend on the temporal component of the meson momentum $p_0$.
Therefore, the integration in the energy $q_0$ on the right-hand side can be done explicitly yielding
\begin{multline}
i\int\frac{dq_0}{2\pi}\gamma_0 S(q_0+M/2,\veq;T)
\chi(\veq;M)S(q_0-M/2,\veq;T)\gamma_0\\
=i(1-n_q-\bar{n}_q)\int\frac{dq_0}{2\pi}\gamma_0S(q_0+M/2,\veq)\chi(\veq;M)S(q_0-M/2,\veq)\gamma_0\\
=(1-n_q-\bar{n}_q)\gamma_0\Bigl[(\Lambda_+\gamma_0)\chi^{[+]}(\Lambda_-\gamma_0)
+(\Lambda_-\gamma_0)\chi^{[-]}(\Lambda_+\gamma_0)\Bigr]\gamma_0,
\label{SSfin}
\end{multline}
where it was used that
\be
i\int\frac{dq_0}{2\pi}\left[\frac{1}{q_0\pm
M/2-E_q+i\epsilon}\right] \left[\frac{1}{q_0\mp
M/2+E_q-i\epsilon}\right]
=\frac{1}{2E_q\mp M}.
\ee
Then it is easy to verify that the following relations hold:
\be
\begin{split}
i\int\frac{dq_0}{2\pi}S(q_0+M/2,\veq)\chi(\veq;M)&S(q_0-M/2,\veq)
\\
&=(\Lambda_+\gamma_0)\chi^{[+]}(\Lambda_-\gamma_0)
+(\Lambda_-\gamma_0)\chi^{[-]}(\Lambda_+\gamma_0),
\label{SS}
\end{split}
\ee
\begin{align}
i\int\frac{dq_0}{2\pi i}S(q_0+M/2,\veq)&\chi(\veq;M)\Delta S(q_0-M/2,\veq;T)\nonumber\\
=&-n_q\left[(\Lambda_-\gamma_0)\chi^{[-]}(\Lambda_+\gamma_0)
+(\Lambda_+\gamma_0)\chi^{[0]}(\Lambda_+\gamma_0)
\right]\label{SdS}\\
&-\bar{n}_q\left[(\Lambda_+\gamma_0)\chi^{[+]}(\Lambda_-\gamma_0)
-(\Lambda_-\gamma_0)\chi^{[0]}(\Lambda_-\gamma_0)\right],\nonumber
\end{align}
\begin{align}
i\int\frac{dq_0}{2\pi i}\Delta S(q_0+M/2,&\veq;T)\chi(\veq;M)S(q_0-M/2,\veq)\nonumber\\
=&-n_q\left[(\Lambda_+\gamma_0)\chi^{[+]}(\Lambda_-\gamma_0)
-(\Lambda_+\gamma_0)\chi^{[0]}(\Lambda_+\gamma_0)
\right]\label{dSS}\\
&-\bar{n}_q\left[(\Lambda_-\gamma_0)\chi^{[-]}(\Lambda_+\gamma_0)
+(\Lambda_-\gamma_0)\chi^{[0]}(\Lambda_-\gamma_0)\right],\nonumber
\end{align}
\be
\int\frac{dq_0}{2\pi i}\Delta S(q_0+M/2,\veq;T)\chi(\veq;M)\Delta S(q_0-M/2,\veq)=0,
\label{dSdS}
\ee
where for convenience we introduced shorthand notations
\be
\chi^{[+]}(\vep;M)=\frac{\chi(\vep;M)}{2E_p-M},\quad
\chi^{[-]}(\vep;M)=\frac{\chi(\vep;M)}{2E_p+M},\quad
\chi^{[0]}(\vep;M)=\frac{\chi(\vep;M)}{M}.
\label{chipm}
\ee

Notice that, by virtue of Eq.~\eqref{SSfin}, the temperature dependence in the resulting thermal bound state equation is again squeezed to the factor $1-n_p-\bar{n}_p$ that multiplies the potential. Quite predictably, the bound state equation shares this feature with the
thermal mass-gap equation \eqref{mgeT}. Then, for brevity, we define
\be
(1-n_q-\bar{n}_q)V(\vep-\veq)\equiv V(\vep,\veq;T)
\ee
and use this shorthand notation in the formulae below.

\begin{figure}[t]
\begin{center}
\includegraphics[width=0.9\textwidth]{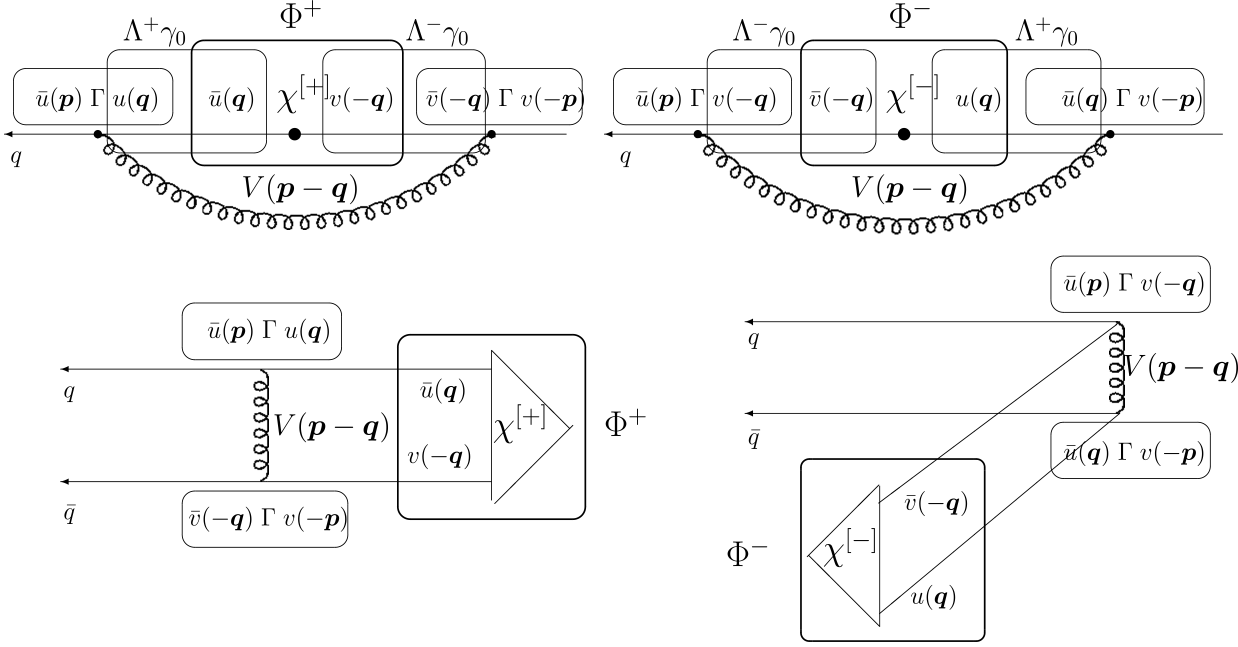}
\end{center}
\caption{First row: a graphical mapping between the Dirac and spin
representations for the meson wave functions. Second row: forward and backward in time motion of the quark--antiquark pair in the meson. Adapted from \cite{Nefediev:2004by}.}
\label{fig:messalp}
\end{figure}

The two structures that emerge in Eq.~\eqref{SSfin} read
\be
\begin{split}
(\Lambda_+\gamma_0)\chi^{[+]}(\Lambda_-\gamma_0)
&=\sum_{s_3,s_4}
u_{s_3}(\vep)\Bigl[\bar{u}_{s_3}(\vep)\chi^{[+]}(\vep;M)v_{-s_4}(-\vep)\Bigr]\bar{v}_{-s_4}(-\vep)\\
&\equiv\sum_{s_3,s_4}
u_{s_3}(\vep)\Phi_{s_3s_4}^{+}(\vep;M)\bar{v}_{-s_4}(-\vep)
\end{split}
\label{phiplus}
\ee
and
\be
\begin{split}
(\Lambda_-\gamma_0)\chi^{[-]}(\Lambda_+\gamma_0)
&=\sum_{s_3,s_4}
v_{-s_3}(-\vep)\Bigl[\bar{v}_{-s_3}(-\vep)\chi^{[-]}(\vep;M)u_{s_4}(\vep)\Bigr]\bar{u}_{s_4}(\vep)\\
&\equiv\sum_{s_3,s_4}
v_{-s_3}(-\vep)\Phi_{s_3s_4}^{-}(\vep;M)\bar{u}_{s_4}(\vep),
\end{split}
\label{phiminus}
\ee
where the definition of the projectors $\Lambda_\pm$ in
Eq.~\eqref{Lambdasdef} was used. This way we introduced the wave functions $\Phi^{[\pm]}$ that are $2\times 2$ matrices in the basis of the quark helicities.
Then, sandwiching the bound state equation \eqref{BSeq} between $\bar{u}_{s_1}(\vep)$ and $v_{-s_2}(-\vep)$ or between $\bar{v}_{-s_1}(-\vep)$ and $u_{s_2}(\vep)$ and
using that, according to the definition of the matrix wave functions $\Phi^{[\pm]}$ in Eqs.~\eqref{phiplus} and \eqref{phiminus},
\be
\begin{split}
\bar{u}_{s_1}(\vep)\chi(\vep;M)v_{-s_2}(-\vep)=
(2E_p-M)\Phi_{s_1s_2}^{+}(\vep;M),\\
\bar{v}_{-s_1}(-\vep)\chi(\vep;M)u_{s_2}(\vep)=
(2E_p+M)\Phi_{s_1s_2}^{-}(\vep;M),
\end{split}
\ee
we arrive at the bound state equation in the form of two coupled equations,
\be
\begin{split}
[2E_p-M]\Phi_{s_1s_2}^+(\vep;M)=\sum_{s_3s_4}\int\frac{d^3q}{(2\pi)^3}
&\left(T^{++}_{s_1s_3s_4s_2}(\vep,\veq)\Phi_{s_3s_4}^+(\veq;M)\right.\\
+&\left.T^{+-}_{s_1s_3s_4s_2}(\vep,\veq)\Phi_{s_3s_4}^-(\veq;M)\right),\\
[2E_p+M]\Phi_{s_1s_2}^-(\vep;M)=\sum_{s_3s_4}\int\frac{d^3q}{(2\pi)^3}
&\left(T^{-+}_{s_1s_3s_4s_2}(\vep,\veq)\Phi_{s_3s_4}^+(\veq;M)\right.\\ +&\left.T^{--}_{s_1s_3s_4s_2}(\vep,\veq)\Phi_{s_3s_4}^-(\veq;M)\right),
\label{Salpeterlev5}
\end{split}
\ee
where the quantities
\be
\begin{split}
&T^{++}_{s_1s_3s_4s_2}(\vep,\veq)=
[\bar{u}_{s_1}(\vep)\gamma_0 u_{s_3}(\veq)][-V(\vep,\veq;T)]
[\bar{v}_{-s_4}(-\veq)\gamma_0 v_{-s_2}(-\vep)],\\
&T^{+-}_{s_1s_3s_4s_2}(\vep,\veq)=
[\bar{u}_{s_1}(\vep)\gamma_0 v_{-s_3}(-\veq)][-V(\vep,\veq;T)]
[\bar{u}_{s_4}(\veq)\gamma_0 v_{-s_2}(-\vep)],\\
&T^{-+}_{s_1s_3s_4s_2}(\vep,\veq)=
[\bar{v}_{-s_1}(-\vep)\gamma_0 u_{s_3}(\veq)][-V(\vep,\veq;T)]
[\bar{v}_{-s_4}(-\veq)\gamma_0 u_{s_2}(\vep)],\\
&T^{--}_{s_1s_3s_4s_2}(\vep,\veq)=
[\bar{v}_{-s_1}(-\vep)\gamma_0 v_{-s_3}(-\veq)][-V(\vep,\veq;T)][\bar{u}_{s_4}(\veq)\gamma_0
u_{s_2}(\vep)]
\label{ampls}
\end{split}
\ee
describe the two quark or antiquark vertices connected via the confining interaction. Obviously, the generalisation of the Lorentz nature of the interaction in Eq.~\eqref{ampls} from $\gamma_0\times\gamma_0$ (see the Hamiltonian in Eq.~\eqref{GNJL}) to the most general form $\Gamma\times\Gamma$ (with $\Gamma$ for a generic Dirac matrix) is straightforward.
This way we have proceeded from the Dirac representation for the meson wave function (the amplitudes $\chi^{[\pm]}$) to its helicity representation (the amplitudes $\Phi^{\pm}$) --- see the graphical mapping scheme presented in the first row in Fig.~\ref{fig:messalp}.
Notice also that the form of the vertices in Eq.~\eqref{ampls} suggests a straightforward physical interpretation of the amplitudes $\Phi^{\pm}$. Indeed, while in the diagonal terms $T^{++}$ and $T^{--}$ each vertex contains either two quark amplitudes $u$ or two antiquark amplitudes $v$, in the off-diagonal terms $T^{+-}$ and $T^{-+}$ each vertex contains one amplitude $u$ and one amplitude $v$. Therefore, the amplitudes $\Phi^+$ and $\Phi^-$ describe the forward and backward in time motion of the quark--antiquark pair in the meson, and the off-diagonal terms $T^{+-}$ and $T^{-+}$ connect these two types of motion. It is important to stress that, due to the instantaneous nature of the interquark interaction, the quark and antiquark in the meson always proceed from their forward to backward in time motion and vice versa \emph{in unison} --- see the second row in Fig.~\ref{fig:messalp}.\label{hl} Then, in the limit of an infinitely heavy antiquark, its backward in time motion is forbidden and, therefore, the light quark attached to it by an instantaneous confining potential will also be forced to travel only forward in time. We shall benefit from this property in the consideration of the spectrum of heavy--light mesons below.

If the off-diagonal terms $T^{+-}$ and $T^{-+}$ are neglected in Eq.~\eqref{Salpeterlev5}, the system  splits into two
decoupled equations related by a trivial symmetry $\{M\leftrightarrow -M,\Phi^+\leftrightarrow\Phi^-\}$ and describing the meson $\bar{q}_1q_2$ and the corresponding ``antimeson'' $\bar{q}_2q_1$, respectively. Then the equation for the meson takes the form of a Salpeter equation in Eq.~\eqref{Salp} (if additionally the dressed quark energy $E_p$ is approximated by the free energy $E_p^{(0)}=\sqrt{\vep^2+m^2}$), so this way we establish a connection between GNJL and a simple potential quark model. Therefore, the presence of the off-diagonal terms in Eq.~\eqref{Salpeterlev5} and the resulting coupling of the forward and backward in time motions of the quark and antiquark constitute an essential difference between these two models and allow GNJL to naturally explain a small (vanishing in the strict chiral limit) mass of the lightest pseudoscalar state in the mesonic spectrum. Indeed, the two wave functions of the pion with the quantum numbers $J^{PC}=0^{-+}$ in the helicity basis can be presented in the form
\be
\Phi_{s_1s_2}^{\pm}(\vep)=\frac{1}{\sqrt{2}}\left(i\sigma_2\right)_{s_1s_2}
Y_{00}(\hat{\vep})\vp_\pi^\pm(p),
\label{vppi}
\ee
where $Y_{00}(\hat{{\bm p}})=1/\sqrt{4\pi}$ is the normalised to unity lowest spherical harmonic, the matrix $i\sigma$ ensures the pseudoscalar quantum numbers of the state (\emph{cf} the anomalous term in Eq.~\eqref{ChiralChar}), and $\vp_\pi^\pm(p)$ are the two radial wave functions. Then it is easy to find that
\be
\begin{split}
T_\pi^{++}(\vep,\veq)=T_\pi^{--}(\vep,\veq)&=-V(\vep,\veq;T)
\left[\cos^2\frac{\vp_p-\vp_q}{2}-\frac{1-(\hat{{\bm p}}\hat{\veq})}{2}\cos\vp_p\cos\vp_q\right],\\
\label{pia}
T_\pi^{+-}(\vep,\veq)=T_\pi^{-+}(\vep,\veq)&=-V(\vep,\veq;T) \left[\sin^2\frac{\vp_p-\vp_q}{2}+\frac{1-(\hat{{\bm
p}}\hat{\veq})}{2}\cos\vp_p\cos\vp_q\right],
\end{split}
\ee
possessing an attractive property,
\be
T_\pi^{++}(\vep,\veq)+T_\pi^{+-}(\veq,\vep)=T_\pi^{-+}(\vep,\veq)+T_\pi^{--}(\veq,\vep)=-V(\vep,\veq;T).
\ee

Let us start from the case of $T=0$. Then, in the strict chiral limit of $m=0$ (and, therefore, $M_\pi=0$), the two equations in the system \eqref{Salpeterlev5} become identical and take the form
\be
2E_p\vp_\pi(p)=\int\frac{d^3q}{(2\pi)^3}[T_\pi^{++}(\vep,\veq)+T_\pi^{+-}(\vep,\veq)]\vp_\pi(q)=
-\int\frac{d^3q}{(2\pi)^3}V({\bm p}-\veq)\vp_\pi(q),
\ee
where $\vp_\pi(p)=\vp_\pi^+(p)=\vp_\pi^-(p)={\cal N}\sin\vp_p$, with ${\cal N}$ for the norm that is irrelevant for the present discussion and will be addressed in detail below. Then passing over to the coordinate space, one readily arrives at Eq.~\eqref{Salp2} with $M=0$ which, as was argued above, is identical to the mass-gap equation \eqref{mge}.
This way the dual nature of the chiral pion as the Goldstone boson of SBCS and the lightest pseudoscalar state in the spectrum of hadrons is firmly established. This connection between the two faces of the pion holds for any $T<\Tch$ but is ruined at $T>\Tch$ when the pion still exists as the lightest pseudoscalar meson in the spectrum, however, it is no more the Goldstone boson since chiral symmetry is not spontaneously broken. We provide a detailed discussion of the properties of the pionic solution to the bound state equation in Eq.~\eqref{BSeq} in Sec.~\ref{sec:pion}.

\subsection{Bogoliubov-like transformation for mesons}

In the previous section, the bound state equation for quark--antiquark mesons in Eq.~\eqref{Salpeterlev5} was derived from the matrix Bethe--Salpeter equation depicted graphically in Fig.~\ref{fig:bseq}.
In particular, it was argued that each mesonic state is described by two wave functions simultaneously that correspond to the forward and backward in time motion of the quark--antiquark pair in the meson. However, their orthogonality and normalisation conditions have not been addressed so far. In this section, we re-derive the bound state equation
\eqref{Salpeterlev5} employing a Hamiltonian approach in \cite{Kalashnikova:1999wt} and demonstrate how the
properties of these two wave functions naturally arise in this case.

In Sec.~\ref{sec:BCS}, the effect of SBCS in the GNJL model was discussed at the level of the dressed non-interacting quarks (BCS level), and the Hamiltonian of the model was diagonalised via a fermionic Bogoliubov-Valatin transformation from bare to dressed quarks --- see Eq.~\eqref{H2diag}. However, proceeding from quarks to mesons implies that the interaction term $:H_4:$ in the Hamiltonian \eqref{H024} ($N_c$-suppressed at the BCS level) must be equally important as the term $:H_2:$ responsible for the quark dressing. Therefore, one needs to proceed beyond the BCS approximation and deal with quark--antiquark states rather than non-interacting dressed quarks. To this end we define two diagonal compound operators,
\be
\begin{split}
&B_{ss'}(\vep,\vep')=\frac{
1}{\sqrt{N_c}}\sum_{\alpha=1}^{N_c} b_{\alpha
s}^\dagger(\vep)b_{s'}^\alpha(\vep'),\\
&D_{ss'}(\vep,\vep')=\frac{
1}{\sqrt{N_c}}\sum_{\alpha=1}^{N_c} d_{\alpha
s}^\dagger(-\vep)d_{\alpha s'}(-\vep'),
\label{operatorsBD}
\end{split}
\ee
``counting'' the number of quarks and antiquarks and two anomalous operators,
\be
\begin{split}
&M^\dagger_{ss'}(\vep,\vep')=\ds\frac{\ds
1}{\ds\sqrt{N_c}}\sum_{\alpha=1}^{N_c} b^\dagger_{\alpha
s'}(\vep')d^\dagger_{\alpha s}(-\vep),\\
&M_{ss'}(\vep,\vep')=\ds\frac{\ds
1}{\ds\sqrt{N_c}}\sum_{\alpha=1}^{N_c} d_{\alpha s}(-\vep)b_{\alpha
s'}(\vep'),
\label{operatorsMM}
\end{split}
\ee
creating and annihilate quark-antiquark pairs with the given momenta and helicities.
In the limit $N_c\to\infty$,
these compound operators obey the standard bosonic commutation relations, with the only non-vanishing commutator being
\be
[M_{ss'}(\vep,\vep')\;M_{\sigma\sigma'}^\dagger(\veq,\veq')]=
(2\pi)^3\delta^{(3)}(\vep-\veq)
(2\pi)^3\delta^{(3)}(\vep'-\veq')\delta_{s\sigma}\delta_{s'\sigma'}.
\label{MMcom}
\ee

At the BCS level, the Hamiltonian (\ref{H2diag}) can be expressed in terms of the operators in Eq.~\eqref{operatorsBD} as
\be
\hat{H}=E_{\rm
vac}+\sqrt{N_c}\sum_{s=\pm 1/2}\int\frac{d^3
p}{(2\pi)^3}E_p[B_{ss}(\vep,\vep)+D_{ss}(\vep,\vep)]+\ldots,
\label{H2diag2}
\ee
where, na{\"i}vly, the factor $\sqrt{N_c}$ in the second contribution on the right-hand side guarantees its dominance over the terms ${\cal O}(N_c^0)$ shown as the ellipsis.
However, beyond the BCS approximation, quarks and antiquarks can not be created or annihilated as isolated objects, so the operators in Eq.~(\ref{operatorsBD}) can not be independent. Then, with the large-$N_c$ logic in mind, we stick to the minimal number of the quark--antiquark pairs produced and require that
each created quark is accompanied by an antiquark
and vice versa. Then one can deduce the following relations between the operators in Eqs.~\eqref{operatorsBD} and \eqref{operatorsMM},
\be
\begin{split}
&B_{ss'}(\vep,\vep')=\frac{
1}{\sqrt{N_c}}\sum_{s''} \int\frac{\ds d^3p''}{
(2\pi)^3}M_{s''s}^\dagger(\vep'',\vep)
M_{s''s'}(\vep'',\vep'),\\
&D_{ss'}(\vep,\vep')=\frac{
1}{\sqrt{N_c}}\sum_{s''} \int\frac{\ds d^3p''}{
(2\pi)^3}M_{ss''}^\dagger(\vep,\vep'')
M_{s's''}(\vep',\vep''),
\label{anzatz}
\end{split}
\ee
that, in the limit $N_c\to\infty$, preserve the commutation relations between the operators $B$ and $D$. In the meantime, if the relations in Eq.~\eqref{anzatz} are substituted into the Hamiltonian in Eq.~\eqref{H2diag2}, the enhancement factor $\sqrt{N_c}$ goes away. Furthermore, it is easy to demonstrate that if the Hamiltonian of the model in Eq.~\eqref{GNJL} is reformulated entirely in terms of the operators $M$ and $M^\dagger$ in Eq.~\eqref{operatorsMM} then both its parts $:H_2:$ and $:H_4:$ contain contributions of the same order in $N_c$ that need to be retained. Then, after simple algebraic transformations, the Hamiltonian (\ref{GNJL}) takes the form \cite{Nefediev:2004by}
\be
H=E_{\rm vac}'+\int \frac{d^3P}{(2\pi)^3}
\hat{\cal H}({\bm P}),
\label{HH1}
\ee
where $E_{\rm vac}'$ is the vacuum energy of the system that generally differs from the energy of the BCS vacuum and (for simplicity, the Hamiltonian density $\cal H$ is taken in the rest frame, with ${\bm P}=0$)
\be
\begin{split}
&{\cal H}\equiv{\cal H}({\bm P}=0)=\sum_{s_1s_2}\int\frac{d^3
p}{(2\pi)^3}
2E_pM_{s_1s_2}^\dagger(\vep,\vep)M_{s_2s_1}(\vep,\vep)
-\frac12\sum_{s_1s_2s_3s_4}\int\frac{d^3p}{(2\pi)^3}\frac{d^3q}{(2\pi)^3}\\
&\times\Bigl[
T^{++}_{s_1s_3s_4s_2}(\vep,\veq)
M^\dagger_{s_2s_1}(\vep,\vep)M_{s_4s_3}(\veq,\veq)
+
T^{+-}_{s_1s_3s_4s_2}(\vep,\veq)
M^\dagger_{s_2s_1}(\veq,\veq)M^\dagger_{s_3s_4}(\vep,\vep)\\[3mm]
&+
T^{-+}_{s_1s_3s_4s_2}(\vep,\veq)
M_{s_1s_2}(\vep,\vep)M_{s_4s_3}(\veq,\veq)
+
T^{--}_{s_1s_3s_4s_2}(\vep,\veq)
M_{s_3s_4}(\vep,\vep)M_{s_1s_2}^\dagger(\veq,\veq)\Bigr],
\end{split}
\label{HH2}
\ee
where the coefficients $T$ introduced in Eq.~\eqref{ampls} have naturally re-appeared. It is easy to see that the operator structure of the Hamiltonian in Eq.~\eqref{HH2},
\be
{\cal H}\sim h_1 M^\dagger M+\frac12 h_2(M^\dagger M^\dagger+MM),
\ee
resembles that of the bosonic toy model in Eq.~\eqref{Haa}. Therefore, the Hamiltonian in Eq.~\eqref{HH2} is subject to a bosonic Bogoliubov-like transformation of the form in Eq.~\eqref{BVa},
\be
m_\nu=\ds\int\frac{d^3p}{(2\pi)^3}\Tr\left[M(\vep,\vep)\Phi^{+\dagger}_\nu(\vep)-
M^\dagger(\vep,\vep)\Phi^-_\nu(\vep)\right],
\label{mMgen}
\ee
where $\nu$ stands for a complete set of the quantum numbers, $\Phi$'s are the matrix wave functions introduced in Eqs.~\eqref{phiplus} and \eqref{phiminus}, and the trace is taken in the quark helicities. The quantity $m_\nu$ in Eq.~\eqref{mMgen} is interpreted as the operator that annihilates the $\nu$-th mesonic state in its rest frame. According to the interpretation of the wave functions $\Phi^\pm$ discussed above, the quark--antiquark pair in the physical meson can move both forward and backward in time, so quite naturally the corresponding contributions come multiplied by the operators $M^\dagger$ and $M$ creating and annihilating a $q\bar{q}$ pair.
To ensure that the operators $m_\nu$ and $m_{\nu'}^\dagger$ obey the standard bosonic commutation relations we require that
\be
\begin{split}
&[m_\nu,m_{\nu'}^\dagger]=
\int\frac{d^3p}{(2\pi)^3}\Tr\left[\Phi^{+\dagger}_\nu(\vep)\Phi^+_{\nu'}(\vep)-
\Phi^{-\dagger}_{\nu'}(\vep)\Phi^-_\nu(\vep)\right]=\delta_{\nu\nu'},\\
&[m_\nu,m_{\nu'}]=
\int\frac{d^3p}{(2\pi)^3}\Tr\left[\Phi^{+\dagger}_\nu(\vep)\Phi^-_{\nu'}(\vep)-
\Phi^{+\dagger}_{\nu'}(\vep)\Phi^-_\nu(\vep)\right]=0,
\end{split}
\label{wfnorm}
\ee
thus imposing the normalisation and orthogonality conditions on the meson wave functions. We stress that the relative ``$-$'' sign between the two contributions on the right-hand side is a direct consequence of the bosonic nature of the above Bogoliubov-like transformation reminiscent of Eq.~\eqref{uvnorm} for the bosonic toy model. As a result of the procedure just outlined, the Hamiltonian of the model is diagonalised in the mesonic sector. The physical vacuum $\ket{\Omega}$ annihilated by the mesonic operators, $m_\nu\ket{\Omega}=0$, differs from the BCS vacuum $\ket{0}$. However, in physical applications this difference can be disregarded and the quantities like, for example, the chiral condensate evaluated in the BCS vacuum can be approximately taken for their physical values.

It should also be noted that, similarly to the BCS level of the model, the diagonalisation of the Hamiltonian in the mesonic sector is performed to the leading order in the $1/N_c$ expansion. The first neglected sub-leading terms contain the meson creation and annihilation operators to the powers 3 and 4 and have the order $1/\sqrt{N_c}$ and $1/N_c$, respectively. These terms describe mesons decays and meson--meson scattering. Then the $N_c$-dependence of the mass and width of a generic mesonic state in the GNJL model is \label{MGamma}
\be
M={\cal O}(N_c^0),\quad \Gamma={\cal O}(N_c^{-1}),
\label{Ncdep}
\ee
in agreement with the common wisdom.

\subsection{The case of the chiral pion}
\label{sec:pion}

Let us check in detail how the above general procedure works for the pion that has the quantum numbers $J=L=S=0$, so the operator $M_{ss'}(\vep,\vep)$ can be written in the form
\be
M_{ss'}(\vep,\vep)=\frac{1}{\sqrt{2}}\left(i\sigma_2\right)_{ss'}Y_{00}(\hat{\vep})M(p),
\label{po}
\ee
where the spin-angular structure is equivalent to the one in the matrix wave function of the pion in Eq.~(\ref{vppi}). On substituting the expression in Eq.~(\ref{po}) into the Hamiltonian in Eq.~(\ref{HH2}), we find
\be
\begin{split}
{\cal H}_\pi&=\int\frac{p^2d
p}{(2\pi)^3}2E_pM^\dagger(p)M(p)-\frac12\int\frac{p^2
dp}{(2\pi)^3}\frac{q^2dq}{(2\pi)^3}
\Bigl[T^{++}_\pi(p,q)M^\dagger(p)M(q)\\
&+T^{+-}_\pi(p,q)M^\dagger(q)M^\dagger(p)
+T^{-+}_\pi(p,q)M(p)M(q)+T^{--}_\pi(p,q)
M^\dagger(q)M(p)\Bigr],
\label{HHpi}
\end{split}
\ee
where the spin traces and angular integrations are done explicitly and this way the four coefficients $T_\pi(p,q)$ are straightforwardly related to those introduced in Eq.~\eqref{ampls}.
The bosonic Bogoliubov-Valatin transformation applied to the operators $M(p)$ and $M^\dagger(p)$ takes the form
\be
\begin{split}
&m_\pi=\ds\int\frac{p^2dp}{(2\pi)^3}\left[M(p)\vp_\pi^+(p)-
M^\dagger(p)\vp_\pi^-(p)\right],\\
&m_\pi^\dagger=\ds\int\frac{p^2dp}{(2\pi)^3}\left[M^\dagger
(p)\vp_\pi^+(p)- M(p)\vp_\pi^-(p)\right],
\end{split}
\label{mMpi}
\ee
where the scalar pion wave functions $\vp_\pi^\pm(p)$ (chosen to be real) and the pion creation and annihilation operators were introduced. Imposing the standard bosonic commutation relation on the above pion creation and annihilation operators, one arrives at the normalisation condition,
\be
[m_\pi,m_\pi^\dagger]=
\int\frac{p^2dp}{(2\pi)^3}\left[\vp_\pi^{+2}(p)-\vp_\pi^{-2}(p)\right]=1,
\ee
with the minus sign between the two contributions on the right-hand side, as was discussed above. Meanwhile, it is more convenient to stick to the standard relativistic normalisation for the pion wave functions,
\be
\int\frac{p^2dp}{(2\pi)^3}\left[\vp_\pi^{+2}(p)-\vp_\pi^{-2}(p)\right]=2M_\pi.
\label{normpi}
\ee

The requirement that the Hamiltonian in Eq.~\eqref{HHpi} is diagonal in terms of the operators $m_\pi$ and $m_\pi^\dagger$ in Eq.~\eqref{mMpi},
\be
{\cal H}_\pi=M_\pi m_\pi^\dagger m_\pi+\ldots,
\label{diagtot}
\ee
with the ellipsis for $N_c$-suppressed terms, leads to the bound state equation \eqref{Salpeterlev5}
with the coefficients in Eq.~\eqref{pia}. In the limit $m\to 0$, its solution normalised as defined in Eq.~\eqref{normpi} reads
\be
\vp_\pi^\pm(p)=\frac{2\sqrt{\pi
N_c\vphantom{\hat{N}}}}{f_\pi}\Bigl(\sin\vp_p \pm
M_\pi\Delta_p+\ldots\Bigr),
\label{vppm}
\ee
where the ellipsis denotes neglected terms of higher orders in the pion mass, the correction function $\Delta_p$ is solution to the equation
\be
2E_p\Delta_p=\sin\vp_p+\int\frac{d^3q}{(2\pi)^3}V(\vep,\veq;T)\Bigl(\sin\vp_p\sin\vp_q+(\hat{\vep}\cdot\hat{\veq}
)\cos\vp_p\cos\vp_q\Bigr)\Delta_q,
\ee
and the pion decay constant is calculated as
\be
f_\pi^2=\frac{N_c}{\pi^2}\int_0^\infty p^2dp\Delta_p\sin\vp_p.
\label{fpidef}
\ee
The result in Eq.~\eqref{vppm} confirms the claim made above (see Eq.~\eqref{Salp2} and the discussion below it) that, in the chiral limit, the mass-gap equation \eqref{mge} is no more than a bound state equation for the pion, with the radial wave function given by $\sin\vp_p$. Also, if one revisits the axial charge operator in Eq.~\eqref{ChiralChar},
then it is easy to see that, as per Eq.~\eqref{vppm},
the second term on the right-hand side indeed creates or annihilates a pion in its rest frame, so $\braket{0|Q_5|\pi(\vep=0)}=if_\pi M_\pi$, in agreement with the criterion for SBCS in Eq.~\eqref{sbcs}. Then the action of the axial charge on a hadronic state appears to be twofold. On the one hand, it flips the parity of the state (the first term on the right-hand side of Eq.~\eqref{ChiralChar}) or creates a pion (the second term on the right-hand side of Eq.~\eqref{ChiralChar}). This observation naturally explains why the relation between the masses of the opposite-parity states in Eq.~\eqref{mpm} fails in the Nambu--Goldstone mode.
It is also easy to verify that the famous Gell-Mann--Oakes--Renner relation \cite{GellMann:1968rz} holds,
\be
f_\pi^2M_\pi^2=-2m\braket{\bar{q}q},
\label{GMOR}
\ee
with the chiral condensate defined in Eq.~\eqref{chircond} --- see, for example, a detailed derivation and discussion in \cite{Nefediev:2004by}.
We can also re-visit the temperature dependence of the chiral condensate and use that (i) the $\pi$-$\pi$ scattering that plays the most essential role for the temperature dependence of $\braket{\bar{q}q}$ at low temperatures scales as $1/N_c$ (see the discussion above Eq.~\eqref{Ncdep}) and (ii) the soft pions physics is controlled by the pion decay constant $f_\pi$ that scales as $\sqrt{N_c}$ (see Eq.~\eqref{fpidef}). Then it is straightforward to conclude that
\be
\braket{\bar{q}q}_{T\to\Tch}/\braket{\bar{q}q}_0=1+{\cal O}(N_c^{-1})=
1+{\cal O}(T^2/f_{\pi}^2)+\ldots,
\label{chc2}
\ee
where the temperature dependence of the correction, $\propto T^2$, trivially comes from the dimensional analysis after the $N_c$-scaling of the correction is established and saturated by the appropriate power of $f_\pi$. Therefore, the GNJL model allows one to qualitatively understand the temperature behaviour of the chiral condensate in Eq.~\eqref{chc}. However, to establish the numerical coefficient in this formula one would need to proceed beyond the leading order in the $1/N_c$ expansion of the model in a self-consistent way. It remains to be seen whether or not this task is feasible.

A comment on the physical interpretation of the pion decay constant $f_\pi$ is in order here. This parameter defines the notion of a ``soft'' pion as a pion with, roughly speaking, the momentum below $f_\pi$. Then, employing the algebra of currents and the PCAC hypothesis in Eq.~\eqref{PCAC}, one can prove that any amplitude involving a soft pion must vanish in the limit of the pion momentum tending to zero. This constitutes the essence of the Adler selfconsistency condition or simply ``Adler zero'' \cite{Adler:1965ga}.
In other words, soft pions decouple from all other states in the spectrum of hadrons. As a side remark, it is instructive to apply this condition to the GNJL model in two-dimensions. As mentioned above (see page~\pageref{tHooft}), it becomes similar to the famous `t~Hooft model for two-dimensional QCD in this case, so the latter theory can be studied along the same lines as discussed here and very similar results can be obtained --- see, for example, the review \cite{Kalashnikova:2001df}. In particular, the Gell-Mann--Oakes--Renner relation in Eq.~\eqref{GMOR} holds in two dimensions as well. However, since the right-hand side in this relation has the same dimension as the mass term in the Lagrangian, $m\bar{q}q$, which is $[m^D]$, with the dimension of the space-time $D=2$, then the two-dimensional pion decay constant on the left-hand side of Eq.~\eqref{GMOR} has to be dimensionless. An inevitable conclusion is that this constant still defines the pion physics but does not set any scale for the soft pions. This way every pion with any momentum is regarded as soft and decouples from other hadrons in the chiral limit. Indeed, this conclusion can be proved analytically employing the properties of the pion solution to the two-dimensional bound state equation \cite{Kalashnikova:2000dw}. Importantly, the $\vp_\pi^-$ component of the pion wave function plays as essential role in the derivation as its $\vp_\pi^+$ component, so the derivation could not be possible in the framework of a simple potential quark model.

Finally, a comment on baryons in the GNJL model is also in order. As was explained above, despite the large-$N_c$ logic employed in the studies of the model, for practical purposes baryons are still regarded as 3-quark hadrons rather than large-$N_c$ solitons in the spirit of the works \cite{Skyrme:1961vq,Witten:1983tx}. The interaction in the model is pair-wise, so only two quarks can interact with each other at a time. Then, if each interaction act results in proceeding from the forward in time motion to the backward in time motion for each interacting particle, then for a quark--antiquark meson it can be schematically expressed as
\be
\bar{q}_1q_2\to q_1\bar{q}_2.
\ee
Thus, if the meson was colourless before the interaction act it remains colourless after the interaction act as well. This scheme does not work for a baryon since, for example, after the interaction of the quarks 2 and 3, one would get
\be
q_1q_2q_3\to q_1\bar{q}_2\bar{q}_3.
\ee
In other words, while the initial state (the three-quark baryon) is colourless, the final state would have open colour. Therefore, the concept of a two-component wave function does not apply to baryons, so from this perspective, the latter appear to be simpler objects than quark-antiquark mesons.

\subsection{Chiral symmetry in excited hadrons}

We are now in a position to investigate the properties of excited mesons in GNJL and, in particular, check the implications of SBCS in their spectrum. A heavy--light system composed of a static antiquark ($\bar{Q}$) and a light quark ($q$) provides a convenient test ground for such investigations since the system is effectively a single-particle one. In this subsection we always work at $T=0$. As was explained above (see the discussion at page \pageref{hl}), the backward in time motion is forbidden in this system for both particles because of the instantaneous nature of the interaction.
To proceed we re-visit the $\bar{Q}q$ meson studied in Sec.~\ref{sec:model}, so our starting point now is the Schwinger--Dyson-type equation \eqref{DS4} written in Minkowski space, \cite{Nefediev:2007pc}
\be
\left(i\gamma_0\frac{\partial}{\partial t}-i\vegam\frac{\partial}{\partial \vex}-m\right)
S(t,\vex,\vey)-\int d^3z {\cal M}(\vex,\vez)S(t,\vez,\vey)=\delta(t)\delta^{(3)}(\vex-\vey),
\label{DS5}
\ee
where the effective mass operator is
\be
{\cal M}(\vex,\vey)=\frac12K(\vex,\vey)\gamma_0\Lambda(\vex,\vey),\quad\Lambda(\vex,\vey)=2i\int\frac{d\omega}{2\pi}S(\omega,\vex,\vey)\gamma_0,
\label{Mop01}
\ee
and the quark kernel $K(\vex,\vey)$ is defined in Eq.~\eqref{KV}. First of all, Eq.~\eqref{DS5} allows one to address the problem of the Lorentz nature of confinement. In particular, it was argued in \cite{Simonov:1997et} that, for a sufficiently light quark, Eq.~(\ref{DS5}) admits linearisation via the substitution
\be
\Lambda(\vex,\vey)\approx\gamma_0\delta^{(3)}(\vex-\vey)+\ldots,
\label{Lm}
\ee
where the ellipsis denotes sub-leading at large distances terms. Then it is easy to see that Eq.~\eqref{DS5} turns to an effective Dirac equation,
\be
\left(i\gamma_0\frac{\partial}{\partial t}-i\vegam\frac{\partial}{\partial \vex}-(m+V(|\vex|))\right)
S(t,\vex,\vey)=\delta(t)\delta^{(3)}(\vex-\vey),
\label{SEscalar}
\ee
where the confining potential is added to the mass term and as such represents a scalar interaction that breaks chiral symmetry. With the approach based on the chiral angle developed above one can get a deeper insight into the nature of confinement in the studied system. Indeed, it is particularly easy now to evaluate the double Fourier transform of
$\Lambda(\vex,\vey)$,
\be
\Lambda(\vep,\veq)=2i\int\frac{d\omega}{2\pi}S(\omega,\vep,\veq)\gamma_0=(2\pi)^3\delta^{(3)}(\vep-\veq)U_p,
\label{iS4}
\ee
where the explicit expressions for the Green function in Eq.~\eqref{Feynman} and projectors in Eq.~\eqref{Lpm} can be used then to arrive at
\be
U_p=\left(\sin\vp_p+(\vec{\gamma}\hat{\vep})\cos\vp_p\right)\gamma_0.
\label{Up4}
\ee
Notice that the expression in Eq.~\eqref{Lm} is then readily reproduced in the limit of the strongest possible effect of SBCS with $\vp_p=\pi/2$ everywhere, so that $U_p=\gamma_0$. On the other hand, for a physical behaviour of the chiral angle shown in Fig.~\ref{fig:vp}, the effective confining interaction is a dynamical, momentum-dependent mixture of the scalar and spatial-vector contributions. No rising with the distance time-vector interaction arises, so no problem with the Klein paradox is encountered as a matter of principle.

The Schwinger--Dyson-like equation \eqref{DS5} can be re-written in the form of a Shr{\"o}dinger-like equation for the wave function $\Psi(\vex)$ as
\be
(\vec{\alpha}\hat{\vep}+\beta m)\Psi(\vex)+\beta\int d^3z {\cal M}(\vex,\vez)\Psi(\vez)=E\Psi(\vex),
\label{DS6}
\ee
where the energy $E$ is counted from the infinite mass of the static antiquark. Notice that alternatively the latter equation can be derived directly from the bound state equation in Eq.~\eqref{Salpeterlev5} adapted for a heavy--light system \cite{Kalashnikova:2005tr}. Then, passing over to the momentum space and using the mass--gap equation \eqref{mge} in the matrix form,
\be
E_pU_p={\vec \alpha}\vep+\beta
m+\frac12\int\frac{d^3k}{(2\pi)^3} V(\vep-\vek)U_k,
\ee
one can re-write Eq.~(\ref{DS6}) as
\be
E_pU_p\Psi(\vep)+\frac12\int\frac{d^3k}{(2\pi)^3}V(\vep-\vek)[U_p+U_k]\Psi(\vek)=E\Psi(\vep),
\label{Se10}
\ee
which is subject to a Foldy--Wouthuysen transformation,
\be
\Psi(\vec{p})=T_p{\psi(\vec{p})\choose 0},
\label{Tpop}
\ee
with the Foldy operator defined in Eq.~\eqref{Tpdef}.
Notice that the vanishing low component of the Foldy-rotated wave function in Eq.~\eqref{Tpop} represents the static antiquark, so the bound state equation for the heavy--light system at hand can be written in the form of a single equation for the light quark described in terms of the wave function $\psi(\vep)$.
Substituting the wave function in the form in Eq.~\eqref{Tpop} to the bound state equation \eqref{Se10} and performing simple algebraic transformations one arrives at the desired would-be single-quark Shr{\" o}dinger-like equation in the form
\be
E_p\psi(\vep)+\int\frac{d^3k}{(2\pi)^3}V(\vep-\vek)\left[C_pC_k+
({\vec \sigma}\hat{\vep})({\vec
\sigma}\hat{\vek})S_pS_k\right]\psi(\vec{k})=E\psi(\vep),
\label{FW4}
\ee
where $\vec{\sigma}$ are the Pauli matrices, $\hat{\vep}$ and $\hat{\vek}$ are the unit vectors for the momenta $\vep$ and $\vek$, respectively, and
\be
C_p=\cos\frac12\left(\frac{\pi}{2}-\vp_p\right),\quad S_p=\sin\frac12\left(\frac{\pi}{2}-\vp_p\right).
\label{CpSp}
\ee

As the first step in our studied of the bound state equation \eqref{FW4} let us set $\vp_p=\pi/2$. It is easy to see then that $C_p=1$, $S_p=0$, and we arrive at a Salpeter-like equation in coordinate space,
\be
[E_p+V(r)]\psi(r)=E\psi(r).
\label{Salp3}
\ee
The above consideration entails several conclusions.
On the one hand, Salpeter equations like \eqref{Salp2} or \eqref{Salp3} imply maximal breaking of chiral symmetry and as such are relevant only for low-lying states where the typical quark momenta are small, so the chiral angle indeed takes values close to $\pi/2$.
On the other hand, a na{\"i}ve substitution $E_p\to E_p^{(0)}=\sqrt{\vep^2+m^2}$ routinely employed in potential quark models to arrive at Eq.~\eqref{Salp} inevitably fails for the pion as the Goldstone boson of SBCS --- see the discussion at page \pageref{negativeEp} above. Finally, although the confining potential in Salpeter equation \eqref{Salp} is added not to the mass term under the square root but rather to the energy, it does not at all imply that the Lorentz nature of confinement is a time-like vector vulnerable to the Klein paradox. On the contrary, Salpeter equation \eqref{Salp3} follows from a Dirac-like equation \eqref{SEscalar} with a purely scalar confining potential that corresponds to the maximal breaking of chiral symmetry.

We now turn to our studies of the bound state equation in Eq.~\eqref{FW4}. In the introductory Subsec.~\ref{sec:genquest} we posed a question of the Regge trajectories behaviour for excited mesons. Equation \eqref{FW4} provides a convenient way to answer this question. Indeed, if the light quark is highly excited (for example, it has the angular momentum $L\gg 1$), so that its mean momentum is large compared with the typical scale associated with the interaction (for example, the interaction scale $K_0$ for the power-like potential in Eq.~\eqref{Vconf}), then $\vp_p\to 0$ (see Fig.~\ref{fig:vp}). In this limit, $C_p\approx S_p\approx \frac{1}{\sqrt{2}}$ and equation \eqref{FW4} becomes
\be
E_p\psi(\vep)+\frac12\int\frac{d^3k}{(2\pi)^3}V(\vep-\vek)\left[1+
({\vec \sigma}\hat{\vep})({\vec
\sigma}\hat{\vek})\right]\psi(\vec{k})=E\psi(\vep).
\label{FW5}
\ee
It is easy to see that this equation is invariant with respect to the substitution
\be
\psi(\vep)\to\psi'(\vep)=({\bm\sigma}\hat{\vep})\psi(\vep),
\ee
where $\psi(\vep)$ and $\psi'(\vep)$ describe a pair of states related by the spatial parity transformation. Therefore, if one climbs sufficiently high in the spectrum, such states become gradually degenerate and approximately fill the multiplets inherent in the Wigner--Weyl realisation of chiral symmetry \cite{LeYaouanc:1984ntu,Glozman:1999tk,Cohen:2001gb,Nowak:2003ra,Swanson:2003ec,Glozman:2007ek}.
The bound state equation \eqref{FW4} was studied in detail in \cite{Kalashnikova:2005tr}. A comprehensive study of the spectrum of excited mesons in the framework of the GNJL model was performed in \cite{Wagenbrunn:2007ie} and effective restoration of chiral symmetry for large excitation numbers was observed explicitly employing the chiral basis of states suggested in \cite{Glozman:2009bt}.

At the end of Sec.~\ref{sec:chreal}, a Goldberger--Treiman relation in Eq.~\eqref{GTh} was derived employing the general condition in Eq.~\eqref{sbcs} for spontaneously broken chiral symmetry. In \cite{Nefediev:2006bm}, this relation was derived in the framework of GNJL for a particular case of heavy--light mesons. To this end a definition of the pion coupling $g_{hh'\pi}$,
\be
\langle h'\pi^a |V|h \rangle=2M i g_{hh'\pi}(h^{'\dag} \tau^a h)(2\pi)^3\delta^{(3)}({\bm P}'+\vep-{\bm P}),
\label{Vnn}
\ee
was used and the matrix element on the left-hand side was evaluated using GNJL generalised to two light quark flavours. As a result, the relation in Eq.~\eqref{GTh} naturally appeared, with the eigenfunction $\psi(\vep)$ and eigenenergy $E$ defined in Eq.~\eqref{FW4} and their counterparts $\psi'(\vep)$ and $E'$ obeying a bound state equation,
\be
E_p\psi'(\vep)+\int\frac{d^3k}{(2\pi)^3}V(\vep-\vek)\left[S_pS_k+
({\vec \sigma}\hat{\vep})({\vec
\sigma}\hat{\vek})C_pC_k\right]\psi'(\vec{k})=E'\psi'(\vep),
\label{FW4pr}
\ee
related to Eq.~\eqref{FW4} through the interchange $C_p\leftrightarrow S_p$. Then the axial charge $G_A$ was found to be
\be
G_A=\frac{1}{2M}\int\frac{d^3p}{(2\pi)^3}\psi^{\prime\dagger}(\vep)\psi(\vep)\cos\vp_p,
\label{GAcalc}
\ee
where the wave functions are normalised as
\be
\int\frac{d^3p}{(2\pi)^3}|\psi(\vep)|^2=\int\frac{d^3p}{(2\pi)^3}|\psi'(\vep)|^2=2M.
\ee
The interested reader can find further details of the derivation in \cite{Nefediev:2006bm}. We skip them here and only mention the conclusions drawn in the cited work concerning this microscopic derivation.
Quite predictably, the component of the pion wave function $\vp_\pi^-$ (denoted as $Y$ in the cited paper) played as essential role in the derivation as the component $\vp_\pi^+$ (denoted there as $X$). Neglecting $\vp_\pi^-$ (like in na{\"i}ve quark models) would result in a severe violation of the Goldberger--Treiman relation \eqref{GTh}. It is also instructive to check the implications of this relation for excited heavy--light mesons. Since, in this limit,
\be
\Delta M=M'-M\too 0,\quad G_A\too 1,
\ee
with $\nu$ for a generic excitation quantum number, then
\be
g_{hh'\pi}\too 0,
\label{decoupl}
\ee
so the Goldberger--Treiman relation entails
a Goldstone boson decoupling from excited
hadrons. The latter property can also be understood employing the notion of dressed quarks and their coupling to the chiral pion \cite{Glozman:2006xq}. To this end we first
use the definition of the axial-vector current in Eq.~\eqref{j5def}, simplified for the case of $N_f=1$ and written in terms of the dressed quark fields,
and assume that the latter obey an effective Dirac equation with some mass $m_q^{\rm eff}$ that arises due to SBCS. Then for the matrix element of the divergence of the axial current between two quark fields we can find
\be
\braket{q(p)|\partial^\mu j_{5\mu}|q(p')}\propto m_q^{\rm eff}(\bar{u}_p\gamma_5 u_{p'}),
\label{eq2}
\ee
where we refrain from quoting the explicit factors on the right-hand side that are irrelevant for the discussion. One the other hand, employing the PCAC conjecture in Eq.~\eqref{PCAC}, we find
\be
\braket{q(p)|\partial^\mu j_{5\mu}|q(p')}=f_\pi m_\pi^2 \langle q(p)|\pi|q(p')\rangle
\propto f_\pi g_\pi(k^2) (\bar{u}_p\gamma_5 u_{p'}),
\label{eq1}
\ee
where $k_\mu=p_\mu-p'_\mu$ and the pion--quark--quark form factor $g_\pi(k^2)$ was introduced. Taking relations \eqref{eq2} and \eqref{eq1} together and for simplicity assuming that the on-shell pion form factor $g_\pi\equiv g_\pi(m_\pi^2)$ absorbs all numerical factors, we arrive at the sought relation,
\be
f_\pi g_\pi=m_q^{\rm eff}=\braket{M_p},
\label{GTr}
\ee
where we used that the effective quark mass can be naturally associated with the averaged value of $M_p$ in Eq.~\eqref{Mpdef2}. Since the wave function of highly excited states is localised at large values of the momentum $p$ where the function $M_p$ decreases fast (see Fig.~\ref{fig:vp}), the averaged value on the right-hand side in Eq.~\eqref{GTr} decreases with the excitation number of the meson $\nu$ and, therefore, so does the pion coupling $g_\pi$.

Concluding this section we emphasise that the GNJL model provides a microscopic description of the effective restoration of chiral symmetry in excited hadrons and related phenomena such as the pion decoupling from such hadrons. Importantly, no changes are implied in the vacuum of the theory that remains chirally broken. What is meant then by the effective chiral restoration in the spectrum of excited hadrons is a progressively fading effect of SBCS on excited hadronic states with the rise of their excitation numbers. As a result, such hadrons approximately fill the multiplets inherent in the Wigner--Weyl mode of the chiral symmetry realisation.

\subsection{Spectrum of light--light mesons}

In this section, we advance with our studies of the GNJL model at finite temperatures and discuss solutions of the bound state equation \eqref{Salpeterlev5} for different quantum numbers of the mesons.
Since the interaction in the Hamiltonian \eqref{GNJL} is flavour-blind, one has a trivial degeneracy of the states with $I=0$ and $I=1$, so the results below can be regarded as corresponding to either of the above isospins. Also, a disclaimer concerning the lightest scalar state in the spectrum of mesons is in order here. In what follows, we shall often refer to it as the ``$\sigma$-meson'', where the quotation marks indicate that it has to be distinguished from the physical $f_0(500)$ meson (former $\sigma$-meson)  observed experimentally \cite{ParticleDataGroup:2024cfk}. Indeed, the nature of the latter is closely related to a strong $S$-wave $\pi$-$\pi$ interaction at low energies that can not be naturally incorporated into the formalism of the GNJL model --- see the discussion at page \pageref{MGamma}. In particular, the pole trajectory of the physical state $f_0(500)$ as a function of the number of colours $N_c$ derived in the unitarised chiral perturbation theory \cite{Pelaez:2006nj,Pelaez:2015qba} is at odds with
the expectations for a generic quark--antiquark state contained in Eq.~\eqref{Ncdep}. Indeed, the real part of its pole (the mass) grows with the number of colours (instead of staying a constant as given in Eq.~\eqref{Ncdep}) and the $N_c$-dependence on its imaginary part (the width) is non-trivial and not consistent with the $1/N_c$ scaling in Eq.~\eqref{Ncdep} either. Thus, the state referred to as the ``$\sigma$-meson'' below is no more than the lightest generic scalar quark--antiquark meson predicted by the employed quark model.

\begin{figure}[t!]
\centering
\includegraphics[width=0.99\textwidth]{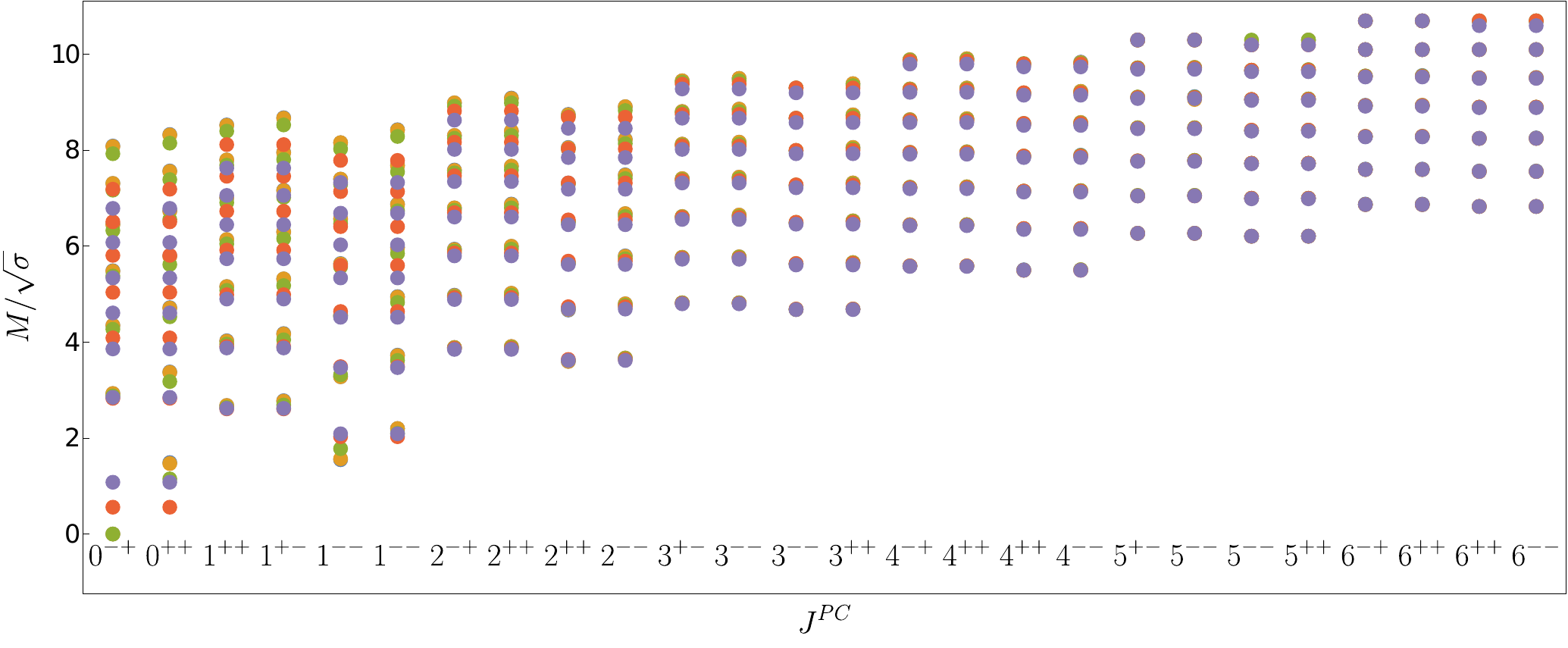}
\caption{The masses of the quark--antiquark mesons with the given quantum numbers $nJ^{PC}$ (for $n=0..6$ and $J=0..6$) found as the eigenvalues of the bound state equation \eqref{Salpeterlev5} \cite{Glozman:2024dzz}. The results for the temperatures $T=0$, $0.5\Tch$, $0.9\Tch$, $1.1\Tch$, and $1.5\Tch$ are shown as blue, yellow, green, red, and violet points, respectively.}
\label{fig:mesonsall}
\end{figure}

\begin{figure}[t!]
\centering
\includegraphics[width=0.99\textwidth]{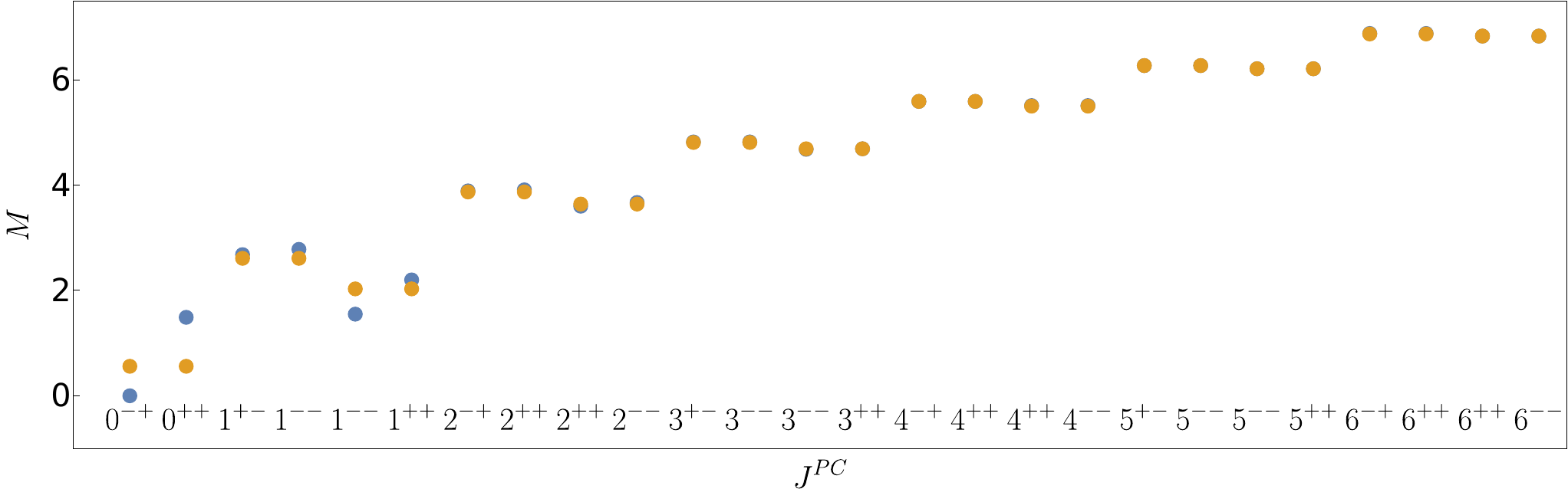}
\caption{The same as in Fig.~\ref{fig:mesonsall} but for the radial excitation number $n=0$ and for the temperatures $T=0$ (blue points) and $T=1.1\Tch$ (yellow points). Adapted from \cite{Glozman:2024dzz}.}
\label{fig:mesons}
\end{figure}

\begin{figure}[t!]
\centering
\includegraphics[width=0.49\textwidth]{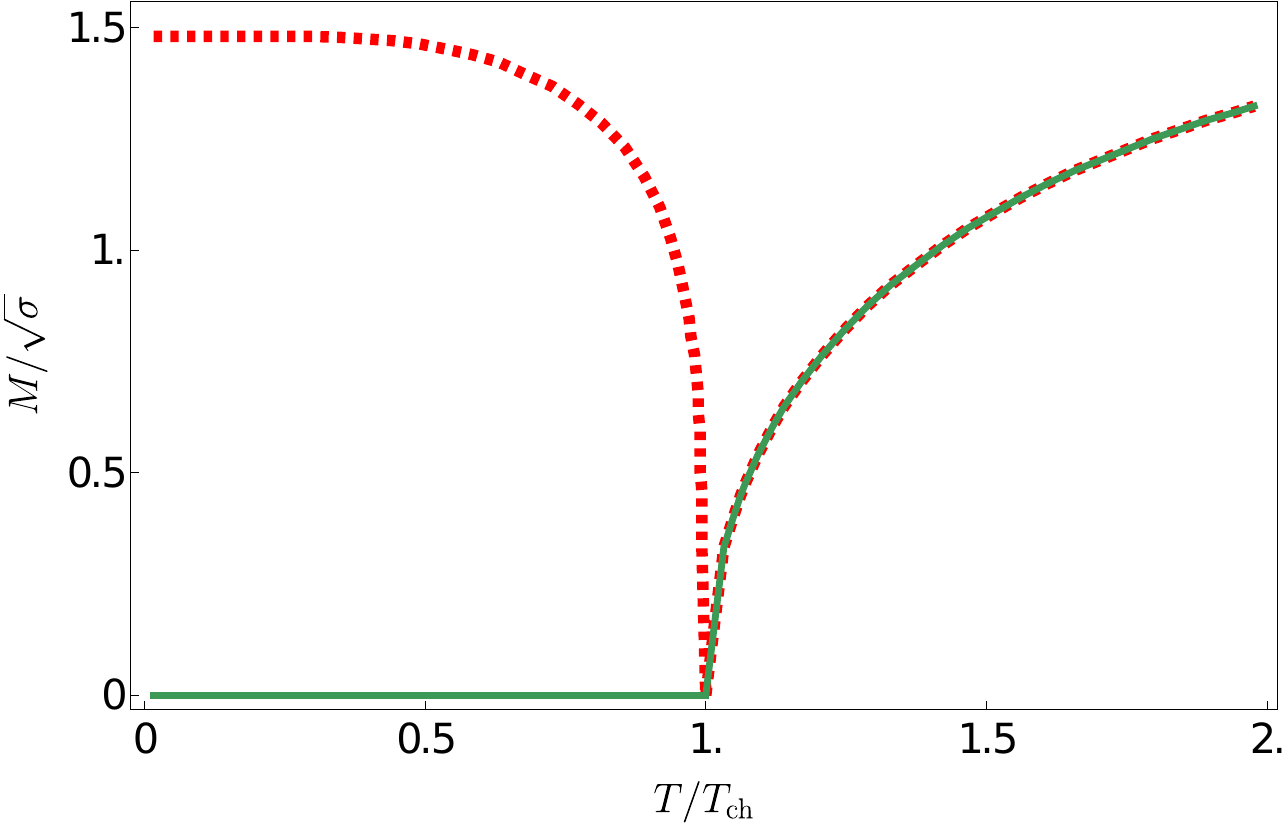}
\includegraphics[width=0.49\textwidth]{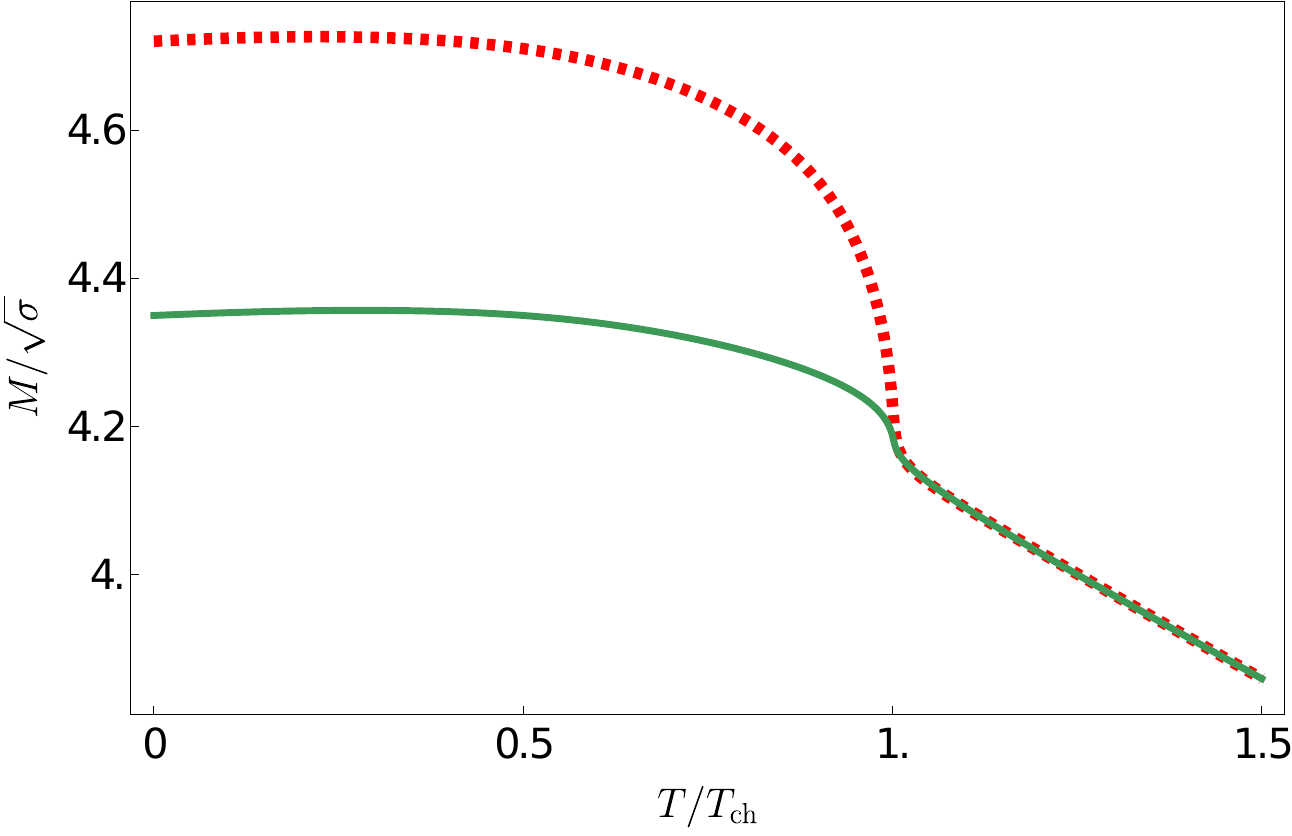}
\caption{The temperature dependence of the masses of pseudoscalar (the green solid line) and scalar (the red dashed line) mesons with the radial excitation number $n=0$ (the left plot) and $n=2$ (the right plot). The plots are based on the results reported in \cite{Glozman:2024dzz}.}
\label{fig:pisigma}
\end{figure}

As our starting point, in Fig.~\ref{fig:mesonsall}, we show the results for the spectrum of the quark--antiquark bound state equation \eqref{Salpeterlev5} obtained in \cite{Glozman:2024dzz}. This figure clearly demonstrates the appearance of well established clusters of degenerate states as one proceeds from low to high total spins $J$'s. To better understand the structure and degeneracies of the spectrum we also provide a simplified plot containing the meson masses evaluated at $T=0$ and $T=1.1\Tch$ and corresponding to the radial excitation number $n=0$ and different quantum numbers $J^{PC}$ --- see Fig.~\ref{fig:mesons}.

We start our discussion from the case of $J=0$ that is exceptional for two reasons: (i) the chiral pion is constrained by SBCS and (ii) the quantum numbers $0^{--}$ and $0^{+-}$ are not accessible in the quark--antiquark scheme, so there exists only one pair of opposite-parity states with $J=0$. The pattern for the mesons with $J=0$ demonstrated by Fig.~\ref{fig:mesons} complies with the most natural expectations. Indeed, at $T=0$, the pion is massless and its mass splitting with the ``$\sigma$-meson'' takes a typical value of the order of the interaction scale $\sqrt{\sigma}$. On the other hand, for $T>\Tch$, when chiral symmetry is restored in the vacuum, these two states become strictly degenerate, both possessing a non-vanishing mass. This pattern is further exemplified in Fig.~\ref{fig:pisigma} where we show the temperature dependence of the masses of the states $0^{-+}$ and $0^{++}$ with the radial excitation numbers $n=0$ (the left plot) and $n=2$ (the right plot). Note that, since radially excited pseudoscalar mesons are not Goldstone bosons and their masses are not constrained by chiral symmetry, then they are massive both below and above the critical temperature $\Tch$. They become degenerate with the respective scalars above $\Tch$.

\begin{figure}
\centering
\includegraphics[width=0.6\textwidth]{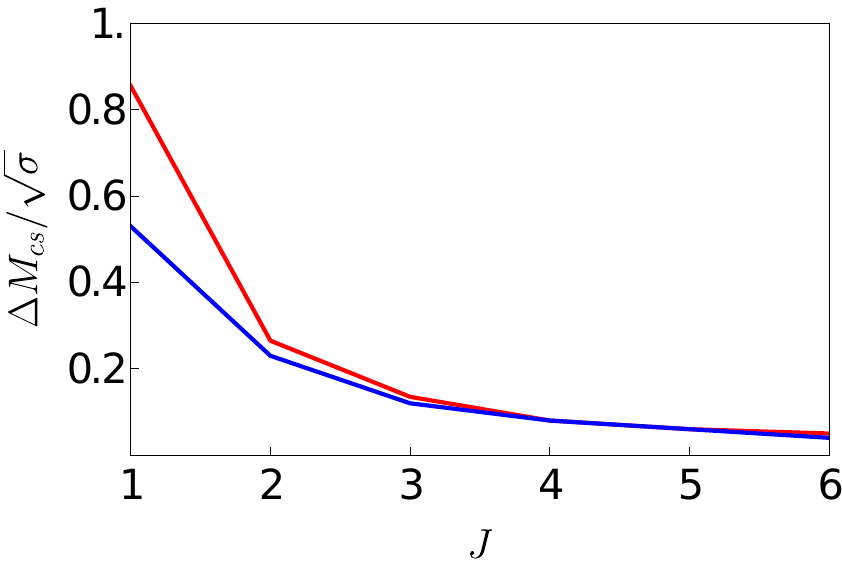}
\caption{The dependence of the splitting between two degenerate chiral doublets within the same $SU(2)_{CS}$ quartet with $n=0$ on the total spin of the mesons $J$  for $T=0$ (the red curve) and $T=1.5\Tch$ (the blue curve).}
\label{fig:dMcs}
\end{figure}

For $J=1$ and $T=0$ one readily observes the following pattern (see Fig.~\ref{fig:mesons}):
\be
M(1^{--};T=0)-M(1^{++};T=0)~\simeq~M(1^{++};T=0)-M'(1^{--};T=0),
\label{M1T0}
\ee
where the prime is used to distinguish between the two different vector states in the given quartet with $J=1$.\footnote{See \cite{Glozman:2007ek} for a complete set of possible chiral multiplets for the quark--antiquark mesons with different total spins $J$.} On the other hand, for $T=1.1\Tch$, the pattern becomes
\be
\begin{split}
&M(1^{++};T>\Tch)-M'(1^{--};T>\Tch)\\
&\hspace*{0.3\textwidth}=M(1^{--};T>\Tch)-M'(1^{+-};T>\Tch)=0,\\
&M(1^{--};T>\Tch)-M(1^{++};T>\Tch)\simeq
M(1^{--};T=0)-M(1^{++};T=0).
\label{M1T11}
\end{split}
\ee
In other words, for $J=1$, we observe an appearance of two chiral doublets formed by the states that become strictly degenerate above the chiral restoration temperature. In the meantime, the splitting between these two doublets still takes a natural value of the order of $\sqrt{\sigma}$ both at $T=0$ and $T>\Tch$. The above pattern is replicated for higher $J$'s with one essential difference that the splitting between the two chiral doublets decreases with $J$, so starting from the angular momentum around 3 or 4, one observes a quartet of approximately degenerate states of the same spin $J$ rather than two split chiral doublets. This pattern is well seen in the plot in Fig.~\ref{fig:mesonsall}.
Therefore, the observed feature of the spectrum of mesons to demonstrate a higher degree of degeneracy than dictated by chiral symmetry alone appears to be a general prediction of the GNJL model. In \cite{Glozman:2024dzz}, this feature of the spectrum was explained by the appearance of an additional, emergent symmetry of the excited states. In particular, it is conjectured in the literature that the spectrum of confined but chirally symmetric hadrons in QCD should exhibit a broad chiral spin symmetry $SU(2)_{CS}$ (including the chiral group as a subgroup) \cite{Glozman:2014mka,Glozman:2015qva} that was
first observed on the lattice through a degeneracy of hadrons at $T=0$ upon an artificial truncation
of the near-zero modes of the Dirac operator \cite{Denissenya:2014poa,Denissenya:2014ywa}.
--- for a recent review see \cite{Glozman:2022zpy}. Naturally, the GNJL model appears to be a well suited investigation tool to approach and study the regime of chirally symmetric but confined hadrons by proceeding above the chiral restoration temperature $\Tch$.
In particular, in Fig.~\ref{fig:dMcs}, we show the dependence of the splitting between the two degenerate chiral doublets within the same $SU(2)_{CS}$ quartet with the radial excitation number $n=0$ on the total spin of the mesons $J$ for $T=0$ and $T=1.5\Tch$.
Since chiral symmetry is restored above $\Tch$, which removes an essential source of the $SU(2)_{CS}$ symmetry violation, then the splitting depicted in Fig.~\ref{fig:dMcs} decreases with $J$ faster for $T>\Tch$.

\subsection{Properties of mesons at finite temperatures}

\begin{figure}[t!]
\centering
\begin{tabular}{ll}
\includegraphics[width=0.49\textwidth]{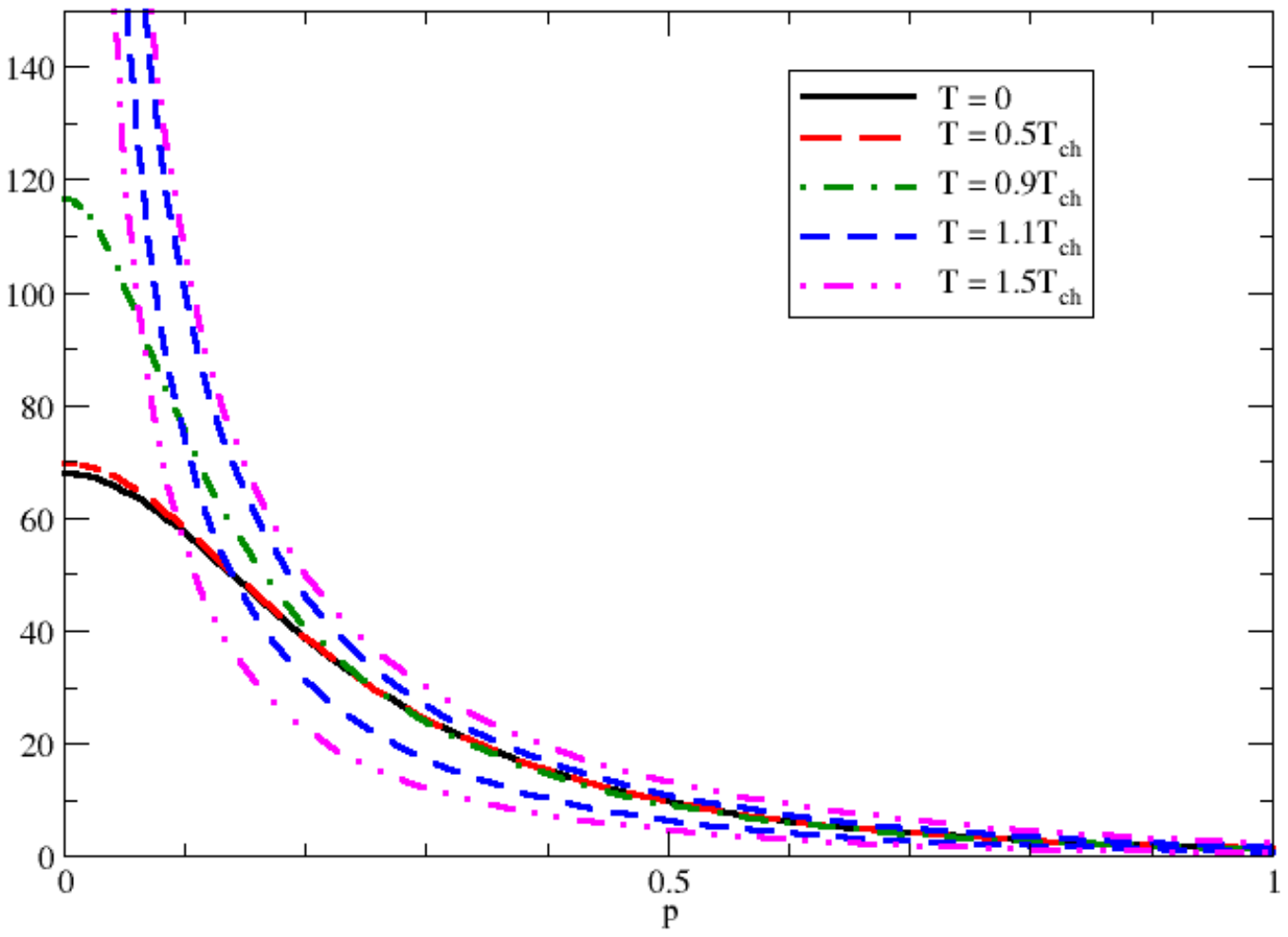}&
\hspace*{-0.07\textwidth}
\includegraphics[width=0.49\textwidth]{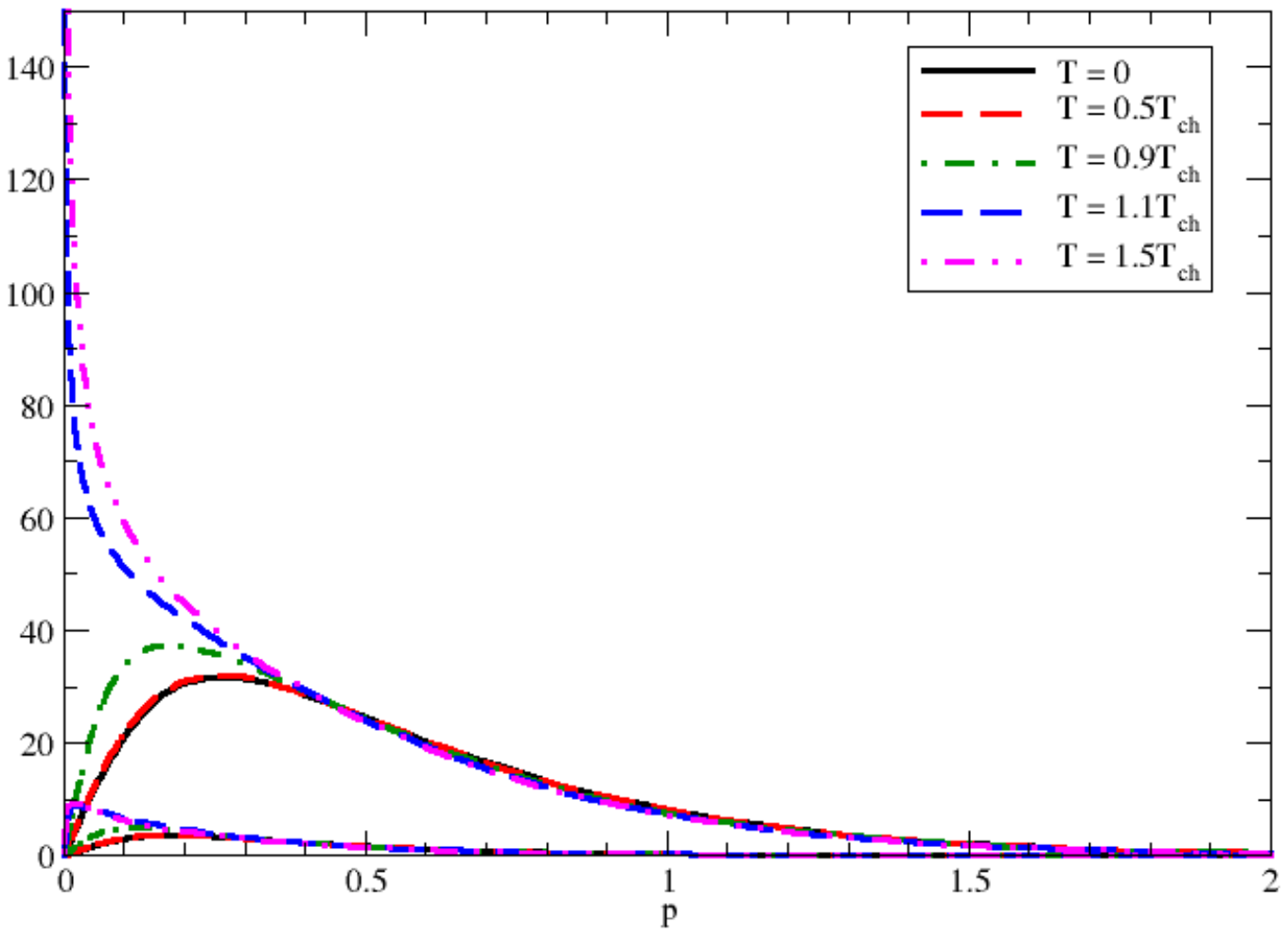}\\[-8mm]
\includegraphics[width=0.53\textwidth]{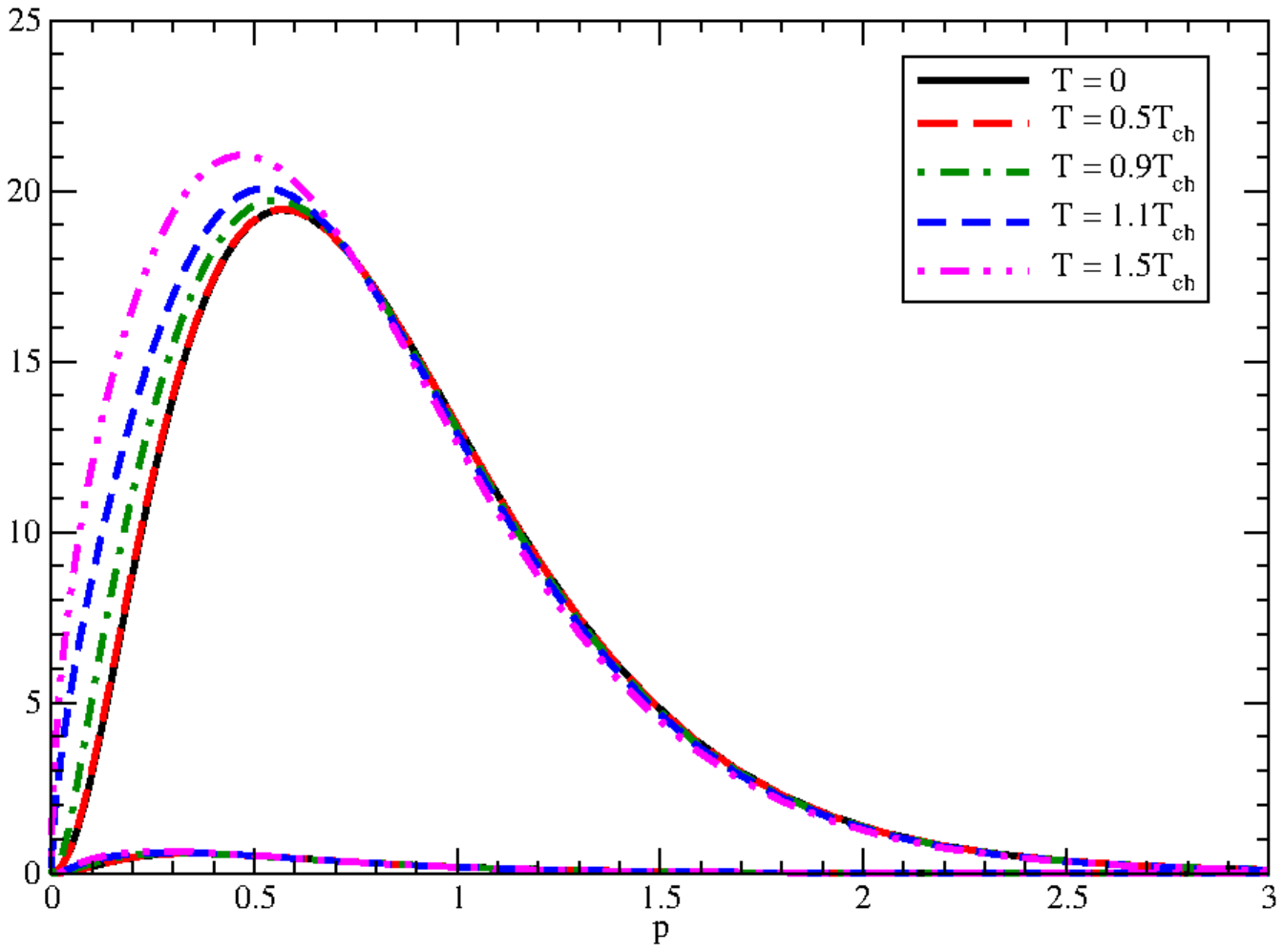}&
\hspace*{-0.07\textwidth}
\includegraphics[width=0.53\textwidth]{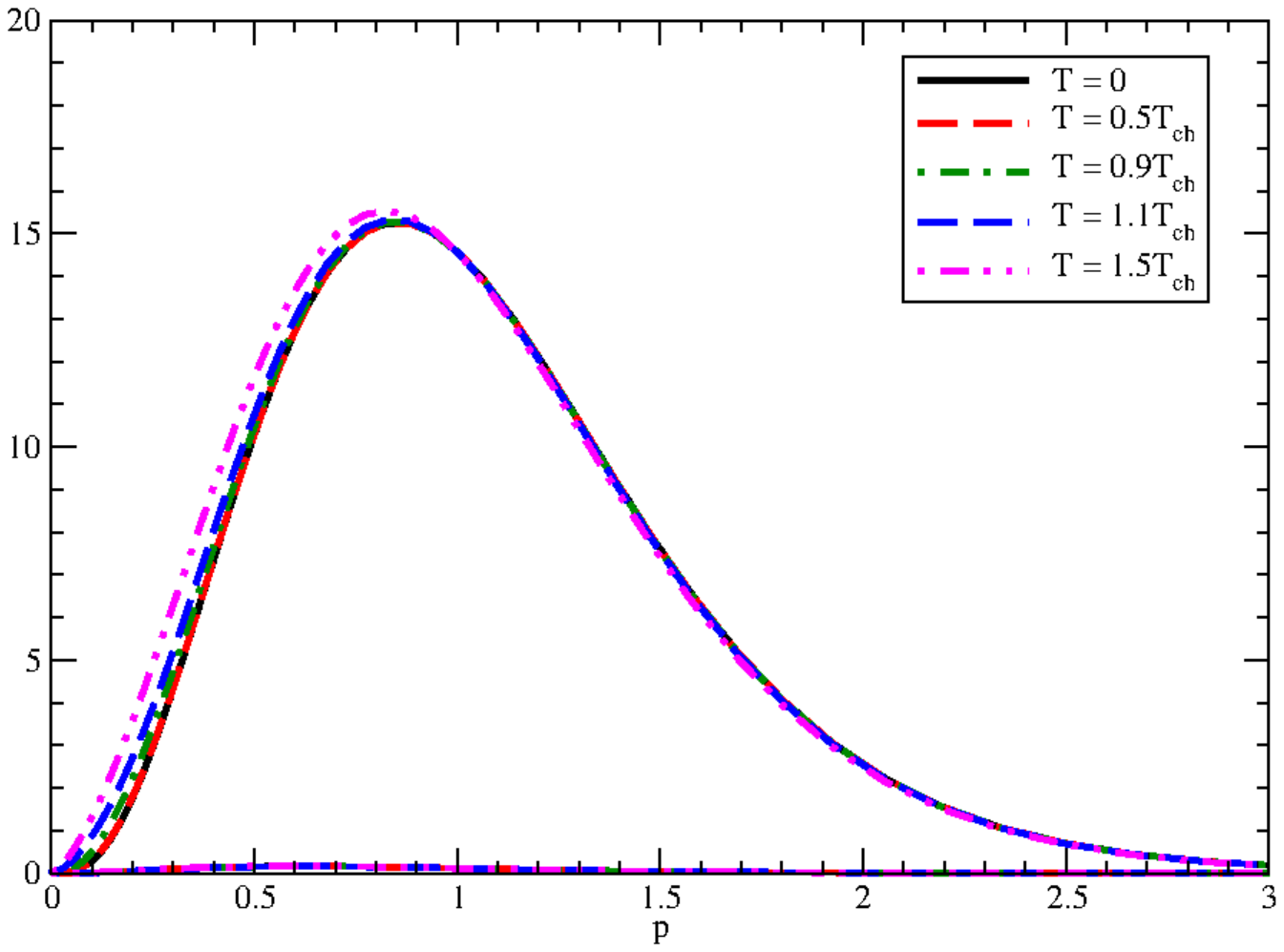}
\end{tabular}
\caption{The ground state ($n=0$) radial wave functions $\vp^\pm(p)$ for $J^{PC}=0^{-+}$ (the upper left plot), $J^{PC}=1^{+-}$ (the upper right plot), $J^{PC}=2^{-+}$ (the lower left plot), and $J^{PC}=3^{+-}$ (the lower right plot) at different temperatures --- see the inlays. For $J^{PC}=0^{-+}$ and $T<\Tch$, the corresponding pseudoscalar meson is the massless Goldstone boson with $\vp_\pi^+(p)=\vp_\pi^-(p)$ (see Eq.~\eqref{vppm} in the chiral limit with $M_\pi\to 0$). For $J^{PC}=0^{-+}$ and $T>\Tch$ as well as for all other sets of quantum numbers and all temperatures, $\vp^+(p)>\vp^-(p)$, so for each temperature there are two curves in the corresponding plots. Adapted from \cite{Glozman:2024dzz}.}
\label{fig:wfs}
\end{figure}

In the previous section, the spectrum of the quark--antiquark mesons in the GNJL model at zero and finite temperatures was investigated and discussed in some detail. In particular, highly excited mesons were demonstrated to approximately fill the multiplets inherent in a broad symmetry group that includes chiral symmetry as a subgroup. A natural question that arises is whether other properties of hadrons change as the temperature exceeds the chiral restoration point at $T=\Tch$.

We start from a discussion of the properties of the wave functions of the low-lying states obtained in \cite{Glozman:2024dzz} and plotted in Fig.~\ref{fig:wfs}. We note that the general pattern explained in Sec.~\ref{sec:BSqq} persists for all mesons at $T<\Tch$. Namely, the chiral pion with $n=0$ and $J^{PC}=0^{-+}$ plays a role of the Goldstone boson of SBCS and possesses the wave functions describing its forward and backward in time motion equal to each other (in the strict chiral limit). For all mesons with different quantum numbers, $\vp^+(p)>\vp^-(p)$ or even $\vp^+(p)\gg\vp^-(p)$, if the excitation of the meson is high enough. All above wave functions are nodeless, localised in momentum, and bounded from above, so they all comply with the natural expectations  for the ground states with $n=0$.

Meanwhile, the situation changes severely for $T>\Tch$. In this case, while the wave functions of the states with higher spins (see the plots in the lower row in Fig.~\ref{fig:wfs}) experience only a little modification, the wave functions of the low-spin states (see the plots in the upper row in Fig.~\ref{fig:wfs}) demonstrate a dramatic change and start to grow fast at low momenta. This growth is tamed by the current quark mass, so in the strict chiral limit of $m\to 0$, both $\vp^+(p)$ and $\vp^-(p)$ become singular at the origin. Importantly, this striking feature of the mesons' wave functions does not spoil their normalisation property since the individual singular contributions from the $\vp^+$ and $\vp^-$ components cancel against each other in the normalisation condition in Eq.~\eqref{wfnorm}. The reason for such behaviour of the wave functions can be readily traced down to the thermal factor $1-n_p-\bar{n}_p$ that multiplies the potential in the bound state equation \eqref{BSeq}, with the Fermi--Dirac distribution functions defined in Eq.~\eqref{nnnew}. Indeed, at $T<\Tch$ where chiral symmetry is spontaneously broken and the effective quark mass $M_p$ does not vanish, this factor tends to a constant in the limit $p\to 0$. On the contrary, for $T>\Tch$ when $M_p=0$ (we stick to the chiral limit of the current quark mass $m=0$), the small-momentum behaviour of the Fermi-Dirac distribution provides
\be
(1-n_p-\bar{n}_p)_{|\mu=0,M_p=0,T\neq 0}\mathop{\propto}_{p\to 0}p,
\label{PB}
\ee
which is nothing else than yet another implication of the Pauli blocking effect. For a non-vanishing angular momentum in the quark--antiquark system, the centrifugal barrier suppresses the wave function near the origin, so the behaviour in Eq.~\eqref{PB} does not affect the wave function appreciably. However, since the centrifugal barrier is not operative for the states with the angular momentum $L=0$, then the mesons with $J=0$ and $J=1$ ``feel'' the above Pauli blocking effect and their wave functions strongly deviate from those at $T<\Tch$.

The above change in the behaviour of the wave functions for the low-spin mesons is expected to result in a related change in observable properties of these states. Of particular interest is the size of the mesons since the properties of the medium formed by such quark--antiquark states depend on this value.
It has to be noted, however, that the definition of the mean radius of a $q\bar{q}$ state in GNJL is not a trivial task. Indeed, in a simple potential quark model like the one in Eq.~\eqref{Salp}, averaging of a physical quantity is naturally defined as its overlap with the probability distribution given by $|\psi(r)|^2$. In the meantime, since each meson in GNJL is described by two wave functions obeying a non-trivial normalisation condition like the one for the pion in Eq.~\eqref{normpi}, then a direct probabilistic interpretation of the two-component meson wave function is obscure. Under such circumstances, a natural quantity to study would be the root-mean-square (r.m.s.) radius $\braket{r^2}^{1/2}$ introduced as \cite{Alkofer:2005ug}
\be
\braket{r^2}=\left.6\frac{\partial F(q^2)}{\partial q^2}\right|_{q^2=0},
\label{eq:r}
\ee
where the form factor of the hadron $h$ is defined through the matrix element of the vector current,
\be
\braket{h(P')|j_\mu(0)|h(P)}=i(P+P')_\mu F(q^2).
\label{eq:F}
\ee

\begin{figure*}[t!]
\centering
\includegraphics[width=0.9\textwidth]{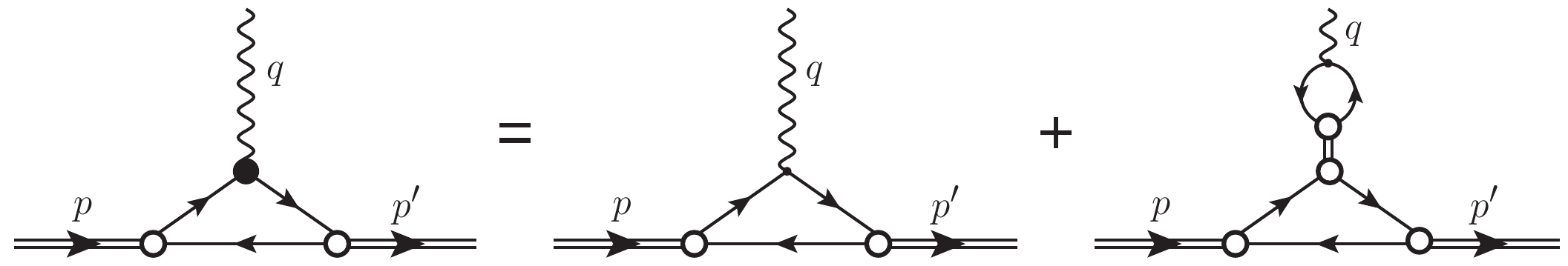}
\caption{The meson-photon interaction. The single, double, and wavy line correspond to the quark (antiquark), meson, and photon, respectively. The last diagram implies a sum over all intermediate vector mesons coupled to the photon through a quark--antiquark loop. Adapted from \cite{Glozman:2024dzz}.}
\label{fig:mesonphoton}
\end{figure*}

\begin{figure}[t!]
\centering
\includegraphics[width=0.9\textwidth]{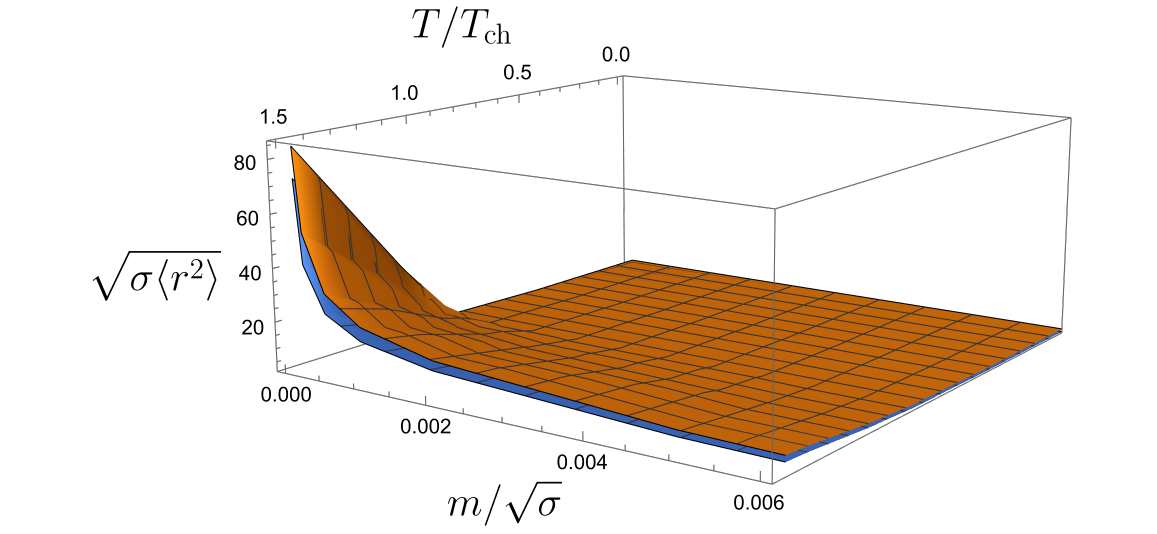}
\caption{3D plot for the r.m.s. radius of the pion (yellow) and ``$\sigma$-meson'' (blue) as function of the quark mass and temperature.}
\label{fig:radius3D}
\end{figure}

Strictly speaking, the conservation law for the vector current in Eq.~\eqref{eq:F} implies a proper dressing of the photon--quark--antiquark vertex with the interaction that is equivalent to summing up all the contributions from the photon conversion to the intermediate vector mesons via a quark--antiquark loop as depicted in the last diagram in Fig.~\ref{fig:mesonphoton}. However, since the main contribution to the r.m.s. radius introduced in Eq.~\eqref{eq:r} comes from small momentum transfers with $q^2\to 0$, then the last contribution in Fig.~\ref{fig:mesonphoton} is suppressed by the squared mass of the lightest vector meson in the spectrum that appears to be of the order $\sigma$ (around 1~GeV$^2$ in the settings of Sec.~\ref{sec:finT}). Then the electromagnetic current in Eq.~\eqref{eq:F} can be approximated by a sum of two single-quark currents (the first diagram on the right-hand side in Fig.~\ref{fig:mesonphoton} plus a similar term for the photon coupled to the other quark line in the loop). The results of the calculations for the chiral pion and ``$\sigma$-meson'' performed in the Breit frame for the quarks in \cite{Glozman:2024dzz} are presented in Fig.~\ref{fig:radius3D}. From this figure one can conclude that both studied states (this conclusion also extends to the states with $J=1$) ``swell'' with the temperature rise above the chiral restoration temperature $\Tch$. Further details of the calculations and a relevant discussion can be found in
\cite{Glozman:2024dzz} while here we only quote an estimate made in the cited work. Namely, at $T=1.5\Tch$ and for the physical value of the current quark mass around few MeV, the lightest $J=0$ states in the spectrum increase their size from 3 to 5 times as compared to that at $T=0$. In other words, proceeding above $\Tch$, one enters a specific phase of the theory with the medium filled with large-size and, therefore, strongly overlapping and tightly packed chirally symmetric but confined hadrons that approximately fill the multiplets of a broad symmetry group. This is a distinct prediction of the GNJL model in Eq.~\eqref{GNJL} extended to finite temperatures as explained in Sec.~\ref{sec:finT}. Needless to say that these results as a matter of principle could not be obtained in the framework of potential quark models that do not incorporate quark spins as their intrinsic ingredients and, therefore, as a matter of principle can not address the effects related to chiral symmetry. It remains to be seen to what extent the above conclusions apply to real QCD that is free of the truncations and shortcomings inherent in GNJL. Lattice calculations are also supposed to shed light on this
problem.

\section{Conclusions}

In this review, we briefly summarised a considerable progress achieved through years and by many authors in the studies of the confined chiral quark model in Eq.~\eqref{GNJL}. In particular, it was discussed in detail how chiral symmetry in the vacuum of the model gets spontaneously broken, what microscopic mechanisms drive this effect, and what implications of the phenomenon of SBCS can be observed in the spectrum of hadrons.

As a next step, the model was generalised to finite temperatures in a way consistent with our present understanding of the thermal properties of QCD. We stress that a rigorous formulation of the model at $T>0$ can only be done in terms of the mesons as ``observable'' degrees of freedom, which \emph{inter alia} implies a systematic account for the $N_c$-suppressed contributions and as such appears to be an involved problem. Then, although a straightforward extension of the model to finites temperatures formulated in terms of dressed quarks meets severe difficulties, a simplified but phenomenologically successful extension is still possible and provides reasonable and potentially verifiable predictions for QCD in the chirally symmetric but confined phase above the chiral restoration temperature, if such a phase indeed exists in QCD.

At every intermediate step, the results obtained for the GNJL model were confronted with the insights that could be gained employing a simple potential quark model and essential differences between the two approaches and their predictions were pinpointed and discussed in detail. As a general conclusion, the confined chiral quark model is argued to be a powerful investigation tool and valuable
source of inspiration concerning various phenomena inherent, or believed to be inherent, in real QCD. Despite their multiple shortcomings, the chiral quark models of this type
still have good potential as a simple and tractable test ground for real QCD that allows one to gain a microscopic description of the studied phenomena and come up with the conclusions and predictions that can be further scrutinised employing more sophisticated exploration methods such as lattice calculations or real experiment.

\section*{Acknowledgments}

I would like to thank all my co-authors in the papers on various aspects of the quark model for a long-term fruitful collaboration and a lot of inspiration I got from them. I would also like to express my gratitude to Robert Wagenbrunn for providing the results of numerical calculations that go beyond the scope of our common publications \cite{Glozman:2024xll,Glozman:2024dzz} and Feng-Kun Guo  for the hospitality during my stay in Beijing where a large part of this review was written. This work was supported by Deutsche Forschungsgemeinschaft (Project No. 525056915) and the CAS President’s International Fellowship Initiative (PIFI) (Grants No.~2024PVA0004\_Y1).

\end{document}